\documentclass{article}
\usepackage[preprint]{neurips_2026}
\usepackage{amsmath}				% For Math
\usepackage{amssymb}
\usepackage{amsthm}
\usepackage{relsize}
\usepackage{mathtools}

\usepackage[utf8]{inputenc} % allow utf-8 input
\usepackage[T1]{fontenc}    % use 8-bit T1 fonts
\usepackage{hyperref}       % hyperlinks
\usepackage{url}            % simple URL typesetting
\usepackage{booktabs}       % professional-quality tables
\usepackage{amsfonts}       % blackboard math symbols
\usepackage{nicefrac}       % compact symbols for 1/2, etc.
\usepackage{microtype}      % microtypography
\usepackage{xcolor}         % colors

\usepackage{graphicx}				% For including figure/image
\usepackage{cancel}					% To use the slash to cancel out stuff in work
\usepackage{array} 
\usepackage{mdframed}
\usepackage{algorithm}
\usepackage{tablefootnote}
\usepackage{algpseudocode}
\usepackage{float}
\makeatletter
\newenvironment{breakablealgorithm}
  {% begin
   \begin{center}
     \refstepcounter{algorithm}% 新增算法編號
     \hrule height.8pt depth0pt \kern2pt% 上方橫線
     \renewcommand{\caption}[2][\relax]{%
       {\raggedright\textbf{\ALG@name~\thealgorithm} ##2\par}%
       \ifx\relax##1\relax
         \addcontentsline{loa}{algorithm}{\protect\numberline{\thealgorithm}##2}%
       \else
         \addcontentsline{loa}{algorithm}{\protect\numberline{\thealgorithm}##1}%
       \fi
       \kern2pt\hrule\kern2pt% 下方橫線
     }
  }{% end
     \kern2pt\hrule\relax
   \end{center}
  }
\makeatother
\usepackage{calligra}
\usepackage{braket}
\usepackage{calrsfs}
\usepackage[mathscr]{euscript}
\newtheorem{definition}{Definition} % [section]
\newtheorem{theorem}{Theorem} %  [section]
\newtheorem*{theorem*}{Theorem}
\newtheorem{lemma}{Lemma}[section]
\newtheorem{fact}{Fact}[section]
\newtheorem{remark}{Remark}
\newtheorem{assumption}{Assumption} 
\newtheorem{corollary}{Corollary} [theorem]
\newtheorem*{corollary*}{Corollary}
\newenvironment{proofsketch}{%
  \begin{proof}[Proof sketch]%
}{%
  \end{proof}%
}
\newcommand{\appendixsection}[1]{%
  \refstepcounter{section}%
  \section*{Appendix~\Alph{section}: #1}%
  \addcontentsline{toc}{section}{Appendix~\Alph{section}: #1}%
}
\DeclareMathOperator{\Tr}{Tr}
\newcommand{\email}[1]{\href{mailto:#1}{#1}}

\makeatletter
\def\BState{\State\hskip-\ALG@thistlm}
\makeatother
\begin{document}
\title{Efficient Matrix Product State Learning in Logarithmic Depth}

\author{
    Chia-Ying Lin\thanks{Department of Physics and Astronomy, Rice University, USA. Email: \email{cl207@rice.edu}.}  \and
	Nai-Hui Chia\thanks{Ken Kennedy Institute and Smalley-Curl Institute, Rice University, USA. Email: \email{nc67@rice.edu}.}  \and
	Shih-Han Hung\thanks{Department of Electrical Engineering and Center for Quantum Science and Engineering, National Taiwan University, Taiwan. Email: \email{shihhanh@ntu.edu.tw}.}
}
\ifdefined\authorNote
\newcommand{\cy}[1]{{\color{magenta} [{CYL:} #1]}}
\newcommand{\nai}[1]{{\color{magenta} [{Nai:} #1]}}
\else
\newcommand{\cy}[1]{}
\newcommand{\nai}[1]{}
\fi

\date{}

\maketitle

\begin{abstract}
    Learning the closest matrix product state (MPS) representation of a quantum state is known to enable useful tools for quantum machine learning and analysis of complex quantum systems.
    In this work, we study the problem of learning MPS in the following setting: given many copies of an input MPS, the task is to recover a classical description of the state. The best known polynomial-time algorithm, introduced by [LCLP10, CPF+10], requires linear circuit depth and $\widetilde O(n^5)$ samples, and has seen no improvement in over a decade. The combination of linear circuit depth and large sample complexity, neither known to be optimal, renders existing algorithms impractical for near-term quantum devices with limited resources.

    We introduce parallel disentangling algorithms for MPS learning. For exact MPS learning, our algorithm runs in polynomial time and uses circuit depth $O(\log n)$ and sample complexity $\widetilde O(n^3)$, improving both the depth and the dependence on the system size $n$. The key idea is to exploit the bounded-rank structure of reduced states on middle blocks of an MPS and organize the disentangling operations in a tree structure.

    We further extend the algorithm to closest MPS learning, improving the sample complexity dependence on $n$ from $n^9$ to $n^7$ and complement the algorithms with an $\Omega(n)$ product-state lower bound. We also investigate MPS learning under hardware constraints, including restricted measurements and geometric connectivity. Under the Learning Parity with Noise (LPN) assumption, we show computational hardness for learning an MPS(2) family with non-adaptive single-qubit measurements. Finally, we show that our algorithm can be implemented with depth $O(q n^{1/q})$ on a $q$-dimensional hypercubic lattice, giving an asymptotic reduction in depth. Together, our work provides a complete characterization of the quantum resources needed for efficient MPS learning.
    
    %We give an efficient algorithm for exact MPS learning with circuit depth  $O(\log n)$ and sample complexity $\widetilde O(n^3)$, improving the prior $O(n)$-depth and $\widetilde O(n^5)$ sample complexity. We also extend the framework to closest-MPS learning, improving the dependence on $n$ from $n^9$ to $n^7$. We complement our algorithms with an information-theoretic lower bound. We show that learning product states, and hence learning MPS, requires $\Omega(n)$ copies. 

    %We further study MPS learning under hardware constraints, including restricted measurements and geometric connectivity. Under the Learning Parity with Noise (LPN) assumption, we show hardness for high-fidelity learning of an MPS(2) family from non-adaptive single-qubit measurements, assuming the learner outputs an efficiently preparable hypothesis. Finally, we show that our algorithm can be implemented with depth $O(q n^{1/q})$ on a $q$-dimensional hypercubic lattice, giving an asymptotic reduction in depth. These results complement our algorithms and provide a complete characterization of the quantum resources needed for efficient MPS learning.

\end{abstract}

%\tableofcontents

\section{Introduction} 
Matrix product state is a fundamental family of quantum states defined as follows: 
\begin{definition}[Matrix Product State with Bond Dimension $D$ (MPS($D$))]\label{def:mps}
Let $n \in \mathbb{N}$ and let $d \in \mathbb{N}$ be the local dimension of each site (for qubits, $d=2$).  
An \emph{$n$-site matrix product state (MPS)\footnote{We adopt the periodic boundary condition (PBC) convention throughout, in which all local tensors are square matrices and the physical amplitudes are obtained by taking a trace over the virtual indices. Open-boundary MPS can be embedded into this framework by a standard padding construction.} of bond dimension $D$} is a pure state $\ket{\psi} \;=\; \sum_{i_1=0}^{d-1} \sum_{i_2=0}^{d-1} \cdots \sum_{i_n=0}^{d-1}
\mathrm{Tr}\!\left[ A^{[1]}_{i_1} A^{[2]}_{i_2} \cdots A^{[n]}_{i_n} \right]
\ket{i_1 i_2 \dots i_n}$ ,
where for each site $k \in \{1,\dots,n\}$ and each physical index $i_k \in \{0,\dots,d-1\}$,  
$A^{[k]}_{i_k}$ is a complex matrix of size $D \times D$.
The integer $D$ is called the \emph{bond dimension} of the MPS.
\end{definition} 
Matrix Product States (MPS) exhibit several noteworthy properties. First, when the bond dimension $D$ is small, an $n$-site qudit MPS requires only $O(dnD^2)$ parameters, in contrast to the $O(d^n)$ parameters needed for a general quantum state. This compressibility in description makes MPS a natural target for efficient learning, testing, and synthesis algorithms. Moreover, for any bipartition of the MPS with open boundary conditions, the Schmidt rank of $|\psi\rangle$ across the cut is at most $D$. This means the bond dimension directly characterizes the degree of entanglement: for instance, $D=1$ corresponds to a product state with no entanglement, while if the bond dimension equals the full Hilbert space dimension, the MPS representation can express all possible quantum states.

Matrix product states (MPS) play an increasingly important role not only in quantum many-body physics, but also in machine learning and quantum computing. In machine learning, MPS and closely related tensor-train models have been used as structured, parameter-efficient representations of high-dimensional data, with applications to neural-network compression, supervised learning, high-order feature modeling, and generative modeling
~\cite{novikov2015tensorizing,stoudenmire2016supervised, novikov2016exponential,han2018unsupervised,sengupta2022tensor}. They have also appeared in quantum machine learning as tensor-network architectures that can be mapped to variational quantum circuits
~\cite{rieser2023tensor}. From a quantum computing perspective, MPS provide a natural class of variational ansatz states for quantum algorithms, including ground-state energy estimation and combinatorial optimization~\cite{cirac2021matrix,wu2024variational,king2025beyond}. Their importance is also deeply rooted in quantum many-body physics. In particular, ground states of gapped one-dimensional local Hamiltonians are known to satisfy an entanglement area law and are therefore provably well approximated by matrix product states with bounded bond dimension
~\cite{hastings2007area,arad2012improved,arad2013area,
landau2015polynomial,soleimanifar2022testing}. This connection bridges physical Hamiltonian systems with computationally tractable descriptions, enabling efficient numerical and algorithmic methods for representing, preparing, and optimizing low-energy quantum states ~\cite{white1992density,daley2004time,vidal2004efficient,
verstraete2004density,orus2008infinite}. Beyond these applications, recent developments suggest that MPS and related tensor-network structures also intersect with quantum cryptography and quantum complexity theory~\cite{aaronson2022quantum}. Thus, understanding the sample complexity, circuit depth, and computational complexity of learning MPS is a timely and fundamental problem, with implications across machine learning, quantum algorithms, many-body physics, and quantum information theory.

In this work, we focus on learning the classical description of an input quantum state which is promised to be an MPS with a specified bond dimension. This problem is known as \emph{MPS learning}. From the perspective of learning theory, MPS learning is a quantum analogue of
learning a structured, low-dimensional representation of a high-dimensional
object. The unknown state lives in a Hilbert space of dimension exponential in
$n$, but the target representation has only $O(ndD^2)$ parameters. %Our results show that this representation can be recovered with polynomial sample complexity and logarithmic quantum depth, giving a resource-efficient learning procedure for a central tensor-network model. 
From a computational-complexity perspective, MPS provides a succinct representation for many quantum states with low entanglement; the relatively low parameter count, $O(dnD^2)$ for bond dimension $D$, admits the possibility of polynomial-time learning algorithms, in contrast to the exponential complexity of general quantum state learning. \emph{MPS learning} is also of significant practical and theoretical interest: for example, efficient algorithms or provable hardness results can be leveraged to certify whether the quantum hardware prepares correct quantum states, to construct cryptographic primitives, and to analyze properties of ground states and Hamiltonians.

The best known algorithms for MPS learning were introduced in~\cite{cramer2010efficient,landoncardinal2010efficient}. However, these algorithms are improper: the output MPS may have a bond dimension strictly larger than that of the input state, and the existence of efficient proper MPS learning algorithm is still an open question~\cite{anshu2023survey}. This means existing MPS learning algorithms cannot directly serve as testing algorithms.  The first efficient MPS testing algorithm was recently given in~\cite{soleimanifar2022testing}. Furthermore, in realistic scenarios where the provided quantum states are likely to be affected by noise due to hardware imperfections or environmental factors, there is growing interest in learning algorithms robust to the noise. Along this line,~\cite{bakshi2025learning} proposes an algorithm that can learn the closest MPS approximation in the presence of noise. Recently, an alternative line of work~\cite{qin2024quantum} studied quantum state tomography for matrix product operator (MPO) states using Haar-random projective measurements. Their main result shows that, given an \(n\)-qudit MPO of bond dimension \(r\), one can stably recover the ground-truth state from \( O(nd^2r^2)\) random measurement settings and a total of \( O(n^3 d^2 r^2 / \epsilon^2)\) samples via a constrained least-squares minimization, achieving \(\epsilon\)-closeness in the Frobenius norm. Although this establishes a theoretical possibility of learning MPS/MPO states with polynomial sample complexity, the approach relies on Haar-random basis measurements, which are not efficiently implementable on quantum hardware.

Despite the fundamental importance and broad applicability of MPS learning, our understanding of its computational complexity remains limited, and the algorithmic frontier has seen little progress over the past decade. The best-known algorithm uses $\widetilde O(n^5)$ copies of samples \cite{landoncardinal2010efficient}\footnote{The original work does not provide a detailed complexity analysis; we follow analysis in~\cite{bakshi2025learning} to get the complexity.}, while the strongest known lower bound, derived from product-state learning lower bounds, stands at only $\Omega(n)$\footnote{See Theorem~\ref{theorem:lowerbound} and Section ~\ref{sec:lowerbound} for a formal proof.}. This leaves a substantial and largely unexplored gap between the achievable upper and lower bounds. In addition, existing algorithms are typically formulated for matrix product states with open boundary conditions (OBC), and therefore do not directly extend to states with periodic boundary conditions. Moreover, all existing algorithms require linear circuit depth, which poses a severe obstacle for near-term quantum devices, where limited gate fidelity renders deep circuits prohibitively noisy. Bridging these gaps and designing shallow-depth, noise-resilient algorithms for MPS learning is therefore an open and pressing problem, carrying significant implications for both quantum algorithm design and practical deployment on emerging quantum hardware.

Along this line, we are driven by the following question:
\begin{center}
\emph{Can we design new MPS learning algorithms whose sample complexity substantially narrows the gap to the known lower bounds, while achieving shallow circuit depth?}
\end{center}

\paragraph{Our results.}
We answer the above questions affirmatively. We introduce a parallel disentangling algorithm for exact MPS learning, reducing the circuit depth from $O(n)$ to $O(\log n)$ and the sample complexity from $\widetilde O(n^5)$ to $\widetilde O(n^3)$. We also extend the algorithm to closest MPS learning, reducing the sample complexity from $\widetilde O(n^9)$ to $\widetilde O(n^7)$. Our algorithms improve the dependence on the system size, at the cost of a worse dependence on the bond dimension $D$; see Table~\ref{table:comparison}. This tradeoff is meaningful in many applications where MPS and tensor-network representations are useful for succinctly representing high-dimensional data because the bond dimension is small. We complement these algorithms with an $\Omega(n)$ sample-complexity lower bound.

Finally, we study MPS learning under hardware constraints. We show that our algorithm has depth $O(q n^{1/q})$ on a $q$-dimensional hypercubic lattice, asymptotically improving previous approaches. We also prove, assuming LPN, that no efficient algorithm can learn the MPS(2) family using only non-adaptive single-qubit measurements. Together, these results characterize the quantum resources needed for efficient MPS learning. Table~\ref{table:comparison} summarizes our results and comparisons with prior work.

%propose an algorithm that employs a circuit of depth $M=O(\log n)$ to learn a description of the target state, improving upon the previous linear-depth construction. In addition, our method reduces the sample complexity from $O(n^5)$ to $O(n^3)$ for exact MPS learning, and from $O(n^9)$ to $O(n^7)$ for closest MPS learning. Furthermore, the time complexity is poly$\left(D,n,\frac{1}{\epsilon},\log(\frac{1}{\delta})\right)$.  Moreover, our framework naturally handles both periodic and open boundary conditions, whereas most prior learning algorithms were formulated for OBC.

\begin{table}[t]
\centering
\small
\setlength{\tabcolsep}{4pt}
\renewcommand{\arraystretch}{1.15}
\resizebox{\linewidth}{!}{%
\begin{tabular}{|c|c|c|}
\hline
\textbf{Problem} & \textbf{Prior work} & \textbf{This work} \\
\hline
Exact MPS learning &
$\widetilde O(n^5D^2/\epsilon^4)$ samples, depth $O(n)$ ~\cite{cramer2010efficient,landoncardinal2010efficient}&
$\widetilde O(n^3D^6/\epsilon^4)$ samples, depth $O(\log n)$ (Theorem~\ref{theorem:1}) \\
%Better when $D=o(\sqrt n)$ \\
\hline
Closest-MPS learning &
$\widetilde O(n^9D^8/\epsilon^8)$ samples ~\cite{bakshi2025learning}&
$\widetilde O(n^7D^{12}/\epsilon^{12})$ samples (Theorem~\ref{theorem:2})\\
%Better in $n$, worse in $D,\epsilon$ \\
\hline
$2$-dimensional lattice &
depth $O(n)$ ~\cite{cramer2010efficient,landoncardinal2010efficient}&
depth $O(\sqrt n)$ (Theorem~\ref{thm:2d})\\
%Routing overhead included \\
\hline
$q$-dimensional hypercubic lattice &
depth $O(n)$ ~\cite{cramer2010efficient,landoncardinal2010efficient}&
depth $O(qn^{1/q})$ (Corollary~\ref{coro:higher-dim})\\
%Hypercubic lattice \\
\hline
Product-state lower bound &
-- &
$\Omega(nd/\epsilon)$ copies (Theorem~\ref{theorem:lowerbound})\\
%For product-state hypotheses \\
\hline
Restricted measurements &
-- &
LPN-hard for MPS$(2)$ (Theorem~\ref{theorem:single} and Corollary~\ref{coro:single})\\
%Efficiently preparable output \\
\hline
\end{tabular}%
}
\caption{Summary of the main results and comparisons to previous works. Here, \(n\) is the number of qudits, \(d\) is the local dimension, \(D\) is the bond dimension, and \(\epsilon\) is the target error.} %When \(D\) is constant or subpolynomial in \(n\), our improved \(n\)-dependence yields an asymptotic reduction in sample complexity.}
\label{table:comparison}
\end{table}

\section{Logarithmic-depth algorithms for MPS learning}

\begin{definition}[Quantum State Learning Problem]
Let $\mathcal{C}$ be a class of quantum states on $n$ qubits. 
The learner has access to independent copies of an unknown state $\rho \in \mathcal{C}$. 
The goal is to output, with high probability, a classical description of a hypothesis state $\hat{\rho}$ such that $ \mathrm{dist}(\rho, \hat{\rho}) \leq \epsilon,$
where $\epsilon > 0$ specifies the allowed error in the chosen distance measure $\mathrm{dist}(\cdot, \cdot)$. 
Typical examples include the trace distance, where the goal is 
\(\|\rho - \hat{\rho}\|_1 \leq \epsilon\), 
and the fidelity, where the goal is 
\(F(\rho, \hat{\rho}) \geq 1 - \epsilon\). \footnote{
Common accuracy metrics in quantum state learning include trace distance and fidelity. 
Fidelity \(F(\rho,\sigma)=\left(\mathrm{Tr}\sqrt{\sqrt{\rho}\sigma\sqrt{\rho}}\right)^2\)
measures state overlap, with larger values indicating greater similarity.}
\end{definition}

We focus on the class
$\mathcal{C}=\mathrm{MPS}(D),$
the set of $n$-qudit matrix product states with bond dimension at most $D$. We measure the quality of the output by fidelity. In particular, when the
algorithm outputs a pure hypothesis state $\ket{\hat{\phi}}$, the fidelity guarantee takes
the form
\(    
    \bra{\hat{\phi}}\rho\ket{\hat{\phi}} \geq 1-\epsilon   
.\)
\begin{theorem}
\label{theorem:1}
Given access to copies of an $n$-qudit matrix product state $\rho$ with bond dimension $D$ and parameters $\epsilon, \delta \in (0,1)$, Algorithm~\ref{alg:A2} outputs a description state $\ket{\hat{\phi}}$, such that, with probability $1-\delta$, $\bra{\hat{\phi}} \rho \ket{\hat{\phi}} \;\geq\; 1 - \epsilon.$
The algorithm requires $N=O(\frac{D^{6}\cdot d^4 \cdot n^3 \cdot \log(n/\delta)}{(\log_d D)^3\epsilon^4})$
copies of $\rho$ and runs in time $\text{poly}\left(D,n,\frac{1}{\epsilon},\log(\frac{1}{\delta})\right)$, and outputs $\ket{\hat{\phi}}$ through a quantum circuit of depth $O(\log n)$.
\end{theorem}

To prove Theorem~\ref{theorem:1}, we introduce the parallel disentangling MPS learning algorithm (Algorithm~\ref{alg:A2}). We provide a high-level proof sketch below and defer the complete proof to Appendix~\ref{appendix-correctnessA2}.

The previous algorithms~\cite{landoncardinal2010efficient,bakshi2025learning} exploit the fact that, for an $\mathrm{MPS}(D)$, any single cut (say between sites $l$ and $l+1$) induces reduced density matrices of rank at most $D$ on the two resulting segments $\{1,\dots,l\}$ and $\{l+1,\dots,n\}$. In contrast, if we isolate a middle block by making two cuts, then the reduced density matrix of that block has rank at most $D^2$. This allows us to go beyond disentangling only one leftmost qudit at a time. Indeed, as long as the block size $k$ satisfies $d^{k-1}\geq D^2$, we can apply disjoint disentangling unitaries to multiple blocks in parallel, thereby disentangling
the leftmost qudit of each block simultaneously.

Our parallel disentangling algorithms starts with the following two observations: (1) Any partition of the middle block and the rest has Schmidt rank at most $D^2$, which allows us to disentangle multiple qudits simultaneously within a single layer. (2) By organizing the disentangling procedure into a binary tree, the circuit depth can be reduced to $M = O(\log(n))$, instead of linear $O(n)$.

To preserve the rank-$D^2$ constraint in each layer, the blocks must be chosen consistently across layers: if a current-layer block contains any qudit inherited from a previous-layer block, then it must contain all qudits inherited from that block. Equivalently, no block from one layer is split across multiple blocks in the next layer. This ensures that, after the disentangling unitaries from earlier layers are applied, each current-layer block still satisfies the same Schmidt-rank constraint, since its qudits have not been mixed with the rest of the system. This consistency condition naturally induces a tree structure. Since the tomography cost scales with the block size, we use a binary tree to keep the blocks as small as possible, choosing block size \(2p\), where \(d^p \geq D^2\).

Due to the binary tree structure, the circuit depth is reduced from linear in \(n\) to
\(M=O(\log n)\). This makes the circuit significantly shallower, which is especially
important for near-term quantum devices where deep circuits are highly susceptible to noise. The tree structure also improves the error accumulation. At layer \(j\), the errors from the \(2^{M-j}\) disjoint blocks contribute \(2\sqrt{2^{M-j}\cdot 2\eta}\) to the error passed to the next layer. Therefore, the total accumulated fidelity error is bounded by
$
1 - F(\ket{\phi},\ket{\phi_{n-k+1}})
\le
2\sum_{j=1}^M \sqrt{2^{M-j}\cdot 2\eta}
=
O\!\left(\sqrt{2^M\cdot 2\eta}\right)
=
O\!\left(\sqrt{n\eta}\right).
$
Thus, to achieve final error at most \(\epsilon\), it suffices to choose the tomography precision \(\eta = O(\epsilon^2/n).\)
Using optimal tomography algorithms, the sample complexity for each block is then
\(O(n^2/\epsilon^4)\). Since the algorithm performs tomography on \(O(n)\)
blocks in total, the overall sample complexity becomes \(O(n^3/\epsilon^4)\).

%As a further remark, our framework also naturally accommodates matrix product states with both periodic and open boundary conditions\footnote{Since any OBC MPS can be viewed as a special case of a PBC MPS via padding additional zeros to extend rows in the first tensor $A^{[1]}_{i_1}$ and columns in the last tensor $A^{[n]}_{i_n}$, our setting generalizes the input class considered in previous work.}, whereas most prior learning algorithms were restricted to open boundary conditions.

\begin{breakablealgorithm}
\caption{Disentangling unitary construction}
\label{alg:A1}
\raggedright
\textbf{Input:}  A number $D^2$ and a description of a $y$-qudit density matrix $\hat{\sigma}\in \mathbb{C}^{d^y \times d^y}$ such that $d^y\geq D^2$.\\
\textbf{Output:} A description of a Disentangling Unitary $U$ for the state $\hat{\sigma}$.\\
\textbf{Procedure:}
\begin{algorithmic}[1]
\State Perform the spectral decomposition of $\hat{\sigma}$:
$$
\hat{\sigma} = \sum_{i=1}^{r} a_i \ket{\hat{\phi}_i}\bra{\hat{\phi}_i},
$$
where \(r\leq d^{y}\) and \( \{ \ket{\hat{\phi}_i}\}_{i=1}^{r} \) are orthonormal eigenvectors of \( \hat{\sigma} \) corresponding to non-zero eigenvalues \( a_1\geq a_2 \geq \dots \geq a_r \geq 0 \).
\State Let \(\ket{\phi_i} = \ket{\hat{\phi}_i}\) for $i=1,\dots,D^2$ and extend \( \{ \ket{\phi_i} \}_{i=1}^{D^2} \) to a complete orthonormal basis \( \{ \ket{\phi_i} \}_{i=1}^{d^{y}} \) of \( (\mathbb{C}^d)^{\otimes y} \) by choosing arbitrary orthonormal vectors \( \ket{\phi_i} \) for \( i = D^2+1, \dots, d^{y} \).
\State Set \(p \coloneq 2 \lceil \log_d{D}\rceil\).
\State Let \( \{ \ket{a_1, \dots, a_{y-p}} \}_{a_i \in \{0, \dots, d - 1\}} \) be the standard basis for the first \( y-p \) qudits, and let \( \{ \ket{j} \}_{j = 1}^{d^{p}} \) be an arbitrary orthonormal basis for the remaining \( p \) qudits. 
\State Define the unitary
$
U = \sum_{a_1, \dots, a_{y-p} = 0}^{d - 1} \sum_{j = 1}^{d^{p}}
\left( \ket{a_1, \dots, a_{y-p}} \otimes \ket{j} \right)
\left\langle \phi_{\mathrm{idx}(a_1, \dots, a_{y-p}, j)} \right|,
$
where the index mapping is defined by \(\mathrm{idx}
(a_1, \dots, a_{y-p}, j) = j + d^{p} \cdot \sum_{l = 1}^{y-p} a_l d^{y-p - l}.
\)
\State \Return the unitary $U$.
\end{algorithmic}
\end{breakablealgorithm}

\begin{breakablealgorithm}
\caption{Parallel disentangling MPS learning}
\label{alg:A2}
\raggedright
\textbf{Input:} Copies of an unknown matrix product states $\rho \in \mathbb{C}^{d^n \times d^n}$ with bond dimension \(D\), error parameter $\epsilon$ and failure probability $\delta$.\\
\textbf{Output:} A description of a quantum state $\ket{\hat{\phi}}\in (\mathbb{C^d})^{\otimes n}$. \\
\textbf{Procedure:}
\begin{algorithmic}[1]
\State Set \(p \coloneq 2 \lceil \log_d{D}\rceil\) and let \( M \) be the smallest positive integer $M'$ such that $2^{M'}p \geq n$.
\State Set $\ell_1 \coloneq \lceil\frac{1}{p} \left( n-2^{M-1}p \right)\rceil$, $s_1\coloneq \text{Mod}\left[n-2^{M-1}p,p\right]$, $k_1 \coloneq 2\ell_1 p-p+s_1$. 
\State Define the support $\mathcal{B}_i^1$ set for the $i$-th disentangling unitaries in the first layer:
\begin{equation}
\label{support}
\mathcal{B}_i^1 \coloneq 
\begin{cases}
\{2(i-1)p+1,\dots, 2ip\} & \text{if } 1\leq i<\ell_1 \\
\{2(\ell_1-1)p+1,\dots, k_1\} & \text{if } i = \ell_1\\
\{k_1+1+(i-\ell_1-1)p,\dots, k_1+(i-\ell_1)p\}& \text{if } \ell_1<i\leq 2^{M-1}
\end{cases}.
\end{equation}
\State Define a function 
$f(j,i)=\begin{cases}
s_1 & \text{when } j=1, i=\ell_1\\
0 & \text{when } j=1, i>\ell_1 \\
p & \text{otherwise}
\end{cases}.$
\State Let $(\rho^0)' = \rho$ and  $\mathcal{B}^1=\bigcup_{i=1}^{2^{M-1}}\mathcal{B}^1_i$.
\For {$i$ from $1$ to $\ell_1$}
    \State Let $\sigma^1_i = \mathrm{Tr}_{\mathcal{B}^1\setminus \mathcal{B}^1_i} [(\rho^0)']$ and let $\hat{\sigma}^1_i$ be the output of the tomography with error $\eta=\frac{(\sqrt{2}-1)^2\epsilon^2}{2^{M+5} }$ and failure probability $\delta/n$ on $O(\frac{D^2 \cdot d^{2p}\log(n/\delta)}{\eta^2})$ copies of $\sigma^1_i$.
    \State Use Algorithm \ref{alg:A1} to generate the disentangling unitary $U^1_i$ from a description of a $\left(p+f(1,i)\right)$-qudit state $\hat{\sigma}^1_i$ and an integer $D^2$.
    \State Let $\widetilde{\mathcal{B}}^1_i$ be the last $p$ qudits in $\mathcal{B}^1_i$.
\EndFor
\State Let $U^1=\bigotimes_{i=1}^{\ell_1}U^1_i$ and $P^1=\bigotimes_{i=1}^{\ell_1}\left( \ket{0^{f(1,i)}}\bra{0^{f(1,i)}}\right)_{\mathcal{B}^1_i\setminus \widetilde{\mathcal{B}}^1_i}\otimes I_{\widetilde{\mathcal{B}}^1_i}$.
\State Apply $U^1$ to all copies of $(\rho^0)'$ and project onto $P^1$ to get copies of
    $
        (\rho^1)' = \mathrm{Tr}_{\bigcup_{i=1}^{\ell_1}\mathcal{B}^1_i\setminus \widetilde{\mathcal{B}}^1_i} \left[ P^1 U^1 (\rho^0)' (U^1)^\dagger P^1 \right];
    $
\State Let $\widetilde{\mathcal{B}}^1_i=\mathcal{B}^1_i$ for $\ell_1<i\leq2^{M-1}$.
\For {$j$ from $2$ to $M$}
\State Let $\mathcal{B}^j=\bigcup_{i=1}^{2^{M-j+1}}\widetilde{\mathcal{B}}^{j-1}_i$.
    \For {$i$ from $1$ to $2^{M-j}$}
        \State Let $\mathcal{B}^j_i = \widetilde{\mathcal{B}}^{j-1}_{2i-1} \cup \widetilde{\mathcal{B}}^{j-1}_{2i} $ and $\sigma^j_i = \mathrm{Tr}_{\mathcal{B}^{j}\setminus \mathcal{B}^j_i} [(\rho^{j-1})']$.
        \State Let $\hat{\sigma}^j_i$ be the output of the tomography with error $\eta=\frac{(\sqrt{2}-1)^2\epsilon^2}{2^{M+5} }$ and failure probability $\delta/n$ on $O(\frac{D^2 \cdot d^{2p}\log(n/\delta)}{\eta^2})$ copies of $\sigma^j_i$.
        \State Use Algorithm \ref{alg:A1} to generate the disentangling unitary $U^j_i$ from a description of $2p$-qudit state $\hat{\sigma}^j_i$ and an integer $D^2$.
        \State Let $\widetilde{\mathcal{B}}^j_i$ be the last $p$ qudits in $\mathcal{B}^j_i$.
    \EndFor
    \State Let $U^j=\bigotimes_{i=1}^{2^{M-j}}U^j_i$ and $P^j=\bigotimes_{i=1}^{2^{M-j}}\left( \ket{0^{f(j,i)}}\bra{0^{f(j,i)}}\right)_{\mathcal{B}^j_i\setminus \widetilde{\mathcal{B}}^j_i}\otimes I_{\widetilde{\mathcal{B}}^j_i}$.
    \State Apply $U^j$ to all copies of $(\rho^{j-1})'$ and project onto $P^j$ to get copies of $(\rho^{j})' = \mathrm{Tr}_{\bigcup_{i=1}^{2^{M-j}}\mathcal{B}^j_i\setminus \widetilde{\mathcal{B}}^j_i} \left[ P^j U^j (\rho^{j-1})' (U^j)^\dagger P^j \right];$
\EndFor
\State Let $(\hat{\rho}^M)'$ be the output of the tomography with error $\tau=\epsilon/4$ and failure probability $\delta/n$ on $O(\frac{D^2 \cdot d^{2p}\log(n/\delta)}{\tau^2})$ copies of $(\rho^M)'$ and let $\ket{\hat{\psi}}$ be the top eigenvector of $(\hat{\rho}^M)'$.
\State \Return the state $\ket{\hat{\phi}}\coloneq(U^1)^\dagger \cdots (U^M)^\dagger \left( \bigotimes_{j=1}^{M}\bigotimes_{i=1}^{2^{M-j}} \ket{0^{f(j,i)}}_{\mathcal{B}^j_i\setminus \widetilde{\mathcal{B}}^j_i}
\otimes \ket{\hat{\psi}} \right)$.
\end{algorithmic}
\end{breakablealgorithm}

%\begin{remark}
%\label{remark:D}
%Our algorithm improves over previous ones when \(D=o(\sqrt n)\).\footnote{Prior work uses \(O(n^5D^2/\epsilon^4)\) samples, whereas our algorithm uses \(O(n^3D^6/\epsilon^4)\) samples. Thus, comparing the leading polynomial dependencies, our algorithm has better sample complexity whenever \(D=o(\sqrt n)\).}
%This includes many physically relevant cases where the bond dimension is constant or grows slowly with \(n\). For example, GHZ states, W states, AKLT states, and one-dimensional cluster states have exact constant-bond-dimension MPS representations. More generally, one-dimensional gapped ground states can be approximated to inverse-polynomial accuracy by MPS with subpolynomial bond dimension for fixed spectral gap~\cite{arad2013area}. Therefore, whenever the approximating bond dimension satisfies \(D=n^{o(1)}\), our sample complexity is asymptotically smaller.
%\end{remark}

We further consider the task of learning when the target state $\rho$ is not necessarily an exact MPS of bond dimension $D$, and may even be mixed. Our goal is to output a state $\ket{\hat{\phi}}$ such that its fidelity with $\rho$ is within additive error $\epsilon$ of the best achievable fidelity achievable  among all MPS of that bond dimension. This problem can be viewed as the agnostic version of MPS learning.

% Note that the setting here differs slightly from the standard notion of agnostic learning, where one assumes the existence of a state in the class of $\rm \text{MPS}(D)$ with fidelity at least $\delta$ with $\rho$, and the learner is allowed to use this promise in the design of the algorithm. Since we make no such assumption, we refer to our task as \emph{learning the closest MPS}. A similar formulation was also studied in \cite{bakshi2025learning}, where the authors employed a sequential unitary circuit of linear depth, following the approach of~\cite{landoncardinal2010efficient}.

\begin{definition}[Closest MPS Learning Problem]
Let $\mathrm{MPS}(D)$ denote the set of $n$-qudit pure matrix product states with bond dimension at most $D$. The learner is given access to independent copies of an arbitrary $n$-qudit state $\rho$, which is not promised to belong to $\mathrm{MPS}(D)$ and may be mixed. Given accuracy and confidence parameters $\epsilon,\delta\in(0,1)$, the goal is to output, with probability at least $1-\delta$, a classical description of a hypothesis state $\ket{\hat{\phi}}$ such that $\bra{\hat{\phi}}\rho\ket{\hat{\phi}}\geq\max_{\ket{\phi}\in \mathrm{MPS}(D)}\bra{\phi}\rho\ket{\phi}-\epsilon.$ 

% Equivalently, the learner outputs a state whose fidelity with $\rho$ competes with the
% best fidelity achievable by any bond-dimension-$D$ MPS.
\end{definition}

\begin{theorem}
\label{theorem:2}
Given access to copies of an $n$-qudit state $\rho$, as well as parameters $\epsilon , \delta\in (0,1) $, and a bond dimension parameter $D$. Algorithm~\ref{alg:A4} in Appendix~\ref{appendix-A4} outputs a description of a state $\ket{\hat{\phi}}$, such that, with probability $1-\delta$,
\(\bra{\hat{\phi}} \rho \ket{\hat{\phi}} \;\geq\; \max_{\ket{\phi}\in \rm \text{MPS}(D)}\bra{\phi} \rho \ket{\phi} - \epsilon\).
The algorithm requires \(N= 
O\!\Bigl(
\frac{ D^{12} n^{7}\, d^{4} (\ln d)^{7}\, \log\!\tfrac{n}{\delta}}
{\epsilon^{12}\! \left[
  \ln(\ln d)
  + \ln(nD^2/\epsilon^2)\right]^{7}}
\Bigr)
\) copies of $\rho$ and runs in time $\text{poly}\left(D,n,\frac{1}{\epsilon},\log(\frac{1}{\delta})\right)$, to output $\ket{\hat{\phi}}$.
\end{theorem}
\begin{proofsketch}

The proof follows the same learning framework as in Theorem~\ref{theorem:1}, with one additional truncation step to handle the fact that $\rho$ need not be an exact bond-dimension-$D$ MPS.  In the exact setting, the reduced
states on the relevant blocks have rank at most $D^2$; in the agnostic setting, they may
have full rank. Algorithm~\ref{alg:A4} therefore increases the block size to
$2p = 2\left\lceil \log_d(1/\eta) \right\rceil$
and truncates each reduced state to its dominant eigenspace above threshold $\eta$. Since the eigenvalues of a reduced density matrix sum to one, this dominant subspace has dimension at most $1/\eta$, giving an effective low-rank substitute for the exact MPS rank constraint. With a suitable choice of $\eta$ in terms of $\epsilon,n$, and $D$, the truncation and estimation errors remain controlled throughout the algorithm, yielding $\bra{\hat{\phi}}\rho\ket{\hat{\phi}}
    \geq
    \max_{\ket{\phi}\in\mathrm{MPS}(D)}
    \bra{\phi}\rho\ket{\phi}
    -\epsilon$.
The full proof is deferred to Appendix~\ref{appendix-correctnessA4}.
\end{proofsketch}

\begin{figure}[h]
\includegraphics[width=\textwidth]{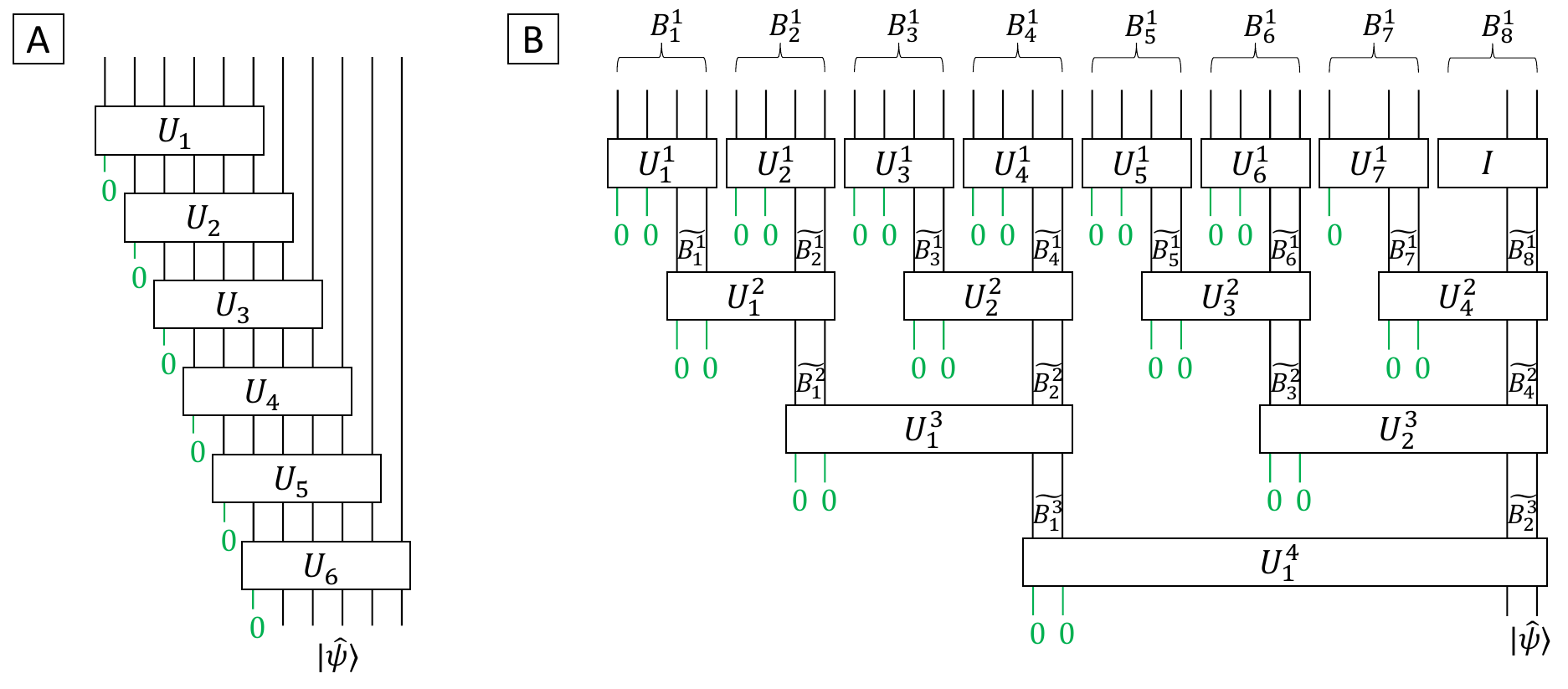}
\caption{
Comparison of circuit structures for MPS tomography.
(A) Previous algorithms~\cite{landoncardinal2010efficient,bakshi2025learning} use a sequential circuit structure.
(B) Our Algorithm~\ref{alg:A2} produces a tree-like circuit structure, shown here for \(n=29\) and \(p=2\).
Each rectangle represents a unitary \(U_i^j\), with \(j\) denoting the layer and \(i\) the unitary index within that layer.
The vertical lines are qudits ordered from left to right, and time flows downward.
The brackets mark the first-layer supports \(\mathcal{B}_i^1\), and the labeled wires mark the surviving sets \(\widetilde{\mathcal{B}}_i^j\) carried to the next layer.
}
\label{fig:compare-algo}
\end{figure}

In the case where the target state is an MPS($D$), the output state of our algorithm can be described by a circuit that consists of $O(n)$ unitaries, each of size $(d^2 D^4) \times (d^2 D^4)$. The total parameters in this circuit representation will be $O(nd^4 D^8)$. %Therefore, it has a succinct description, and one can generate the input state efficiently using this description.  
This is an improper learning algorithm, as in prior work: converting the circuit
description into an explicit one-dimensional MPS representation may increase
the bond dimension. Nevertheless, the output remains succinct and can be used
to efficiently prepare the hypothesis state.

%To further reconstruct the MPS representation---the matrices $A^{[1]}_{i_1},\ldots,A^{[n]}_{i_n}$ in Definition~\ref{def:mps}---from the quantum circuit of our algorithm. We can group the unitaries in the tree structure of different layers together to reconstruct matrices. This approach gives an MPS representation with bond dimension $d^{M+1}D^{2(M+1)}$, larger than that of the previous results; however, our algorithms provide better sample and time complexity and have achieved logarithmic circuit depth. \nai{remove this paragraph?}

% Also, as mentioned above, the circuit representation has the advantage of succinctness and efficient reproducibility as MPS representation. 

% If one wishes to improve the bond dimension further, we propose a second approach as follows: 
% Run our learning algorithm first with $O(n^3)$ copies of the input states, followed by using the learned circuit representation to prepare $O(n^5)$ copies states, and then apply the algorithm of~\cite{landoncardinal2010efficient} to our output. This approach requires only $O(n^3)$ copies of the input state and can obtain a bond dimension at most $dD$. However, since it also runs the algorithm in~\cite{landoncardinal2010efficient}, it uses more time and linear depth. 

\section{Implementation on Nearest-Neighbor Lattices}
The logarithmic-depth implementation of Algorithm~\ref{alg:A2} assumes an abstract circuit model in which the required qudits can be acted on jointly regardless of their geometric locations. In many realistic quantum architectures, however, gates are constrained by an underlying interaction geometry and can only be applied between neighboring qudits. It is therefore natural to ask how the depth of our construction changes under such locality constraints.

In this section, we show that the circuit generated by Algorithm~\ref{alg:A2} can still be implemented efficiently on nearest-neighbor architectures. In particular, when the qudits are arranged on a two-dimensional square lattice, the circuit admits an implementation with depth \(O(\sqrt{n})\). More generally, on a \(q\)-dimensional hypercubic lattice, the depth becomes \(O(q\,n^{1/q})\).

\begin{figure}[h]
\centering

\begin{minipage}[t]{0.5\textwidth}
\centering
\includegraphics[width=\textwidth]{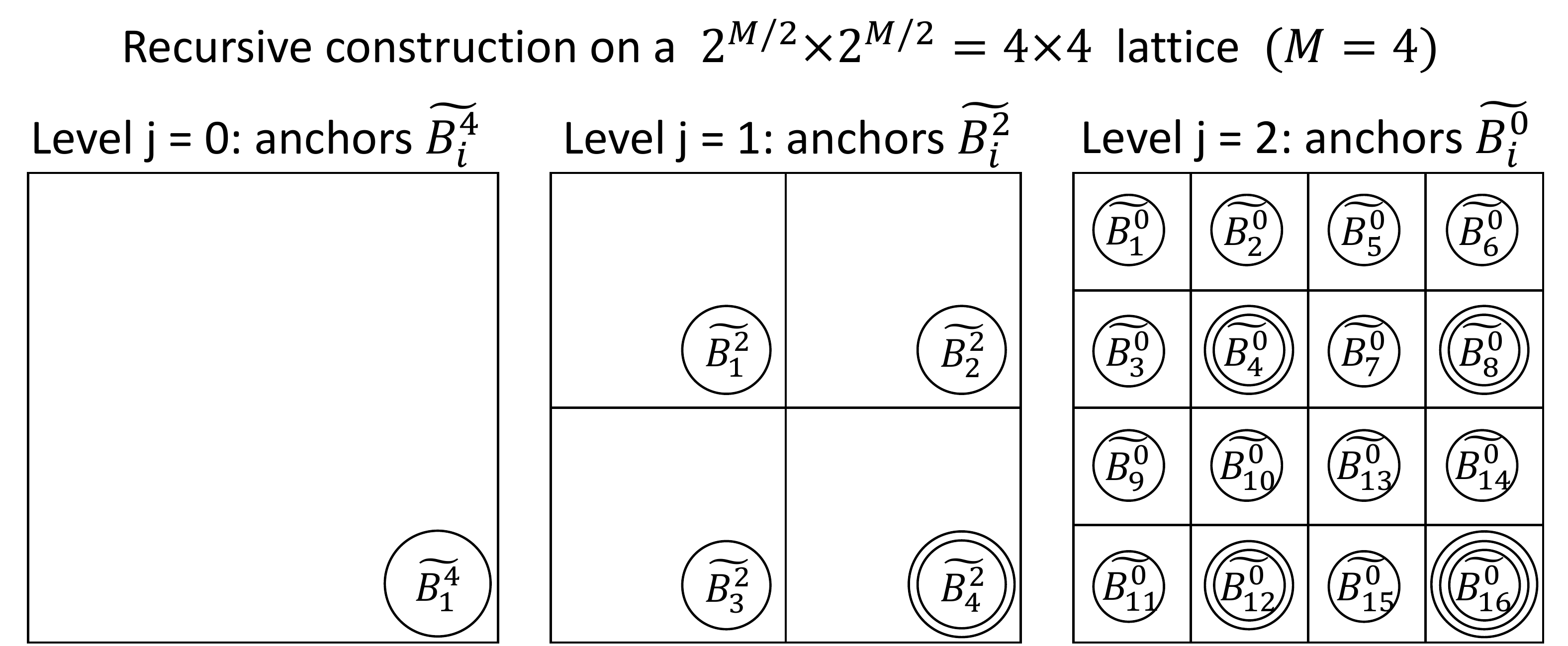}
\caption{Illustration of the recursive block structure on a $2^{M/2} \times 2^{M/2}$ square lattice with $M=4$. At level $j$, each block of size $2^{M/2-(j-1)} \times 2^{M/2-(j-1)}$ is partitioned into four subblocks, each of size $2^{M/2-j} \times 2^{M/2-j}$. In each subblock, the bottom-right site is designated as the anchor $\widetilde{\mathcal{B}}^{M-2j}_i$ (circled). These anchors form a coarser lattice for the next level.}
\label{fig:recursive-proof-2d}
\end{minipage}
\hfill
\begin{minipage}[t]{0.48\textwidth}
\centering
\includegraphics[width=\textwidth]{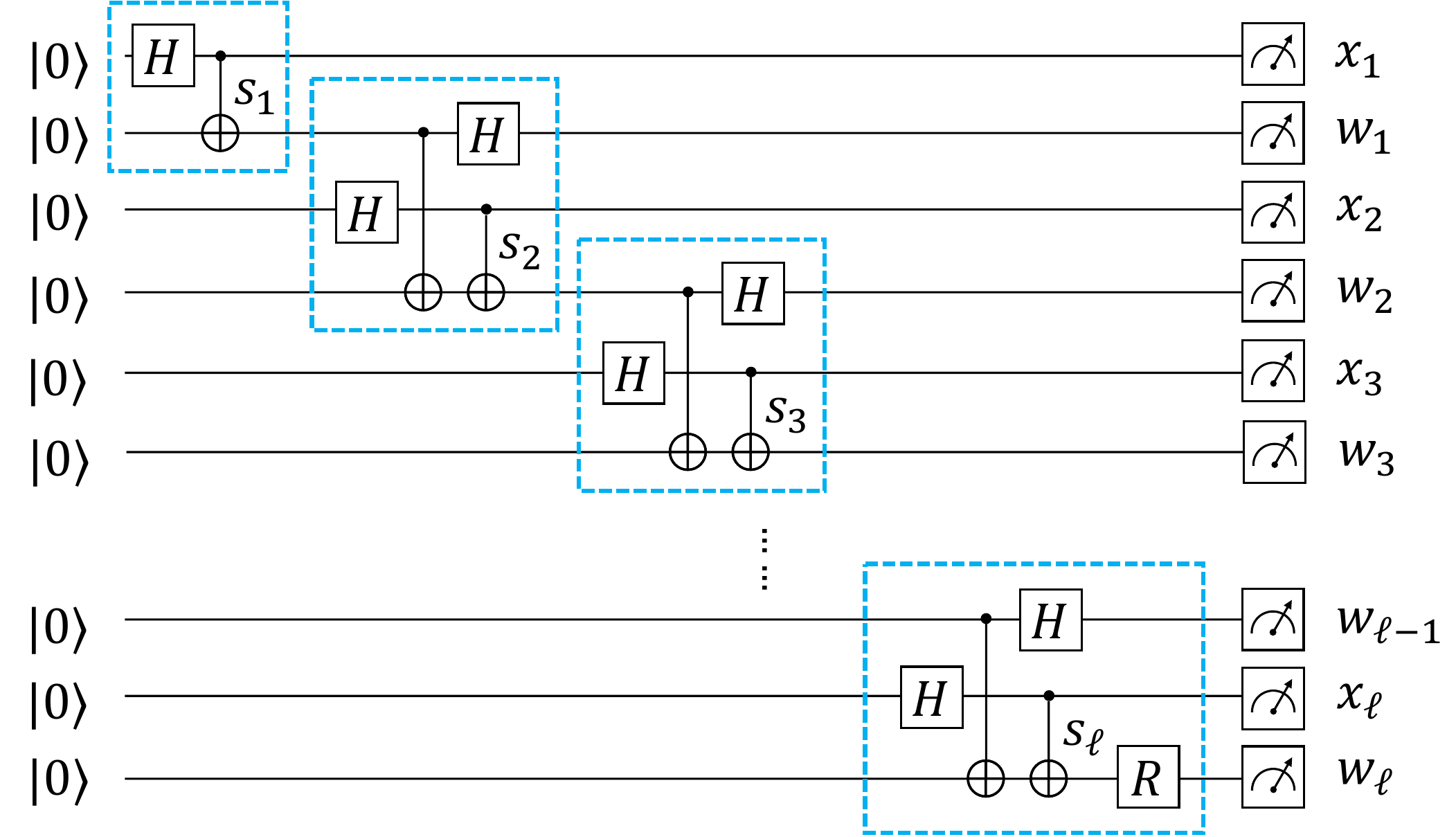}
\caption{Quantum circuit for generating an LPN-hard MPS(2) family. The odd-indexed qubits are initialized to $\ket{0}$ and put into a uniform superposition over $x=(x_1,\dots,x_\ell)\in\{0,1\}^\ell$ by Hadamard gates. The circuit then sequentially computes the prefix parities $z_i=\bigoplus_{j=1}^i x_j s_j$. For each $i<\ell$, a final Hadamard gate makes $w_i$ uniformly random and independent of the other measured outputs. On the last parity register, where $z_\ell=\langle x,s\rangle \pmod 2$, a rotation $R=R_y(2\arcsin(\sqrt\tau))$ is applied so that the measured bit satisfies $w_\ell=\langle x,s\rangle\oplus e$ with $e\sim \mathrm{Bernoulli}(\tau)$. Thus each measurement sample $(x,w_\ell)$ is distributed as a Search-LPN$_{\tau,\ell}$ sample.}
\label{Fig-single-qubit}
\end{minipage}
\end{figure}

\begin{theorem}
\label{thm:2d}
If quantum gates are restricted to nearest-neighbor interactions on a square lattice, the depth of the quantum circuit generated by Algorithm~\ref{alg:A2} is $O(\sqrt{n})$.
\end{theorem}
\begin{proofsketch}
View each patch of \(p\) qudits as one coarse-grained lattice site and arrange the \(n/p\) patches on a square lattice of side length \(\sqrt{n/p}\). As illustrated in Figure~\ref{fig:recursive-proof-2d}, we impose a recursive \(2\times 2\) block structure on the lattice, where in each subblock the bottom-right corner is chosen as the anchor that survives to the next coarser level. At each level, two successive iterations of Algorithm~\ref{alg:A2} are implemented by first applying the disentangling unitaries horizontally and then vertically, so that within every \(2\times 2\) group only the anchor site remains active. 
Under nearest-neighbor connectivity, before each such disentangling step we use nearest-neighbor SWAP gates to route the qubits within each block to bring the relevant pair of active sites adjacent; if the current block has side length \(L\), this requires depth \(O(L)\). Since all blocks at the same level are disjoint, these SWAP operations can be performed in parallel across blocks. Hence the coarse-grained depth is bounded by
$\sum_{j=1}^{O(\log(n/p))} O\!\left(\frac{\sqrt{n/p}}{2^{j-1}}\right)=O(\sqrt{n/p})$. Since each coarse-grained site is a physical patch of \(p\) qudits occupying area \(\Theta(p)\), routing within a patch contributes an additional factor \(O(\sqrt p)\). Thus the total physical depth is \(O(\sqrt n)\). We defer the full routing details to Appendix~\ref{appendix-higher-dim}.
\end{proofsketch}

\begin{corollary}
\label{coro:higher-dim}
Suppose quantum gates are restricted to nearest-neighbor interactions on a $q$-dimensional hypercubic lattice. Then the quantum circuit generated by Algorithm~\ref{alg:A2} can be implemented with depth $O\!\left(q\,n^{1/q}\right).$
\end{corollary}

\begin{proofsketch}
The proof is a direct \(q\)-dimensional analogue of Theorem~\ref{thm:2d}. View each patch of \(p\) qubits as one lattice site and arrange the \(n/p\) patches on a \(q\)-dimensional hypercubic lattice of side length \((n/p)^{1/q}\). We impose a recursive \(2^q\)-ary block structure: at each level, every block is partitioned into \(2^q\) subblocks, and the maximal-coordinate corner of each subblock is chosen as the anchor that survives to the next coarser level. Each level is implemented by applying the disentangling operations successively along the \(q\) coordinate directions, using nearest-neighbor SWAP gates to route active sites within each block; if the current block side length is \(L\), this costs depth \(O(qL)\), while disjoint blocks are processed in parallel. Summing over the geometrically decreasing side lengths gives $\sum_{j=1}^{O(\log(n/p))} O\!\left(q\,\frac{(n/p)^{1/q}}{2^{j-1}}p^{1/q}\right)
=
O\!\left(q\,n^{1/q}\right)$. The full proof is deferred to Appendix~\ref{appendix-higher-dim}.
\end{proofsketch}
%\begin{figure}[H]
%\includegraphics[width=0.8\textwidth]{Fig-2d-lattice}
%\caption{Illustration of the recursive block structure on a $2^{M/2} \times 2^{M/2}$ square lattice with $M=4$. At level $j$, each block of size $2^{M/2-(j-1)} \times 2^{M/2-(j-1)}$ is partitioned into four subblocks, each of size $2^{M/2-j} \times 2^{M/2-j}$. In each subblock, the bottom-right site (corresponding to the maximal-coordinate corner in each block) is designated as the anchor $\widetilde{\mathcal{B}}^{M-2j}_i$ (circled). These anchors form a coarser lattice for the next level.}
%\label{fig:recursive-proof-2d}
%\end{figure}

\section{Lower bounds and hardness for learning MPS}
We prove a sample-complexity lower bound for MPS learning by considering product states, namely MPS(1). To the best of our knowledge, this is the first lower bound of this form for the problem.

\begin{theorem}
\label{theorem:lowerbound}
For integer $n\geq 1$, $d\geq 2$, and $\epsilon\in[0,2/3]$, consider learning a product state $\rho \;=\; \bigotimes_{i=1}^n \rho_i$,
where each $\rho_i$ is a state on a $d$-dimensional Hilbert space. Any quantum algorithm that outputs a product state $\hat\rho$ satisfying $F(\rho,\hat\rho)\geq 1-\epsilon$ must use at least $\Omega(nd/\epsilon)$ copies of $\rho$. 
\end{theorem}
\begin{proofsketch}
    %Our proof is reducing state tomography to product state learning. Briefly, state tomography has a sample complexity lower bound XXX; we show that if we can learn product state with $o(nd/\epsilon)$ samples, then the state tomography lower bound will be violated.\nai{Polish} 
    It suffices to prove the lower bound for pure product states. We argue by reduction from single-qudit state tomography. The known tomography lower bound says that learning an arbitrary $d$-dimensional quantum state to fidelity at least $1-\epsilon'$ requires $\Omega(d/\epsilon')$ copies. If there were an algorithm that learned $n$-qudit product states to fidelity at least $1-\epsilon$ using $o(nd/\epsilon)$ copies, then by embedding a single unknown $d$-dimensional state into one tensor factor of an otherwise known product state, one could learn a single-qudit state to accuracy $\epsilon'=\epsilon/n$ using $o(d/\epsilon')$ copies. This would contradict the tomography lower bound, and therefore product-state learning requires $\Omega(nd/\epsilon)$ copies. We defer the formal proof to Appendix~\ref{sec:lowerbound}.
\end{proofsketch}

We next prove a computational hardness result for learning MPS under non-adaptive single-qubit measurements. We show that efficiently learning MPS(2) in this model would imply an efficient algorithm for Search-LPN (Definition~\ref{def:LPN}) and thus falsify the LPN assumption (Assumption~\ref{assumption:lpn}).

 \begin{theorem}
 \label{theorem:single}
Assume the $\mathrm{LPN}_{\tau,\ell}$ assumption holds for some constant 
$\tau \in (0,1/2-\gamma)$, where $\gamma>0$ is non-negligible. Let 
$\{|\psi_s\rangle : s\in\{0,1\}^{\ell}\}$ be the family of MPS$(2)$ states 
defined in Figure~\ref{Fig-single-qubit}. Then there is no polynomial-time algorithm 
that, using only computational-basis single-qubit measurements on 
$\operatorname{poly}(\ell)$ copies of $|\psi_s\rangle$, outputs a classical 
description of a hypothesis state $\widehat{\rho}$ from which $\widehat{\rho}$ 
can be efficiently prepared\footnote{To the best of our knowledge, all existing MPS learning algorithms, including those presented in this work, share this property. Specifically, they produce classical descriptions from which polynomial-sized quantum circuits can be efficiently generated to prepare the corresponding states} and $
F(\widehat{\rho},|\psi_s\rangle\langle\psi_s|)\ge 1-\epsilon$
with non-negligible probability.
 \end{theorem}
 \begin{proofsketch}
     By the construction in Figure~\ref{Fig-single-qubit}, measuring one copy of $\ket{\psi_s}$ in the computational basis produces a uniformly random $x\in\mathbb Z_2^\ell$ and uniformly random $w_1,w_2,\dots,w_{\ell-1}\in \mathbb Z_2$ together with a bit $w_\ell=\langle x,s\rangle\oplus e$, where $e\sim\mathrm{Bernoulli}(\tau)$. Note that, for each measured copy, the pair $(x,w_{\ell})$ gives exactly one Search-LPN$_{\tau,\ell}$ sample with secret $s$, while $w_1,w_2,\dots,w_{\ell-1}$ are uniformly random binary strings independent of $x$ and $w_{\ell}$. Then, we show that $s$ can be efficiently recovered from the classical description that efficiently prepares $\widehat{\rho}$. 
     Therefore, if a polynomial-time learner could recover $s$ from $q=\mathrm{poly}(\ell)$ such samples with non-negligible probability, then it would solve Search-LPN$_{\tau,\ell}$ with non-negligible probability. This contradicts the Search-LPN$_{\tau,\ell}$ assumption. Therefore no such learner exists. The formal proof is deferred to Appendix~\ref{sec:single-qubit}.
 \end{proofsketch}

Finally, we show the following Corollary and defer the proof to Appendix~\ref{sec:single-qubit}.
 \begin{corollary}
 \label{coro:single}
 Assume the $\mathrm{LPN}_{\tau,\ell}$ assumption holds for 
$\tau \in (0,1/2-\gamma)$, where $\gamma>0$ is non-negligible. Then there is no polynomial-time non-adaptive algorithm 
that, using only single-qubit measurements on 
$\operatorname{poly}(\ell)$ copies of $|\psi_s\rangle$, outputs a classical 
description of a hypothesis state $\widehat{\rho}$ from which $\widehat{\rho}$ 
can be efficiently prepared and such that $
F(\widehat{\rho},|\psi_s\rangle\langle\psi_s|)\ge 1-\epsilon$
with non-negligible probability.% for $\epsilon$ satisfying the same property of Theorem~\ref{theorem:single}
 \end{corollary}

\subsection{Discussion and open questions}
\label{sec:discussion}
%\nai{revise}We present a parallel disentangling MPS learning algorithm that, given a state promised to be an MPS with bond dimension \(D\), uses \(O(n^3 D^6/\epsilon^4)\) copies, runs in \(\poly(n,D)\) time, and has circuit depth \(O(\log n)\). By contrast, the state-of-the-art algorithms of \cite{landoncardinal2010efficient,cramer2010efficient} have depth \(O(n)\), run in \(\poly(n,D)\) time, and require \(O(n^5 D^2/\epsilon^4)\) copies. 

We introduce a parallel disentangling algorithm for exact MPS learning, reducing the circuit depth from $O(n)$ to $O(\log n)$ and the sample complexity from $\widetilde O(n^5)$ to $\widetilde O(n^3)$. We also extend the algorithm to closest MPS learning, reducing the sample complexity from $\widetilde O(n^9)$ to $\widetilde O(n^7)$. Our algorithms improve the dependence on the system size $n$~\cite{landoncardinal2010efficient,cramer2010efficient,bakshi2025learning}, at the cost of a worse dependence on the bond dimension $D$ (and $\epsilon$ for closest MPS learning). Our parallel disentangling algorithms run in polynomial time; for comparison, a sample-efficient but time-inefficient approach in \cite{qin2024quantum} achieves \(O(n^3 D^2/\epsilon^2)\) copies but relies on Haar-random measurements. We also establish a sample-complexity lower bound of \(\Omega(n/\epsilon)\) for learning product states. Together, our algorithms and lower bound provide the first systematic view of the sample-complexity landscape for efficient MPS learning.

% This tradeoff is natural in many applications where MPS and tensor-network representations are useful for succinctly representing high-dimensional data because the bond dimension is small.
 
The main tradeoff in our algorithms is that the improved dependence on the system size \(n\) comes at the cost of a worse dependence on the bond dimension \(D\). This tradeoff is meaningful from a learning-theoretic perspective: the bond dimension is a model-complexity parameter, and tensor-network methods are most useful in the low-complexity regime where $D$ is small compared with the full Hilbert-space dimension. Quantitatively, our algorithm wins previous ones whenever \(D=o(\sqrt n)\).\footnote{Prior work uses \(O(n^5D^2/\epsilon^4)\) samples, whereas our algorithm uses \(O(n^3D^6/\epsilon^4)\) samples. Thus, comparing the leading polynomial dependencies, our algorithm has better sample complexity whenever \(D=o(\sqrt n)\).} In particular, for any subpolynomial bond dimension \(D=n^{o(1)}\), our sample complexity is asymptotically smaller in \(n\). This is precisely the regime in which tensor-network methods serve as compact models for high-dimensional quantum states. It includes constant-bond-dimension families, such as GHZ states, W states, AKLT states, and one-dimensional cluster states, as well as one-dimensional gapped ground states that admit inverse-polynomial-accuracy MPS approximations with subpolynomial bond dimension for fixed spectral gap~\cite{arad2013area}.
% \begin{remark}
% \label{remark:D}
% Our algorithm improves over previous ones when \(D=o(\sqrt n)\).\footnote{Prior work uses \(O(n^5D^2/\epsilon^4)\) samples, whereas our algorithm uses \(O(n^3D^6/\epsilon^4)\) samples. Thus, comparing the leading polynomial dependencies, our algorithm has better sample complexity whenever \(D=o(\sqrt n)\).}
% This includes many physically relevant cases where the bond dimension is constant or grows slowly with \(n\). For example, GHZ states, W states, AKLT states, and one-dimensional cluster states have exact constant-bond-dimension MPS representations. More generally, one-dimensional gapped ground states can be approximated to inverse-polynomial accuracy by MPS with subpolynomial bond dimension for fixed spectral gap~\cite{arad2013area}. Therefore, whenever the approximating bond dimension satisfies \(D=n^{o(1)}\), our sample complexity is asymptotically smaller.
% \end{remark}

We also consider implementing our algorithms on quantum devices that have constraints on connectivity and measurements. In particular, we show that even with connectivity constraints, our algorithm can still outperform existing algorithms in depth and sample complexity; on the other hand, we showed that there exists no polynomial-time algorithm if the learner can only apply non-adaptive single-qubit measurement under the LPN assumption. These results provide a detailed characterization of needed quantum resources for MPS learning.

%Understanding whether such algorithmic primitives can be incorporated into quantum feature learning or certification of low-entanglement structure is an interesting future direction.

A gap still remains between our upper bound and lower bounds; closing it is the main open question. Promising directions include:

\begin{itemize}
  \item \textbf{Closing the gap in $n$.} Can we design an efficient algorithm with sample complexity \(O(n)\), or else prove an \(\Omega(n^3)\) lower bound?
  \item \textbf{Closing the gap in \(D\).} The best known lower bound is \(\Omega(\sqrt{D})\) via MPS testing~\cite{soleimanifar2022testing,aaronson2022quantum}. Can we obtain a tighter lower bound specific to MPS learning?
  \item \textbf{Proper learning.} All existing algorithms, including ours, fail to output a one-dimensional tensor-network representation with bond dimension exactly \(D\). Is there an efficient algorithm that achieves this, or can we prove a no-go theorem under plausible computational assumptions? (See also the open problem in~\cite{anshu2023survey}.)
\end{itemize} 

%Furthermore, we generalize our method to learn, for an arbitrary input state, the closest MPS of a specified bond dimension. This ``closest MPS'' learning algorithm runs in \(\poly(n,D)\) time and depth and uses \(O(n^7 D^{12}/\epsilon^{12})\) copies. Prior to our work, the best-known algorithm required \(O(n^9 D^8/\epsilon^8)\) copies~\cite{bakshi2025learning}. Thus, our algorithm improves the dependence on \(n\) while worsening the dependence on \(D\) and \(\epsilon\). On the lower-bound side, beyond our $\Omega(n)$ sample-complexity lower bound, it is known that learning the closest product state to within error smaller than \(n^{-4}\) is NP-hard~\cite{bakshi2025learning}. In light of these results, a natural goal is to obtain a more fine-grained characterization of the computational complexity of the problem.

Our algorithms have potential applications to quantum machine learning. MPS and related tensor-network models have been used as expressive but structured ansatz classes for high-dimensional data. In this light, logarithmic-depth MPS learning can be viewed as a step toward resource-efficient procedures for extracting compact tensor-network descriptions from quantum data.

\bibliographystyle{alpha}         % 其他樣式：ieeetr, unsrt, alpha, apalike, etc.
\bibliography{references}         % 不用加 .bib 副檔名

\appendix
\appendixsection{Background Knowledge} 
\label{section: preliminary}
\subsection{Matrix Product State}
\begin{definition}
Consider a quantum system consisting of $n$ sites, each associated with a local Hilbert space $\mathbb{C}^d$. A generic state in the global Hilbert space $(\mathbb{C}^d)^{\otimes n}$ can be written as
\[
\ket{\psi} = \sum_{i_1, i_2, \ldots, i_n = 1}^d 
    c_{i_1 i_2 \cdots i_n} \; \ket{i_1} \otimes \ket{i_2} \otimes \cdots \otimes \ket{i_n},
\]
where the coefficients $c_{i_1 i_2 \cdots i_n}$ are complex amplitudes. In the MPS representation, these amplitudes are expressed as products of matrices. Specifically, a state $\ket{\psi}$ is an MPS with bond dimension $D$ if
\[
c_{i_1 i_2 \cdots i_n}
= \mathrm{Tr}\!\left(A^{[1]}_{i_1} A^{[2]}_{i_2} \cdots A^{[n]}_{i_n}\right),
\]
where for each site $j$, the tensors $\{A^{[j]}_{i}\}_{i=1}^d$ are $D \times D$ complex matrices (with the first and last possibly rectangular if open boundary conditions are used). The parameter $D$ is referred to as the bond dimension of the MPS, and controls the amount of entanglement that the state can carry.
\end{definition}
MPS admit a natural graphical interpretation in terms of tensor networks. Each tensor $A^{[j]}$ is represented as a node with one physical index (corresponding to the local dimension $d$) and two virtual indices (corresponding to the bond dimension $D$). Connecting the virtual indices along the chain yields a one-dimensional tensor network, and contracting all virtual indices reproduces the global state. This representation makes it clear that, for a fixed bond dimension $D$, the number of parameters required to specify an MPS grows only linearly with the system size $n$, rather than exponentially as in the full Hilbert space. More precisely, each site tensor contains $d \cdot D^2$ complex entries (except at the boundaries for open boundary conditions), so a chain of $n$ sites requires approximately
\[
\text{Number of parameters} \sim n \, d \, D^2.
\]
This shows that the bond dimension $D$ controls both the expressiveness of the MPS and the amount of entanglement it can represent, while the linear scaling with $n$ allows efficient description of one-dimensional quantum states.
\begin{lemma}
\label{lemma: preliminary1}
Let $\ket{\psi}$ be a Matrix Product State (MPS) on $n$ sites with periodic boundary conditions and bond dimension $D$. For any block $\mathcal{B}_j$ of $k$ contiguous sites, the reduced density matrix
\[
\rho_j = \mathrm{Tr}_{[n] \setminus \mathcal{B}_j} (\ket{\psi}\bra{\psi})
\]
has rank at most $D^2$.
\end{lemma}

\begin{proof}
Let $\mathcal C := [n]\setminus\mathcal B_j$ denote the complement of the block. 
Represent the periodic MPS $\ket\psi$ by contracting local tensors around the ring. 
Cutting the ring at the two virtual bonds that border the contiguous block $\mathcal B_j$ exposes two virtual indices (one on each cut), denoted by $\alpha,\beta\in\{1,\dots,D\}$. 
By contracting all physical indices inside $\mathcal B_j$ while leaving the two boundary virtual indices open, we obtain vectors
\[
\{\,\ket{\Phi_{\alpha\beta}}_{\mathcal B_j}\in\mathbb H_{\mathcal B_j}:\alpha,\beta\in[D]\,\}.
\]
Similarly, contracting all tensors in the complement $\mathcal C$ (with the two virtual indices open) gives vectors
\[
\{\,\ket{E_{\beta\alpha}}_{\mathcal C}\in\mathbb H_{\mathcal C}:\alpha,\beta\in[D]\,\}.
\]
Thus the global state can be written as
\[
\ket\psi \;=\; \sum_{\alpha,\beta=1}^{D} \ket{\Phi_{\alpha\beta}}_{\mathcal B_j}\otimes\ket{E_{\beta\alpha}}_{\mathcal C}.
\]

Taking the partial trace over $\mathcal C$ yields
\[
\rho_j
=\mathrm{Tr}_{\mathcal C}(\ket\psi\bra\psi)
=\sum_{\alpha,\beta,\alpha',\beta'=1}^{D}
\braket{E_{\beta'\alpha'}|E_{\beta\alpha}}\;
\ket{\Phi_{\alpha\beta}}_{\mathcal B_j}\!\bra{\Phi_{\alpha'\beta'}}.
\]
Therefore, the support of $\rho_j$ is contained in 
$\mathrm{span}\{\ket{\Phi_{\alpha\beta}}:\alpha,\beta\in[D]\}$, 
which has dimension at most $D^2$. 
Hence
\[
\mathrm{rank}(\rho_j)\le D^2,
\]
as claimed.
\end{proof}

\begin{lemma}\label{lemma:preliminary2}
Let $[n]=\{1,2,\ldots,n\}$ denote the set of sites, and let $\mathcal{B}_j\subseteq[n]$ be a contiguous block of qudits. Denote by $\mathbb{H}_{\mathcal{B}_j}$ and $\mathbb{H}_{[n]\setminus\mathcal{B}_j}$ the Hilbert spaces associated with $\mathcal{B}_j$ and its complement, respectively. 
Let $\ket{\psi}\in\mathbb{H}_{\mathcal{B}_j}\otimes\mathbb{H}_{[n]\setminus\mathcal{B}_j}$ be a Matrix Product State (MPS) on $n$ sites with periodic boundary conditions and bond dimension $D$.
Let $U_{\mathcal{B}_j}$ and $U_{[n]\setminus\mathcal{B}_j}$ be unitary operators acting on $\mathcal{B}_j$ and its complement, respectively. Then
\[
\rho'_{\mathcal{B}_j}
:=\mathrm{Tr}_{[n]\setminus\mathcal{B}_j}\!\left[(U_{\mathcal{B}_j}\otimes U_{[n]\setminus\mathcal{B}_j})\,\ket{\psi}\bra{\psi}\,(U_{\mathcal{B}_j}^\dagger\otimes U_{[n]\setminus\mathcal{B}_j}^\dagger)\right]
\]
has rank at most $D^2$.
\end{lemma}

\begin{proof}
Write $\mathcal{C}:=[n]\setminus\mathcal{B}_j$. For any operator $X$ on $\mathbb{H}_{\mathcal B_j}\otimes\mathbb{H}_{\mathcal C}$, the partial trace satisfies
\begin{align}
\mathrm{Tr}_{\mathcal C}\!\big[(I_{\mathcal B_j}\otimes U_{\mathcal C})X(I_{\mathcal B_j}\otimes U_{\mathcal C}^\dagger)\big]
&=\mathrm{Tr}_{\mathcal C}[X], \label{eq:pt-env-inv}\\
\mathrm{Tr}_{\mathcal C}\!\big[(U_{\mathcal B_j}\otimes I_{\mathcal C})X(U_{\mathcal B_j}^\dagger\otimes I_{\mathcal C})\big]
&=U_{\mathcal B_j}\,\mathrm{Tr}_{\mathcal C}[X]\,U_{\mathcal B_j}^\dagger. \label{eq:pt-sim}
\end{align}
Applying \eqref{eq:pt-env-inv}--\eqref{eq:pt-sim} to $X=\ket{\psi}\bra{\psi}$ yields
\[
\rho'_{\mathcal B_j}
=U_{\mathcal B_j}\,\rho_{\mathcal B_j}\,U_{\mathcal B_j}^\dagger,\qquad
\rho_{\mathcal B_j}:=\mathrm{Tr}_{\mathcal C}\big[\ket{\psi}\bra{\psi}\big].
\]
Hence $\mathrm{rank}(\rho'_{\mathcal B_j})=\mathrm{rank}(\rho_{\mathcal B_j})$. By Lemma~\ref{lemma: preliminary1}, we know that $\rho_{\mathcal B_j}$ has rank at most $D^2$. Since unitary conjugation preserves rank, it follows that
\[
\mathrm{rank}(\rho'_{\mathcal B_j})
=\mathrm{rank}(\rho_{\mathcal B_j})
\le D^2.
\]
\end{proof}

\subsection{Quantum State Learning Problem Statement}

The task of learning quantum states can be phrased in a general framework. 
A learning algorithm, often referred to as the \emph{learner}, is provided with multiple identical copies of an unknown quantum state $\rho$. 
It is assumed that $\rho$ belongs to a known family of states $\mathcal{C}$, which specifies the scope of the learning task. 
By performing measurements, the learner gathers information about $\rho$ and aims to produce a classical description of a hypothesis state $\hat{\rho}$ that approximates $\rho$ with respect to a chosen accuracy measure. 
The formulation naturally involves three components: the class of candidate states $\mathcal{C}$, the measurement strategies available to the learner, and the metric used to assess accuracy. 
Performance is quantified in terms of \emph{sample complexity}, the number of copies of $\rho$ that are required, and \emph{time complexity}, the total amount of quantum or classical computation needed.

\begin{definition}[Quantum State Learning Problem]
Let $\mathcal{C}$ be a class of quantum states on $n$ qubits. 
The learner has access to independent copies of an unknown state $\rho \in \mathcal{C}$. 
The goal is to output, with high probability, a classical description of a hypothesis state $\hat{\rho}$ such that
\[
    \mathrm{dist}(\rho, \hat{\rho}) \leq \epsilon ,
\]
where $\epsilon > 0$ specifies the allowed error in the chosen distance measure $\mathrm{dist}(\cdot, \cdot)$. 
Typical examples include the trace distance, where the goal is 
\(\|\rho - \hat{\rho}\|_1 \leq \epsilon\), 
and the fidelity, where the goal is 
\(F(\rho, \hat{\rho}) \geq 1 - \epsilon\).
\footnote{
Two commonly used metrics for accuracy in quantum state learning are the \emph{trace distance} and the \emph{fidelity}. The trace distance is defined as \(\|\rho - \sigma\|_1 = \mathrm{Tr}\!\left[ \sqrt{(\rho - \sigma)^\dagger (\rho - \sigma)} \right],\) and quantifies the maximal statistical distinguishability between $\rho$ and $\sigma$. 
A smaller trace distance indicates higher similarity.

The fidelity is defined as \(F(\rho, \sigma) = \left( \mathrm{Tr}\!\left[ \sqrt{\sqrt{\rho}\, \sigma \sqrt{\rho}} \,\right] \right)^2 ,\) and measures the overlap between two quantum states.  A larger fidelity indicates higher similarity. These two quantities are closely related by the Fuchs--van de Graaf inequalities: \(1 - \sqrt{F(\rho, \sigma)} \;\leq\; \tfrac{1}{2}\|\rho - \sigma\|_1 \;\leq\; \sqrt{1 - F(\rho, \sigma)} .\)
Hence, bounding the trace distance and bounding the fidelity are essentially equivalent up to constant factors.}
\end{definition}

The most general setting corresponds to \emph{full quantum state tomography}, where the class of candidate states $\mathcal{C}$ is the set of all $n$-qubit states, and the goal is to reconstruct an unknown $\rho \in \mathcal{C}$ up to small trace distance. Because the dimension of the Hilbert space grows exponentially with $n$, this task requires exponentially many copies of the state in the worst case. The optimal bounds for the sample complexity of Quantum State Tomography is finally obtained by \cite{haah2016sample} and  \cite{o2016efficient}. 
\begin{theorem}[Theorem 1 in \cite{anshu2024survey}]
The sample complexity of quantum state tomography up to trace distance $\delta$ is \(O\!\left(\frac{4^n}{\delta^2}\right).\)
If the state is promised to have rank $r$, then the sample complexity of quantum state tomography up to infidelity\footnote{The infidelity is defined as $1-F$.}
$\epsilon$ is \(\widetilde{\Theta}\!\left(\frac{2^n r}{\epsilon}\right).\)
\end{theorem}

This limitation motivates the study of whether more structured families of states can be learned efficiently. Here, one can consider restricting the class of candidate states $\mathcal{C}$ to specific families, such as stabilizer states, $t$-doped stabilizer states, product states, matrix product states, or Gibbs states at certain temperature regimes, with the goal of reconstructing an unknown $\rho \in \mathcal{C}$ up to small trace distance (or high fidelity).

In this work, we focus on learning matrix product states (MPS) with a given bond dimension.
Following the framework of the general Quantum State Learning Problem, we define the MPS learning problem as follows.

\begin{definition} [Matrix Product State Learning Problem]
Let $\text{MPS}(D)$ denote a class of $n$-qubit states that can be represented as matrix product states of bond dimension at most $D$. 
The learner has access to independent copies of an unknown state $\rho \in \text{MPS}(D)$. 
The goal is to output, with high probability, a classical description of a hypothesis state $\ket{\hat{\phi}}$ such that
\[\bra{\hat{\phi}}\rho\ket{\hat{\phi}}\ge 1-\epsilon\]
The learner's performance is measured in terms of the number of copies of $\rho$ used (\emph{sample complexity}) 
and the total computational cost (\emph{time complexity}).
\end{definition}

\subsection{State Tomography}
\begin{lemma}[Rank-constrained sub-normalized tomography]\label{lemma:subnorm-tomo-rank}
Let $\rho$ be a quantum state on $r$ qudits of local dimension $d$, and suppose that $\mathrm{rank}(\rho) \le D^{2}$.  Let $\Pi = \ket{0^i}\!\bra{0^i}\otimes I^{\otimes (r-i)}$ and define $\mu = \mathrm{Tr}[\Pi\rho]$.  Let $\eta,\delta > 0$. Then there exists an algorithm that, using 
\[O\!\left(\frac{\mu\,D^{2}\,d^{\,r-i}}{\eta^{2}}
\log\!\frac{1}{\delta}
\right)\]
copies of $\rho$, produces an estimate $\hat\sigma$ of $\sigma := \Pi\rho\Pi$ such that $\|\sigma - \hat\sigma\|_{1} \le \eta$ with failure probability at most $\delta$. Moreover, the algorithm runs in time $\mathrm{poly}(d^{\,r-i},1/\eta,\log(1/\delta))$.
\end{lemma}
\begin{proof}
Take $m$ copies of $\rho$ and perform the two-outcome measurement $\{\Pi, I-\Pi\}$ on each copy. The number $m'$ of successful outcomes $\Pi$ is distributed as $\mathrm{Binomial}(m,\mu)$, with expectation $\mu m$. Each successful trial produces the normalized state $\rho|_\Pi := \Pi\rho\Pi/\mu$. We estimate $\sigma = \mu\,\rho|_\Pi$ by \(\hat\sigma := \tfrac{m'}{m}\,\hat{\rho}|_\Pi\), where $\hat{\rho}|_\Pi$ is an estimate of $\rho|_\Pi$. Then
\[
\|\sigma - \hat\sigma\|_1
= \mu \big\| \rho|_\Pi - (m'/(\mu m))\,\hat{\rho}|_\Pi\big\|_1
\le \mu\Big( \|\rho|_\Pi - \hat{\rho}|_\Pi\|_1
+ |1 - m'/(\mu m)|\Big).
\]
To ensure $\|\sigma-\hat\sigma\|_1 \le \eta$, it suffices that $\|\rho|_\Pi - \hat{\rho}|_\Pi\|_1 \le \eta/(2\mu)$ and $|1 - m'/(\mu m)| \le \eta/(2\mu)$, each holding with failure probability at most $\delta/2$.

For the first requirement, the algorithm then traces out the first $i$ qudits and performs tomography on the remaining $(r-i)$-qudit states with error $\eta/(2\mu)$ and failure probability $\delta/2.$ Note that the state on which we perform tomography has rank at most $D^{2}$. By the result of O’Donnell and Wright \cite{o2016efficient}, there exists an algorithm that, with failure probability at most $\delta/2$, produces an estimate $\hat{\rho}|_\Pi$ within trace distance $\eta/(2\mu)$ using
\[
O\!\left(
\frac{D^{2}\, d^{\,r-i}\,\mu^{2}}{\eta^{2}}
\log\frac{1}{\delta}
\right)
\]
post-selected copies. Therefore, it suffices that $m' \ge c\, D^{2} d^{\,r-i}\mu^{2}\log(1/\delta)/\eta^{2}$ for some constant $c$. For the second requirement, Hoeffding’s inequality yields
$$\Pr\!\left[\,\big|m'-\mu m\big|\ge \tfrac{\eta}{2} m \,\right]
\;\le\; 2\exp\!\left(-\tfrac{\eta^{2}}{2} m\right).$$
Hence, if $m = \Omega\!\big(\eta^{-2}\log(1/\delta)\big)$, then with probability at least $1-\delta/2$ we have $|1-m'/(\mu m)| \le \eta/(2\mu)$.  

Combining the two requirements and applying a union bound, we find that the trace-norm error is at most $\eta$ except with probability $\delta$. Since in expectation $\mu m$ copies survive the measurement $\Pi$, we need to start with $m=O\!\left(\frac{\mu\,D^{2}\,d^{\,r-i}}{\eta^{2}}\log\!\frac{1}{\delta}
\right)$ copies of $\rho$. Finally, the reconstruction can be implemented in $\mathrm{poly}(d^{\,r-i},1/\eta,\log(1/\delta))$ time, establishing the claimed runtime bound.
\end{proof}

\begin{lemma}[Sub-normalized tomography]
\label{lemma:subnorm-tomo-rank-agnostic}
Let $\rho$ be a quantum state on $r$ qudits of local dimension $d$. Let $\Pi = \ket{0^i}\!\bra{0^i}\otimes I^{\otimes (r-i)}$ and define $\mu = \mathrm{Tr}[\Pi\rho]$. Let $\eta,\delta > 0$. Then there exists an algorithm that, using 
\[O\!\left(\frac{\mu\,d^{2(r-i)}}{\eta^{2}}
\log\!\frac{1}{\delta}
\right)\]
copies of $\rho$, produces an estimate $\hat\sigma$ of $\sigma := \Pi\rho\Pi$ such that $\|\sigma - \hat\sigma\|_{1} \le \eta$ with failure probability at most $\delta$. Moreover, the algorithm runs in time $\mathrm{poly}(d^{\,r-i},1/\eta,\log(1/\delta))$.
\end{lemma}
\begin{proof}
The argument follows the same structure as the previous proof, except that we now invoke the sample complexity result in \cite{o2016efficient} for general state without any rank constraint: for arbitrary states $\rho \in \mathbb{C}^{d\times d}$, given parameters $\epsilon,\delta \in (0,1)$, their result guarantees that \(O\!\left(\frac{ d^{2}}{\eta^{2}}\log\frac{1}{\delta}
\right)\) copies suﬃce to obtain an $\epsilon$-accurate estimate through the state tomography in the standard trace distance with failure probability at most $\delta$. This suffices to establish the claim.
\end{proof}

\appendixsection{Correctness of Algorithm \ref{alg:A2}}
\label{appendix-correctnessA2}
In this section, we prove the correctness of our algorithm. Let $\rho$ be an $n$-qudit unknown state that is guaranteed to be a matrix product state of bond dimension $D$. Taking $\rho$ as input, Algorithm \ref{alg:A2} aims to learn and output a description of a state $\ket{\hat{\phi}}$ that is sufficiently close to $\rho$.

To analyze the correctness of the algorithm, we rely on several structural properties of matrix product states. In particular, due to the binary tree construction in Algorithm~\ref{alg:A2}, each reduced density matrix $\sigma^j_i$ is guaranteed to have rank at most $D^2$, as established in Lemma \ref{lemma: preliminary1} and\ \ref{lemma:preliminary2}.  These facts will serve as the foundation for the sequence of lemmas following Lemma \ref{lemma:tomography-error}. We set 
\[
p \coloneq 2 \lceil \log_d D \rceil
\]
to account for the number of qudits required to support a density matrix of rank at most $D^2$. This strategy in turn reduces the sample complexity required to learn a description of the unknown MPS($D$). In contrast, for the agnostic setting—where the input is not assumed to be an MPS and may even be a mixed state—we will later adopt a different strategy.

\begin{definition}[Step-wise Reconstructed State]
\label{definition:rhoj}
Let $(\rho^0)'=\rho$ be the initial state before any iteration (see line 8 in Algorithm \ref{alg:A2}). Define
\[
\rho^{0} \coloneq (\rho^0)' = \rho.
\]For each $j = 1, \dots, M$, define the cumulative unitary
\[
E^{j} \coloneq U^{j} U^{j-1} \cdots U^{1}.
\]
Let $(\rho^j)'$ denote the subnormalized post-measurement state of the remaining qudits after \(j\) iterations of Algorithm~\ref{alg:A2}. Define
\[
\rho^{j} \coloneq (E^{j})^{\dagger} \left[ \bigotimes_{a=1}^{j} P^a \otimes (\rho^j)' \right] E^{j},
\]
where $P^a$ and $U^j$ are as in Algorithm \ref{alg:A2}. 

We call $\rho^j$ the \emph{step-wise reconstructed state} corresponding to the $j$-th layer of the circuit, representing the state that would have been reconstructed had the algorithm stopped after the $j$-th iteration (i.e., line 20).
\end{definition}

\begin{definition}[Level-$j$ transformed and projected states]
\label{definition:psij}
Fix \(j\in\{1,\dots,M\}\). For any pure state \(\ket{\phi}\) on an \(n\)-qudit system define the \emph{level-$j$ transformed state}
\[
    \ket{\phi^{j}} \coloneqq E^j \ket{\phi},
\]
where \(E^j\) is the cumulative unitary introduced above. Define the \emph{level-$j$ projector}
\[
    \Pi^{(j)} \coloneqq \bigotimes_{a=1}^{j} P^a,
\]
with \(P^a\) as in Algorithm~\ref{alg:A2}. The \emph{level-$j$ projection} of \(\ket{\phi^{j}}\) is \(\Pi^{(j)}\ket{\phi^{j}}\). 

\(\Pi^{(j)}\ket{\phi^{j}}\) has the product form
\[
    \Pi^{(j)}\ket{\phi^{j}} 
    = \left( \bigotimes_{a=1}^j \bigotimes_{b=1}^{2^{M-a}} \ket{0^{f(a,b)}}_{\mathcal{B}^a_b\setminus\widetilde{\mathcal{B}}^a_b} \right)
      \otimes \ket{\psi^j},
\]
we call \(\ket{\psi^j} \in \mathbb{H}_{\mathcal{\widetilde{B}}^j_1}\otimes\dots\otimes\mathbb{H}_{\widetilde{\mathcal{B}}^j_{2^{M-j}}}\) the \emph{projected (residual) state of \(\ket{\phi}\) at level \(j\)}.  By convention, we define the projected state at level $0$ to be the input state itself, \(\ket{\psi^0} \coloneq \ket{\phi}.\)
\end{definition}

\begin{fact}
\label{fact:fidelity-rep}
For any state $\ket{\phi}$ and for each $j= 0,\dots, M,$ we have $\bra{\phi}\rho^{j}\ket{\phi}=\bra{\psi^{j}}(\rho^{j})'\ket{\psi^{j}}$.
\end{fact}

\begin{definition}[Top-Eigenspace $W^j_i$ of the Estimated Reduced State]
\label{definition:Wji}
Fix \(j\in\{1,\dots,M\}\) and \(i \in \{1,\dots, 2^{M-j}\}\). Let \[\sigma^j_i \;=\; \mathrm{Tr}_{\mathcal{B}^j \setminus \mathcal{B}^j_i} \big[ (\rho^{j-1})' \big]\] be the reduced state obtained by tracing out all qudits in \(\mathcal{B}^j \setminus \mathcal{B}^j_i\) (see Algorithm~\ref{alg:A2}). Let \(\hat{\sigma}^j_i\) denote the estimation of \(\sigma^j_i\) produced by the tomography procedure. 

We define \(W^j_i\) to be the subspace spanned by the eigenvectors of \(\hat{\sigma}^j_i\) corresponding to its \(d^{\,p}\) largest eigenvalues. We then denote by \(\Pi_{W^j_i}\) the orthogonal projector onto \(W^j_i\).
\end{definition}

\begin{fact}
\label{fact:U-PiW-relation}
Let $U^j$ and  $P^j$ as operators from Algorithm \ref{alg:A2}. By the definition of $W^j_i$ and $\Pi_{W^j_i}$, we have
\[(U^j)^\dagger P^j U^j=\Pi_{W^j_1}\otimes\dots\otimes\Pi_{W^j_{2^{M-j}}}.\]
\end{fact}

\begin{fact}
\label{fact:decompose-projector}
For each $j = 1, \dots, M$, the orthogonal complement of 
$\Pi_{W^j_1} \otimes \cdots \otimes \Pi_{W^j_{2^{M-j}}}$ can be decomposed as
\[
    I - \bigotimes_{k=1}^{2^{M-j}} \Pi_{W^j_k}
    = \sum_{i=1}^{2^{M-j}} \left( \bigotimes_{k=1}^{i-1} \Pi_{W^j_k} \right) 
      \otimes \left( I_{\mathcal{B}^j_i} - \Pi_{W^j_i} \right) 
      \otimes \left( \bigotimes_{k=i+1}^{2^{M-j}} I_{\mathcal{B}^j_k} \right),
\]
where $I_{\mathcal{B}^j_i}$ denotes the identity operator on the qudit $\mathcal{B}^j_i$. 

Here, we adopt the convention that the tensor product over an empty index set (i.e., when the upper limit is less than the lower limit) is defined to be the identity operator on the trivial (one-dimensional) Hilbert space.
\end{fact}

\begin{lemma}
\label{lemma:tomography-error}
Let \(\sigma\) be a density matrix with \(\operatorname{rank}(\sigma)\leq D^2\), and let \(\hat{\sigma}\) be a rank-\(n\) density matrix, where \(n \geq D^2\). Let \(W\) be the span of the largest \(D^2\) eigenvectors of \(\hat{\sigma}\), and let \(\Pi_W\) denote the orthogonal projector onto \(W\). If these two matrices satisfy \(\lVert \sigma - \hat{\sigma} \rVert_{1} \leq \eta \), then
\[
\operatorname{Tr}\!\big[(I-\Pi_W)\,\sigma\big] \le 2\eta.
\]
\end{lemma}
\begin{proof}
Write the spectral decomposition of \(\hat{\sigma}\) as
\[
\hat{\sigma} = \sum_{i=1}^n \hat{\sigma}_i \ket{v_i}\bra{v_i},
\]
with eigenvalues sorted in descending order \(\hat{\sigma}_1 \geq \hat{\sigma}_2 \geq \dots \geq \hat{\sigma}_n \geq 0\).  
Define the best rank-\(D^2\) approximation of \(\hat{\sigma}\) as
\[
\hat{\sigma}_{\mathrm{tr}} = \sum_{i=1}^{D^2} \hat{\sigma}_i \ket{v_i}\bra{v_i}.
\]
By the Eckart–Young–Mirsky theorem, \(\hat{\sigma}_{\mathrm{tr}}\) minimizes the trace norm error among all rank-\(D^2\) matrices. Since \(\sigma\) has rank at most \(D^2\),
\[
\|\hat{\sigma}_{\mathrm{tr}} - \hat{\sigma}\|_1 \leq \|\sigma - \hat{\sigma}\|_1 \leq \eta.
\]
Note that $\|I-\Pi_W\|_\infty=1$. We compute
\[
\operatorname{Tr}[(I-\Pi_W)\sigma] = \operatorname{Tr}\big[(I-\Pi_W)(\sigma-\hat{\sigma})\big] + \operatorname{Tr}\big[(I-\Pi_W)\hat{\sigma}\big].
\]
For the first term, by Hölder (duality of Schatten norms),
\[
\big|\operatorname{Tr}\big[(I-\Pi_W)(\sigma-\hat{\sigma})\big]\big| \le \|I-\Pi_W\|_\infty \,\|\sigma-\hat{\sigma}\|_1 \le \eta.
\]
For the second term we use positivity of $\hat{\sigma}-\hat{\sigma}_{\mathrm{tr}}$ (indeed $\hat{\sigma}-\hat{\sigma}_{\mathrm{tr}}=\sum_{i>D^2}\hat{\sigma}_i\ket{v_i}\bra{v_i}\succeq0$) to get
\[
\operatorname{Tr}[(I-\Pi_W)\hat{\sigma}] = \operatorname{Tr}\big[\hat{\sigma}-\hat{\sigma}_{\mathrm{tr}}\big] = \|\hat{\sigma}-\hat{\sigma}_{\mathrm{tr}}\|_1 \le \eta.
\]
Combining the two estimates yields
\[
\operatorname{Tr}[(I-\Pi_W)\sigma] \le \eta + \eta = 2\eta,
\]
as desired.
\end{proof}

\begin{lemma}
\label{lemma:bound-fidelity-varphirho'}
Fix \(j \in \{1,\dots, M\}\). For each \(i = 1, \dots, 2^{M-j}\), let \(\mathbb{H}_{\mathcal{B}^j_i}\) be the Hilbert space associated with \(\mathcal{B}^j_i\), where \(\mathcal{B}^j_i\) is defined in Algorithm \ref{alg:A2}. Let \(W^j_i \subseteq \mathbb{H}_{\mathcal{B}^j_i}\) be a subspace defined in Definition \ref{definition:Wji}, and let \(\Pi_{W^j_i}\) denote the orthogonal projector onto \(W^j_i\). Let \( \mathcal{B}^j \coloneq \bigcup_{k=1}^{2^{M-j}} \mathcal{B}^j_k\) be the collection of all qudits under consideration at step \(j\). Let \((\rho^{j-1})'\) denote the state on \(\mathbb{H}_{\mathcal{B}^j}\) defined in Algorithm~\ref{alg:A2}, and let \(\ket{\varphi}\) be a unit vector in the same Hilbert space. Define the product projector \(\Pi_{W^j} \coloneq \bigotimes_{i=1}^{2^{M-j}} \Pi_{W^j_i}.\)
Suppose that
\[
    (I - \Pi_{W^j}) \ket{\varphi} = \ket{\varphi},
\]
and that for each \(i = 1, \dots, 2^{M-j}\),
\[
    \operatorname{Tr}\!\Big[(I - \Pi_{W^j_i})\,
    \mathrm{Tr}_{\,(\mathcal{B}^{j} \setminus \mathcal{B}^j_i)}\big[(\rho^{j-1})'\big]\Big] \leq 2\eta.
\]
Then
\[
    \bra{\varphi} (\rho^{j-1})' \ket{\varphi} \leq 2 \eta \, 2^{M-j}.
\]
\end{lemma}

\begin{proof}
According to Fact~\ref{fact:decompose-projector}, any unit vector \(\ket{\varphi}\) satisfying \((I - \Pi_{W^j}) \ket{\varphi} = \ket{\varphi}\) can be decomposed as
\[
    \ket{\varphi} = \sum_{i=1}^{2^{M-j}} \ket{\varphi_i},
\]
where
\[
    \ket{\varphi_i} \coloneq 
    \left( \bigotimes_{k=1}^{i-1} \Pi_{W^j_k} \right) 
    \otimes \left( I_{\mathcal{B}^j_i} - \Pi_{W^j_i} \right) 
    \otimes \left( \bigotimes_{k=i+1}^{2^{M-j}} I_{\mathcal{B}^j_k} \right) 
    \ket{\varphi},
\]
and we set \(p_i \coloneq \|\ket{\varphi_i}\|^2\), so that \(\sum_{i=1}^{2^{M-j}} p_i = 1\).

With this decomposition, we have
\begin{align}
\label{eq:phirho'phi}
    &\bra{\varphi} (\rho^{j-1})' \ket{\varphi}
    = \sum_{i=1}^{2^{M-j}} \bra{\varphi_i} (\rho^{j-1})' \ket{\varphi_i}
    + \sum_{\substack{i,k=1 \\ i \neq k}}^{2^{M-j}}
        \bra{\varphi_i} (\rho^{j-1})' \ket{\varphi_k}\nonumber\\
    &=\sum_{i=1}^{2^{M-j}} p_i\frac{\bra{\varphi_i}}{\|\bra{\varphi_i}\|} (\rho^{j-1})' \frac{\ket{\varphi_i}}{\|\ket{\varphi_i}\|} 
    + \sum_{\substack{i,k=1 \\ i \neq k}}^{2^{M-j}}
         \sqrt{p_ip_k}\frac{\bra{\varphi_i}}{\|\bra{\varphi_i}\|} \sqrt{(\rho^{j-1})'}\sqrt{(\rho^{j-1})'} \frac{\ket{\varphi_k}}{\|\ket{\varphi_k}\|}\nonumber\\
    &\leq \sum_{i=1}^{2^{M-j}} p_i\frac{\bra{\varphi_i}}{\|\bra{\varphi_i}\|} (\rho^{j-1})' \frac{\ket{\varphi_i}}{\|\ket{\varphi_i}\|} 
    + \sum_{\substack{i,k=1 \\ i \neq k}}^{2^{M-j}}
         \sqrt{p_ip_k}\left\|\frac{\bra{\varphi_i}}{\|\bra{\varphi_i}\|} \sqrt{(\rho^{j-1})'}\right\|\left\|\sqrt{(\rho^{j-1})'} \frac{\ket{\varphi_k}}{\|\ket{\varphi_k}\|}\right\|\nonumber\\
    &=\sum_{i=1}^{2^{M-j}} p_i\frac{\bra{\varphi_i}}{\|\bra{\varphi_i}\|} (\rho^{j-1})' \frac{\ket{\varphi_i}}{\|\ket{\varphi_i}\|} 
    + \sum_{\substack{i,k=1 \\ i \neq k}}^{2^{M-j}}
         \sqrt{p_ip_k}\sqrt{\frac{\bra{\varphi_i}}{\|\bra{\varphi_i}\|} (\rho^{j-1})'\frac{\ket{\varphi_i}}{\|\ket{\varphi_i}\|}}\sqrt{ \frac{\bra{\varphi_k}}{\|\bra{\varphi_k}\|}(\rho^{j-1})'\frac{\ket{\varphi_k}}{\|\ket{\varphi_k}\|}}.
\end{align}
For each \(i\), since \( \frac{\ket{\varphi_i}}{\|\varphi_i\|}\) is supported on the subspace where the \(\mathcal B^j_i\)-part lies in \((W^j_i)^\perp\), we have
\[
\frac{\bra{\varphi_i}}{\|\varphi_i\|} (\rho^{j-1})' \frac{\ket{\varphi_i}}{\|\varphi_i\|}
\le  \operatorname{Tr}\!\Big[(I-\Pi_{W^j_i})\,
\mathrm{Tr}_{(\mathcal{B}^{j} \setminus \mathcal{B}^j_i)}[(\rho^{j-1})']\Big]
\le 2\eta.
\]
Now, let's substitute it back to Equation \eqref{eq:phirho'phi}:
\begin{align}
    \bra{\varphi} (\rho^{j-1})' \ket{\varphi} 
&\leq 2\eta\left(\sum_{i=1}^{2^{M-j}} p_i\right)
    + 2\eta   \left(\sum_{\substack{i,k=1 \\ i \neq k}}^{2^{M-j}}\sqrt{p_ip_k}\right) \nonumber \\
&= 2\eta   + 2\eta   \sum_{i=1}^{2^{M-j}}\sqrt{p_i} \left[\left(\sum_{k=1}^{2^{M-j}}\sqrt{p_k}\right)-\sqrt{p_i}\right]\nonumber\\
&= 2\eta    
    + 2\eta    \left[\left(\sum_{k=1}^{2^{M-j}}\sqrt{p_k}\right)^2-\sum_{i=1}^{2^{M-j}}p_i\right] \nonumber\\
&\leq 2\eta    
    + 2\eta    \left(2^{M-j}-1\right)\nonumber\\
&=2\eta   2^{M-j},
\end{align}
where in the last inequality we use the Cauchy--Schwarz inequality.
\end{proof}

\begin{lemma}
\label{lemma:tomography-error-pass}
Fix $j \in \{1,\dots,M\}$. For each \(i = 1, \dots, 2^{M-j}\), let \(\mathbb{H}_{\mathcal{B}^j_i}\) denote the Hilbert space associated with \(\mathcal{B}^j_i\), where \(\mathcal{B}^j_i\) is defined in Algorithm \ref{alg:A2}. Let \(W^j_i \subseteq \mathbb{H}_{\mathcal{B}^j_i}\) be a subspace defined in Definition \ref{definition:Wji}, and denote by \(\Pi_{W^j_i}\) the orthogonal projector onto \(W^j_i\). Let \(\mathcal{B}^j \coloneqq \bigcup_{k=1}^{2^{M-j}} \mathcal{B}^j_k\) be the set of all qudits under consideration at step \(j\). Let \((\rho^{j-1})'\) denote the state on \(\mathbb{H}_{\mathcal{B}^j}\) defined in Algorithm~\ref{alg:A2}, and let \(\rho^j\) be the state given in Definition~\ref{definition:rhoj}.

Suppose that, for each \(i = 1, \dots, 2^{M-j}\), the tomography procedure succeeds in the sense that
\[
    \operatorname{Tr}\!\Big[(I - \Pi_{W^j_i})\,
    \mathrm{Tr}_{\,(\mathcal{B}^{j} \setminus \mathcal{B}^j_i)}\big[(\rho^{j-1})'\big]\Big] \leq 2\eta.
\]
Then, for any pure state \(\ket{\phi}\in (\mathbb{C}^d)^{\otimes n}\),
\[
    \big| \bra{\phi} \rho^{j-1} \ket{\phi}
    - \bra{\phi} \rho^{j} \ket{\phi} \big|
    \;\leq\;
    2 \sqrt{\,2 \eta \,   \, 2^{M-j}} \, .
\]
\end{lemma}

\begin{proof}
According to Definition \ref{definition:rhoj} and \ref{definition:psij}, we can express the fidelity $\bra{\phi} \rho^{j} \ket{\phi} $ as
\begin{align}
    \bra{\phi} \rho^{j} \ket{\phi} 
    &= \bra{\phi}  (E^j)^\dagger \left[\left(\bigotimes_{a=1}^{j}P^a\right)\otimes(\rho^{j})' \right]E^j\ket{\phi} \nonumber\\
    &= \bra{\phi^{j-1}}  (U^j)^\dagger \left[\left(\bigotimes_{a=1}^{j}P^a\right)\otimes(\rho^{j})' \right]U^j\ket{\phi^{j-1}} \nonumber\\
    &= \bra{\phi^{j-1}}  \left(\bigotimes_{a=1}^{j-1}P^a\right) (U^j)^\dagger\left[ P^j\otimes(\rho^{j})' \right]U^j \left(\bigotimes_{a=1}^{j-1}P^a\right)\ket{\phi^{j-1}} \nonumber\\
   &=    \bra{\psi^{j-1}} (U^j)^\dagger\left[ P^j\otimes(\rho^{j})'\right] U^j \ket{\psi^{j-1}}
\end{align}
In Algorithm \ref{alg:A2}, we define \((\rho^{j})' = \mathrm{Tr}_{\bigcup_{i=1}^{2^{M-j}}\mathcal{B}^j_i\setminus \widetilde{\mathcal{B}}^j_i} \left[ P^j U^j (\rho^{j-1})' (U^j)^\dagger P^j \right]\)
where $P^j=\bigotimes_{i=1}^{2^{M-j}}\left( \ket{0^{f(j,i)}}\bra{0^{f(j,i)}}\right)_{\mathcal{B}^j_i\setminus \widetilde{\mathcal{B}}^j_i}\otimes I_{\widetilde{\mathcal{B}}^j_i}$ is a zero-state projection operators acting on a collection of qudits  \({\bigcup_{i=1}^{2^{M-j}}\mathcal{B}^j_i\setminus \widetilde{\mathcal{B}}^j_i}\) . In other words, we have the relation:
\begin{equation}
\label{eq:rhoj'}
    P^{j}\otimes(\rho^{j})' =  P^j U^j (\rho^{j-1})' (U^j)^\dagger P^j.
\end{equation}
Use Equation \eqref{eq:rhoj'} and Fact \ref{fact:U-PiW-relation}, we can rewrite the fidelity $\bra{\phi} \rho^{j} \ket{\phi} $ as 
\begin{equation}
    \bra{\phi} \rho^{j} \ket{\phi} =\bra{\psi^{j-1}}\Pi_{W^j_1}\otimes\dots\otimes\Pi_{W^j_{2^{M-j}}}(\rho^{j-1})' \;\Pi_{W^j_1}\otimes\dots\otimes\Pi_{W^j_{2^{M-j}}}\ket{\psi^{j-1}}
\end{equation}
For brevity, we denote \(\Pi_{W^j} \coloneq \bigotimes_{i=1}^{2^{M-j}} \Pi_{W^j_i}.\) Then together with Fact \ref{fact:fidelity-rep}, this implies
\begin{align}
      \left| \bra{\phi} \rho^{j-1} \ket{\phi}
    - \bra{\phi} \rho^{j} \ket{\phi} \right|
    &= \left|\bra{\psi^{j-1}} (\rho^{j-1})' \ket{\psi^{j-1}}-\bra{\psi^{j-1}}  \Pi_{W^j}(\rho^{j-1})' \,\Pi_{W^j}\ket{\psi^{j-1}}\right| \nonumber\\
&\leq \left|\bra{\psi^{j-1}}  (\rho^{j-1})' \,(I-\Pi_{W^j})\ket{\psi^{j-1}}\right|+\left|\bra{\psi^{j-1}}  (I-\Pi_{W^j})(\rho^{j-1})' \,\Pi_{W^j}\ket{\psi^{j-1}}\right| \nonumber\\
&\leq \left\|\ket{\psi^{j-1}}\right\|  \left\|(\rho^{j-1})' \,(I-\Pi_{W^j})\ket{\psi^{j-1}} \right\|+
\left\|\Pi_{W^j}\ket{\psi^{j-1}}\right\| \left\|(\rho^{j-1})' \,(I-\Pi_{W^j})\ket{\psi^{j-1}} \right\| \nonumber \\
&\leq 2 \left\|(\rho^{j-1})' \,(I-\Pi_{W^j})\ket{\psi^{j-1}} \right\|
\end{align}
Let $\ket{\widetilde{\varphi}}=(I-\Pi_{W^j})\ket{\psi^{j-1}}$ and define the normalized vector $\ket{\varphi}\coloneq \frac{\ket{\widetilde{\varphi}}}{\|\ket{\widetilde{\varphi}}\|}$. Then
\begin{align}
2 \left\|(\rho^{j-1})' \,(I-\Pi_{W^j})\ket{\psi^{j-1}} \right\|
&= 2 \left\|(\rho^{j-1})' \,\ket{\varphi} \right\| \,\big\|\ket{\widetilde{\varphi}}\big\| \nonumber\\
&\leq 2 \big\|\ket{\widetilde{\varphi}}\big\| \sqrt{\bra{\varphi}\, [(\rho^{j-1})']^2 \,\ket{\varphi}} \nonumber\\
&\leq 2 \big\|\ket{\widetilde{\varphi}}\big\| \sqrt{\bra{\varphi} (\rho^{j-1})' \ket{\varphi}} \nonumber\\
&\leq 2 \sqrt{\bra{\varphi} (\rho^{j-1})' \ket{\varphi}} 
\end{align}
Notice that $(I-\Pi_{W^j})\ket{\varphi}=\ket{\varphi}$. By Lemma \ref{lemma:bound-fidelity-varphirho'}, we conclude that 
\begin{align}
\left| \bra{\phi} \rho^{j-1} \ket{\phi}
    - \bra{\phi} \rho^{j} \ket{\phi}\right|
&\leq 2 \sqrt{\bra{\varphi} (\rho^{j-1})' \ket{\varphi}} 
\leq2\sqrt{2\eta  2^{M-j}}
\end{align}
\end{proof}

\begin{theorem*}[Restatement of Theorem \ref{theorem:1}]
Given access to copies of an $n$-qudit matrix product state $\rho$ with bond dimension $D$ and parameters $\epsilon, \delta \in (0,1)$, Algorithm~\ref{alg:A2} outputs a description of a state $\ket{\hat{\phi}}$, such that, with probability $1-\delta$,
\[
\bra{\hat{\phi}} \rho \ket{\hat{\phi}} \;\geq\; 1 - \epsilon
\]
The algorithm requires $N=O(\frac{D^{6}\cdot d^4 \cdot n^3 \cdot \log(n/\delta)}{(\log_d D)^3\epsilon^4})$
copies of $\rho$ and runs in time $\text{poly}\left(D,n,\frac{1}{\epsilon},\log(\frac{1}{\delta})\right)$.
\end{theorem*}
\begin{proof}
From Lemma~\ref{lemma: preliminary1} and Lemma \ref{lemma:preliminary2}, it follows that each reduced density matrix $$\sigma^j_i = \mathrm{Tr}_{\mathcal{B}^{j}\setminus \mathcal{B}^j_i} [(\rho^{j-1})']$$
defined in Algorithm~\ref{alg:A2} has rank at most $D^2$. Suppose that the state tomography procedure succeeds for all $\sigma^j_i$ throughout the algorithm. Then, for each $\sigma^j_i$, the tomography produces an estimate $\hat{\sigma}^j_i$ satisfying
\[
\|\sigma^j_i - \hat{\sigma}^j_i\|_1 \;\leq\; \frac{(\sqrt{2}-1)^2 \epsilon^2}{2^{M+5}  }.
\]
In the next step, Algorithm \ref{alg:A2} invokes Algorithm \ref{alg:A1} to select the $D^2$ largest eigenvectors from the estimated density matrix $\hat{\sigma}^j_i$ and constructs a disentangling unitary accordingly. Let $\widetilde{W}^j_i$ denote the subspace spanned by the $D^2$ largest eigenvectors. From Lemma \ref{lemma:tomography-error}, we know that 
\[\operatorname{Tr}\!\Big[(I - \Pi_{\widetilde{W}^j_i})\,
    \sigma^j_i\Big]  \leq \frac{(\sqrt{2}-1)^2 \epsilon^2}{2^{M+4}  }.
\]
Since $\widetilde{W}^j_i$ is contained in the space spanned by the $d^p$ largest eigenvectors of $\hat{\sigma}^j_i$, we have $\widetilde{W}^j_i \subseteq W^j_i$, where $W^j_i$ is defined in Definition \ref{definition:Wji}. Equivalently, their orthogonal complements satisfy
$$(W^j_i)^\perp \subseteq (\widetilde{W}^j_i)^\perp$$
Therefore, it follows that
\[\operatorname{Tr}\!\Big[(I - \Pi_{W^j_i})\,
    \sigma^j_i\Big]  \leq \frac{(\sqrt{2}-1)^2 \epsilon^2}{2^{M+4}  }.
\]
Let $\ket{\phi} \in (\mathbb{C}^d)^{\otimes n}$ be a pure state on an $n$-qudit system. Applying Lemma \ref{lemma:tomography-error-pass}, we get
\[ \big| \bra{\phi} \rho^{j-1} \ket{\phi}
    - \bra{\phi} \rho^{j} \ket{\phi} \big|
    \;\leq\;
    \frac{(\sqrt{2}-1)\epsilon}{2\sqrt{2^{j}}} \, .\]
Using the triangular inequality, we derive
\[\big| \bra{\phi} \rho^{0} \ket{\phi}
    - \bra{\phi} \rho^{M} \ket{\phi} \big|
    \;\leq\;
    \sum_{j=1}^{M}\frac{(\sqrt{2}-1)\epsilon}{2\sqrt{2^{j}}} \leq\frac{\epsilon}{2},\]
which implies 
\begin{equation}
\label{proof:t6}
\bra{\phi} \rho^{M} \ket{\phi} \geq \bra{\phi} \rho \ket{\phi} -\frac{\epsilon}{2}. 
\end{equation}
Recall that $\rho^0\coloneq \rho$ by Definition \ref{definition:rhoj}. For any state $\ket{\phi} \in (\mathbb{C}^d)^{\otimes n}$, we define its level-$j$ projected states $\ket{\psi^{j}} \in \mathbb{H}_{\mathcal{\widetilde{B}}^j_1}\otimes\dots\otimes\mathbb{H}_{\widetilde{\mathcal{B}}^j_{2^{M-j}}}$ through Definition \ref{definition:psij}. By Fact~\ref{fact:fidelity-rep}, we have
\begin{equation}
\label{proof-t1}
\bra{\phi}\rho^{M}\ket{\phi} = \bra{\psi^{M}}(\rho^{M})'\ket{\psi^{M}}.
\end{equation}
On the other hand, for the reconstructed state \(\ket{\hat{\phi}} = (E^j)^\dagger \left( \bigotimes_{j=1}^{M}\bigotimes_{i=1}^{2^{M-j}} \ket{0^{f(j,i)}}_{\mathcal{B}^j_i\setminus \widetilde{\mathcal{B}}^j_i}
\otimes \ket{\hat{\psi}} \right),\) 
since 
\((\rho^M)'=
\left( \bigotimes_{j=1}^{M}\bigotimes_{i=1}^{2^{M-j}} \bra{0^{f(j,i)}}_{\mathcal{B}^j_i\setminus \widetilde{\mathcal{B}}^j_i} \right)
E^M\rho(E^M)^\dagger 
\left( \bigotimes_{j=1}^{M}\bigotimes_{i=1}^{2^{M-j}} \ket{0^{f(j,i)}}_{\mathcal{B}^j_i\setminus \widetilde{\mathcal{B}}^j_i} \right)
\), we have
\begin{equation}
\bra{\hat{\phi}}\rho\ket{\hat{\phi}}
=
\bra{\hat{\psi}} (\rho^M)'\ket{\hat{\psi}}
\label{proof-t2}
\end{equation}
If the final tomography on $(\rho^M)'$ succeeds, it yields an estimate $(\hat{\rho}^M)'$ such that 
$\|(\rho^M)' - (\hat{\rho}^M)'\|_1 \leq \epsilon/4$. By H\"older's inequality for Schatten $p$-norms, for any state $\ket{\varphi} \in \mathbb{H}_{\mathcal{\widetilde{B}}^M_1}$,
\begin{equation}
\label{proof-t3}
\left| \bra{\varphi}(\rho^M)'\ket{\varphi} - \bra{\varphi}(\hat{\rho}^M)'\ket{\varphi} \right| \leq \epsilon/4.    
\end{equation}
Applying this bound gives
\begin{equation}
\label{proof-t4}
\bra{\hat{\psi}}(\rho^{M})'\ket{\hat{\psi}}
\geq \bra{\hat{\psi}}(\hat{\rho}^{M})'\ket{\hat{\psi}} - \epsilon/4
\geq \bra{\psi^M}(\hat{\rho}^{M})'\ket{\psi^M} - \epsilon/4
\geq \bra{\psi^M}(\rho^{M})'\ket{\psi^M} - \epsilon/2.
\end{equation}
Here, we use Equation \eqref{proof-t3} in the first and last inequalities, while the second step follows from the fact that $\ket{\hat{\psi}}$ is the largest eigenvector of $(\hat{\rho}^M)'$. Combining with Equations \eqref{proof-t1} and \eqref{proof-t2}, we obtain
\begin{equation}
\label{proof:t5}
\bra{\hat{\phi}}\rho\ket{\hat{\phi}}
\geq \bra{\phi}\rho^{M}\ket{\phi}-\frac{\epsilon}{2}.
\end{equation}
Together with Equation \eqref{proof:t6},  the desired inequality  is established:
\begin{equation}
\label{proof-t7}
\bra{\hat{\phi}}\rho\ket{\hat{\phi}}
\geq \bra{\phi} \rho \ket{\phi} -\epsilon
\end{equation}
% Finally, applying Lemma~\ref{lemma:Monotonicity-lost} to Equation \eqref{proof-t7}, we obtain
% \begin{equation}
% \bra{\hat{\phi}}\rho^{0}\ket{\hat{\phi}} \geq \bra{\hat{\phi}}\rho^{M}\ket{\hat{\phi}} \geq \bra{\phi} \rho^{0} \ket{\phi} -\epsilon.
% \end{equation}
% Since $\rho^0\coloneq \rho$ by Definition \ref{definition:rhoj}, the desired inequality  is established:
% \begin{equation}
% \bra{\hat{\phi}}\rho\ket{\hat{\phi}} \geq \bra{\phi} \rho \ket{\phi} -\epsilon.
% \end{equation}
Notice that this inequality holds for any pure state $\ket{\phi}\in (\mathbb{C}^d)^{\otimes n}$. In particular, by choosing $\ket{\phi}$ to be our target MPS($D$) state, we obtain the maximal fidelity $\bra{\phi}\rho\ket{\phi}=1$.

Next, we compute the sample complexity of Algorithm \ref{alg:A2}. From Lemma \ref{lemma:subnorm-tomo-rank}, taking the dimension $d^{r-i}$ to be $d^{2p}$, the error parameter to be $\eta=\frac{(\sqrt{2}-1)^2 \epsilon^2}{2^{M+5}} $, and the failure probability to be $\delta/n$, we can perform tomography on each sub-normalized state $\sigma^j_i$ using $O(\frac{D^2 \cdot d^{2p}\log(n/\delta)}{\eta^2})=O(\frac{D^2 \cdot d^{2p} \cdot 2^{2M}\log(n/\delta)}{\epsilon^4})$ copies of sample and runtime that is polynomial in the same parameters. In total, the algorithm (line 12 and 25) performs sub-normalized tomography $\sum_{j=1}^M 2^{M-j}=2^M-1$ times. Recall that $d^{2p}\le D^4 d^4$ and $2^M\le 2n/p$. As a result, the tomography on all $\sigma^j_i$ requires $O(\frac{D^{2}\cdot d^{2p}\cdot 2^{3M} \cdot \log(n/\delta)}{\epsilon^4})=O(\frac{D^{6}\cdot d^4 \cdot n^3 \cdot \log(n/\delta)}{(\log_d D)^3\epsilon^4})$ copies of sample.

The last call of tomography takes the dimension $d^{r-i}$ to be $d^{p}$, the error parameter to be $\eta=\epsilon/4$, and the failure probability to be $\delta/n$. The copies of sample needed for the last call of tomography is $O(\frac{D^2 \cdot d^{p}\log(n/\delta)}{\epsilon^2})<O(\frac{D^2 \cdot d^{2p} \cdot 2^{2M}\log(n/\delta)}{\epsilon^4})$ copies of sample. The overall algorithm then requires
$$O(\frac{D^{6}\cdot d^4 \cdot n^3 \cdot \log(n/\delta)}{(\log_d D)^3\epsilon^4})$$
copies of $\rho$  and runs in $\mathrm{poly}(n, D,1/\epsilon,\log(1/\delta))$ time.
\end{proof}
\appendixsection{Learning the closest matrix product states}
\label{appendix-A4}
In this section, we consider the learning task where the target state $\rho$ is not necessarily an exact matrix product state (MPS) of bond dimension $D$, as assumed in the previous sections, and may even be a mixed state. We define the maximum fidelity between $\rho$ and and the set $\text{MPS}(D)$ of MPS of bond dimension $D$ as
$$\max_{\ket{\phi}\in \text{MPS}(D)} \bra{\phi}\rho\ket{\phi}.$$
Our goal is to output a description of state $\ket{\hat{\phi}}$ such that, given an error parameter $\epsilon > 0$  and access to copies of $\rho$, this state is close enough to the closest matrix product state of bond dimension $D$ in terms of fidelity
$$\bra{\hat{\phi}}\rho\ket{\hat{\phi}} \geq \max_{\ket{\phi}\in \text{MPS}(D)} \bra{\phi}\rho\ket{\phi}-\epsilon.$$
We remark that this task is slightly different from the standard notion of \emph{agnostic learning}. In the problem statement of agnostic learning, besides an error parameter $\epsilon > 0$ and access to copies of $\rho$, the learner should also given additional information $\delta$ that promises $\bra{\phi}\rho\ket{\phi}\ge \delta$ for some $\ket{\phi}\in \text{MPS}(D)$. In other words, learners could also use $\delta$ to design their algorithm. Since our setting does not require such additional information, we refer to our algorithm as the learning of the closest matrix product state, rather than agnostic learning of matrix product states.

\begin{breakablealgorithm}
\caption{Disentangling unitary construction (threshold $\eta$)}
\label{alg:A3}
\raggedright
\textbf{Input:} A threshold $\eta\ge 0$ and a description of a $y$-qudit density matrix $\hat{\sigma}\in\mathbb{C}^{d^y\times d^y}$.\\
\textbf{Output:} A description of a Disentangling Unitary $U$ for the state $\hat{\sigma}$ (constructed from eigenvectors whose eigenvalues exceed $\eta$).\\
\textbf{Procedure:}
\begin{algorithmic}[1]
\State Perform the spectral decomposition of $\hat{\sigma}$:
\[
\hat{\sigma}=\sum_{i=1}^{r} a_i \ket{\hat{\phi}_i}\bra{\hat{\phi}_i},
\]
where \(r\leq d^{y}.\)
\State Define the index set of selected eigenvalues:
\[
S \coloneq \{\, i\in\{1,\dots,r\} \;|\; a_i > \eta \,\}.
\]
\State Let \(m\coloneq |S|\). Note that $m < 1/\eta$ because $\hat{\sigma}$ is trace $1$.
\State Reindex the selected eigenvectors as \(\{\ket{\phi_i}\}_{i=1}^{m}\), where \(\ket{\phi_i}\coloneq\ket{\hat{\phi}_{s_i}}\) for \(s_i\in S\).
\State Extend \(\{\ket{\phi_i}\}_{i=1}^{m}\) to a full orthonormal basis \(\{\ket{\phi_i}\}_{i=1}^{d^{y}}\) of $(\mathbb{C}^d)^{\otimes y}$ by choosing arbitrary orthonormal vectors for indices \(i=m+1,\dots,d^{y}.\)
\State Set \(t \coloneq \left\lceil \log_d m \right\rceil.\)
\State Let \( \{ \ket{a_1, \dots, a_{y-t}} \}_{a_i \in \{0, \dots, d - 1\}} \) be the standard basis for the first \( y-t \) qudits, and let \( \{ \ket{j} \}_{j = 1}^{d^{t}} \) be an arbitrary orthonormal basis for the remaining \( t \) qudits. Define the unitary
\[
U = \sum_{a_1, \dots, a_{y-t} = 0}^{d - 1} \sum_{j = 1}^{d^{t}}
\left( \ket{a_1, \dots, a_{y-t}} \otimes \ket{j} \right)
\left\langle \phi_{\mathrm{idx}(a_1, \dots, a_{y-t}, j)} \right|,
\]
where the index mapping is defined by \(\mathrm{idx}
(a_1, \dots, a_{y-t}, j) = j + d^{t} \cdot \sum_{l = 1}^{y-t} a_l d^{y-t- l}.
\)
\State \Return the unitary $U$.
\end{algorithmic}
\end{breakablealgorithm}

\begin{breakablealgorithm}
\caption{Learning the closest matrix product states}
\label{alg:A4}
\raggedright
\textbf{Input:} Copies of an unknown matrix product states $\rho \in \mathbb{C}^{d^n \times d^n}$ such that there exists matrix product state with bond dimension at most $D$ with fidelity $\theta$ with $\rho$, error parameter \(\epsilon = \sqrt{\frac{64\,n\,D^2}{(\sqrt{2}-1)^2 m d^m}}\) for some large enough integer $m$ (see Remark \ref{remark1}), and failure probability $\delta$.\\
\textbf{Output:} A description of a quantum state $\ket{\hat{\phi}}\in (\mathbb{C^d})^{\otimes n}$. \\
\textbf{Procedure:}
\begin{algorithmic}[1]
\State Define $p$ implicitly by the relation $p = \lceil\log_d(1/\eta)\rceil$,  where $\eta \coloneq \frac{(\sqrt{2}-1)^2 \epsilon^2 p}{64 D^2 n}$.
\State Solve for $p$ to obtain $p = \left\lceil \tfrac{1}{\ln d}\, W\!\left(\ln d \cdot \tfrac{64 n D^2}{(\sqrt{2}-1)^2 \epsilon^2}\right) \right\rceil$ (see Appendix \ref{appendix-1} for derivation).
\State Set \( M \) be the smallest positive integer $M'$ such that $2^{M'}p \geq n$.
\State Set $\ell_1 \coloneq \lceil\frac{1}{p} \left( n-2^{M-1}p \right)\rceil$, $s_1\coloneq \text{Mod}\left[n-2^{M-1}p,p\right]$, $k_1 \coloneq 2\ell_1 p-p+s_1$. 
\State Define the support $\mathcal{B}_i^1$ set for the $i$-th disentangling unitaries in the first layer:
\begin{equation}
\label{support}
\mathcal{B}_i^1 \coloneq 
\begin{cases}
\{2(i-1)p+1,\dots, 2ip\} & \text{if } 1\leq i<\ell_1 \\
\{2(\ell_1-1)p+1,\dots, k_1\} & \text{if } i = \ell_1\\
\{k_1+1+(i-\ell_1-1)p,\dots, k_1+(i-\ell_1)p\}& \text{if } \ell_1<i\leq 2^{M-1}
\end{cases}.
\end{equation}
\State Define a function 
$$f(j,i)=\begin{cases}
s_1 & \text{when } j=1, i=\ell_1\\
0 & \text{when } j=1, i>\ell_1 \\
p & \text{otherwise}
\end{cases}.$$
\State Let $(\rho^0)' = \rho$.
\State Let $\mathcal{B}^1=\bigcup_{i=1}^{2^{M-1}}\mathcal{B}^1_i$.
\For {$i$ from $1$ to $\ell_1$}
    \State Let $\sigma^1_i = \mathrm{Tr}_{\mathcal{B}^1\setminus \mathcal{B}^1_i} [(\rho^0)']$.
    \State Let $\hat{\sigma}^1_i$ be the output of the tomography with error $\eta$ and failure probability $\delta/n$ on $O\!\left(\frac{d^{4p}}{\eta^{2}}
\log\!\frac{n}{\delta}
\right)$ copies of $\sigma^1_i$.
    \State Use Algorithm \ref{alg:A3} to generate the disentangling unitary $U^1_i$ from a description of a $\left(p+f(1,i)\right)$-qudit state $\hat{\sigma}^1_i$ and a parameter $\eta$.
    \State Let $\widetilde{\mathcal{B}}^1_i$ be the last $p$ qudits in $\mathcal{B}^1_i$.
\EndFor
\State Let $U^1=\bigotimes_{i=1}^{\ell_1}U^1_i$.
\State Let $P^1=\bigotimes_{i=1}^{\ell_1}\left( \ket{0^{f(1,i)}}\bra{0^{f(1,i)}}\right)_{\mathcal{B}^1_i\setminus \widetilde{\mathcal{B}}^1_i}\otimes I_{\widetilde{\mathcal{B}}^1_i}$.
\State Apply $U^1$ to all copies of $(\rho^0)'$ and project onto $P^1$ to get copies of
    \[
        (\rho^1)' = \mathrm{Tr}_{\bigcup_{i=1}^{\ell_1}\mathcal{B}^1_i\setminus \widetilde{\mathcal{B}}^1_i} \left[ P^1 U^1 (\rho^0)' (U^1)^\dagger P^1 \right];
    \]
\State Let $\widetilde{\mathcal{B}}^1_i=\mathcal{B}^1_i$ for $\ell_1<i\leq2^{M-1}$.
\For {$j$ from $2$ to $M$}
\State Let $\mathcal{B}^j=\bigcup_{i=1}^{2^{M-j+1}}\widetilde{\mathcal{B}}^{j-1}_i$.
    \For {$i$ from $1$ to $2^{M-j}$}
        \State Let $\mathcal{B}^j_i = \widetilde{\mathcal{B}}^{j-1}_{2i-1} \cup \widetilde{\mathcal{B}}^{j-1}_{2i} $.
        \State Let $\sigma^j_i = \mathrm{Tr}_{\mathcal{B}^{j}\setminus \mathcal{B}^j_i} [(\rho^{j-1})']$.
        \State Let $\hat{\sigma}^j_i$ be the output of the tomography with error $\eta$ and failure probability $\delta/n$ on $O\!\left(\frac{d^{4p}}{\eta^{2}}
\log\!\frac{n}{\delta}
\right)$ copies of $\sigma^j_i$.
        \State Use Algorithm \ref{alg:A3} to generate the disentangling unitary $U^j_i$ from a description of $2p$-qudit state $\hat{\sigma}^j_i$ and a parameter $\eta$.
        \State Let $\widetilde{\mathcal{B}}^j_i$ be the last $p$ qudits in $\mathcal{B}^j_i$.
    \EndFor
    \State Let $U^j=\bigotimes_{i=1}^{2^{M-j}}U^j_i$.
    \State Let $P^j=\bigotimes_{i=1}^{2^{M-j}}\left( \ket{0^{f(j,i)}}\bra{0^{f(j,i)}}\right)_{\mathcal{B}^j_i\setminus \widetilde{\mathcal{B}}^j_i}\otimes I_{\widetilde{\mathcal{B}}^j_i}$.
    \State Apply $U^j$ to all copies of $(\rho^{j-1})'$ and project onto $P^j$ to get copies of
    \[
        (\rho^{j})' = \mathrm{Tr}_{\bigcup_{i=1}^{2^{M-j}}\mathcal{B}^j_i\setminus \widetilde{\mathcal{B}}^j_i} \left[ P^j U^j (\rho^{j-1})' (U^j)^\dagger P^j \right];
    \]
\EndFor
\State Let $(\hat{\rho}^M)'$ be the output of the tomography with error $\tau=\epsilon/4$ and failure probability $\delta/n$ on $O\!\left(\frac{d^{2p}}{\tau^{2}}
\log\!\frac{n}{\delta}
\right)$ copies of $(\rho^M)'$.
\State Let $\ket{\hat{\psi}}$ be the top eigenvector of $(\hat{\rho}^M)'$.
\State \Return the state $\ket{\hat{\phi}}\coloneq(U^1)^\dagger \cdots (U^M)^\dagger \left( \bigotimes_{j=1}^{M}\bigotimes_{i=1}^{2^{M-j}} \ket{0^{f(j,i)}}_{\mathcal{B}^j_i\setminus \widetilde{\mathcal{B}}^j_i}
\otimes \ket{\hat{\psi}} \right)$.
\end{algorithmic}
\end{breakablealgorithm}

\begin{remark}
\label{remark1}
In the construction above, the integer \(m\) is introduced via the relation
\[
m d^{m} = \frac{64\,n\,D^2}{(\sqrt{2}-1)^2\,\epsilon^2},
\]
which ensures that the self-consistent equation in Algorithm \ref{alg:A4}: line 1 (or Equation \eqref{eq:main-ceil} in Appendix \ref{appendix-Existence}) admits the unique integer solution \(p=m\).
The admissible range of \(m\) is not arbitrary, however.
Since the accuracy parameter must satisfy \(0<\epsilon \le 1\), the equality above implies the constraint (see Appendix \ref{appendix-Existence} for more a detailed discussion)
\[
m d^{m} \;\ge\; \frac{64\,n\,D^2}{(\sqrt{2}-1)^2}.
\]
This condition guarantees that the chosen \(\epsilon\) lies within the valid range. 
On the other hand, the algorithm operates on an \(n\)-qudit system, so in order to make our algorithm efficient, we require that \[
m = \Theta(\log_d n).
\]
In other words, if one wants an accuracy of $\epsilon=\Theta(1/\mathrm{poly}(n))$ but happens to choose a specific $\epsilon$ that makes the self-consistent equation unsolvable (see Appendix \ref{appendix-Existence} for the discussion on the existence condition for an integer solution $p$), one can always find another value $\epsilon'=\sqrt{\frac{64\,n\,D^2}{(\sqrt{2}-1)^2 m d^m}}=\Theta(1/\mathrm{poly}(n))$ for some $m=\Theta(\log_d n)$ such that $\epsilon' < \epsilon$, which ensures that the self-consistent equation becomes solvable.

%it is only meaningful to consider \(p\le n\);
%if \(p>n\), the procedure would involve more blocks than the available sites and thus becomes ill-defined.
%Combining these two considerations, one finds that \(m\) cannot scale faster than \(\log n\).
%In particular, choosing
%\[
%m = \Theta(\log n)
%\]
%ensures that (i) the integer solution \(p=m\) exists, (ii) the accuracy parameter \(\epsilon\) remains within \((0,1]\),
%and (iii) the algorithm operates within its valid regime \(p\le n\).
\end{remark}

\appendixsection{Correctness of Algorithm \ref{alg:A4}}
\label{appendix-correctnessA4}

\begin{definition}[Step-wise Reconstructed State]
\label{definition:rhoj-agnostic}
Let $(\rho^0)'=\rho$ be the initial state before any iteration (see line 9 in Algorithm \ref{alg:A4}). Define
\[
\rho^{0} \coloneq (\rho^0)' = \rho.
\]For each $j = 1, \dots, M$, define the cumulative unitary
\[
E^{j} \coloneq U^{j} U^{j-1} \cdots U^{1}.
\]
Let $(\rho^j)'$ denote the subnormalized post-measurement state of the remaining qudits after \(j\) iterations of Algorithm~\ref{alg:A4}. Define
\[
\rho^{j} \coloneq (E^{j})^{\dagger} \left[ \bigotimes_{a=1}^{j} P^a \otimes (\rho^j)' \right] E^{j},
\]
where $P^a$ and $U^j$ are as in Algorithm \ref{alg:A4}. 

We call $\rho^j$ the \emph{step-wise reconstructed state} corresponding to the $j$-th layer of the circuit, representing the state that would have been reconstructed had the algorithm stopped after the $j$-th iteration (i.e., line 21).
\end{definition}

\begin{definition}[Level-$j$ transformed and projected states]
\label{definition:psij-agnostic}
Fix \(j\in\{1,\dots,M\}\). For any pure state \(\ket{\phi}\) on an \(n\)-qudit system define the \emph{level-$j$ transformed state}
\[
    \ket{\phi^{j}} \coloneqq E^j \ket{\phi},
\]
where \(E^j\) is the cumulative unitary introduced above. Define the \emph{level-$j$ projector}
\[
    \Pi^{(j)} \coloneqq \bigotimes_{a=1}^{j} P^a,
\]
with \(P^a\) as in Algorithm~\ref{alg:A4}. The \emph{level-$j$ projection} of \(\ket{\phi^{j}}\) is \(\Pi^{(j)}\ket{\phi^{j}}\). 

\(\Pi^{(j)}\ket{\phi^{j}}\) has the product form
\[
    \Pi^{(j)}\ket{\phi^{j}} 
    = \left( \bigotimes_{a=1}^j \bigotimes_{b=1}^{2^{M-a}} \ket{0^{f(a,b)}}_{\mathcal{B}^a_b\setminus\widetilde{\mathcal{B}}^a_b} \right)
      \otimes \ket{\psi^j},
\]
we call \(\ket{\psi^j} \in \mathbb{H}_{\mathcal{\widetilde{B}}^j_1}\otimes\dots\otimes\mathbb{H}_{\widetilde{\mathcal{B}}^j_{2^{M-j}}}\) the \emph{projected (residual) state of \(\ket{\phi}\) at level \(j\)}.  By convention, we define the projected state at level $0$ to be the input state itself, \(\ket{\psi^0} \coloneq \ket{\phi}.\)
\end{definition}

\begin{fact}
\label{fact:fidelity-rep-agnostic}
For any state $\ket{\phi}$ and for each $j= 0,\dots, M,$ we have $\bra{\phi}\rho^{j}\ket{\phi}=\bra{\psi^{j}}(\rho^{j})'\ket{\psi^{j}}$.
\end{fact}

\begin{definition}[Top-Eigenspace $W^j_i$ of the Estimated Reduced State]
\label{definition:Wji-agnostic}
Fix \(j\in\{1,\dots,M\}\) and \(i \in \{1,\dots, 2^{M-j}\}\). Let \[\sigma^j_i \;=\; \mathrm{Tr}_{\mathcal{B}^j \setminus \mathcal{B}^j_i} \big[ (\rho^{j-1})' \big]\] be the reduced state obtained by tracing out all qudits in \(\mathcal{B}^j \setminus \mathcal{B}^j_i\) (see Algorithm~\ref{alg:A4}). Let \(\hat{\sigma}^j_i\) denote the estimation of \(\sigma^j_i\) produced by the tomography procedure. 

We define \(W^j_i\) to be the subspace spanned by the eigenvectors
of \(\hat{\sigma}^j_i\) corresponding to its eigenvalues exceeding $\eta$. We then denote by \(\Pi_{W^j_i}\) the orthogonal projector onto \(W^j_i\).
\end{definition}

\begin{fact}
\label{fact:U-PiW-relation-agnostic}
Let $U^j$ and  $P^j$ as operators from Algorithm \ref{alg:A4}. By Definition \ref{definition:Wji-agnostic} of $W^j_i$ and $\Pi_{W^j_i}$, we have
\[(U^j)^\dagger P^j U^j=\Pi_{W^j_1}\otimes\dots\otimes\Pi_{W^j_{2^{M-j}}}.\]
\end{fact}

\begin{fact}
\label{fact:decompose-projector-agnostic}
For each $j = 1, \dots, M$, the orthogonal complement of 
$\Pi_{W^j_1} \otimes \cdots \otimes \Pi_{W^j_{2^{M-j}}}$ can be decomposed as
\[
    I - \bigotimes_{k=1}^{2^{M-j}} \Pi_{W^j_k}
    = \sum_{i=1}^{2^{M-j}} \left( \bigotimes_{k=1}^{i-1} \Pi_{W^j_k} \right) 
      \otimes \left( I_{\mathcal{B}^j_i} - \Pi_{W^j_i} \right) 
      \otimes \left( \bigotimes_{k=i+1}^{2^{M-j}} I_{\mathcal{B}^j_k} \right),
\]
where $I_{\mathcal{B}^j_i}$ denotes the identity operator on the qudit set $\mathcal{B}^j_i$. 

Here, we adopt the convention that the tensor product over an empty index set (i.e., when the upper limit is less than the lower limit) is defined to be the identity operator on the trivial (one-dimensional) Hilbert space.
\end{fact}

\begin{lemma}[Monotonicity of Step-wise Reconstructed States]
\label{lemma:Monotonicity-lost-agnostic}
For each \( j = 1, \dots, M \), let \( \rho^j \) be the step-wise reconstructed state defined in Definition \ref{definition:rhoj-agnostic}. Then the sequence \( \rho^1, \dots, \rho^M \) satisfies the operator monotonicity property:
\[
\rho^j \preceq \rho^{j-1}, \quad \text{for all } j = 2, \dots, M.
\]
That is, each reconstructed state \( \rho^j \) is dominated by the previous state \( \rho^{j-1} \) in the semidefinite order.
\end{lemma}
\begin{proof}
In line 32 in algorithm \ref{alg:A4}, we define \[
        (\rho^{j})' = \mathrm{Tr}_{\bigcup_{i=1}^{2^{M-j}}\mathcal{B}^j_i\setminus \widetilde{\mathcal{B}}^j_i} \left[ P^j U^j (\rho^{j-1})' (U^j)^\dagger P^j \right],\]
where $P^j=\bigotimes_{i=1}^{2^{M-j}}\left( \ket{0^{f(j,i)}}\bra{0^{f(j,i)}}\right)_{\mathcal{B}^j_i\setminus \widetilde{\mathcal{B}}^j_i}\otimes I_{\widetilde{\mathcal{B}}^j_i}$ is a zero-state projection operators acting on a collection of qudits  \({\bigcup_{i=1}^{2^{M-j}}\mathcal{B}^j_i\setminus \widetilde{\mathcal{B}}^j_i}\) . In other words, we have the relation:
\begin{equation}
\label{eq:rhoj'-agnostic}
    P^{j}\otimes(\rho^{j})' =  P^j U^j (\rho^{j-1})' (U^j)^\dagger P^j.
\end{equation}
    Using the general property of post-selection (i.e., for any PVM element \( \Pi \), \( \Pi \rho \Pi \preceq \rho \)), we obtain the operator inequality:\[ P^{j}\otimes(\rho^{j})' \preceq  U^j (\rho^{j-1})' (U^j)^\dagger. \]
Tensoring both sides with \( \bigotimes_{a=1}^{j-1} P^a \), we have:
\[ \left(\bigotimes_{a=1}^{j}P^a\right)\otimes(\rho^{j})' \preceq  \left(\bigotimes_{a=1}^{j-1}P^a\right) \otimes U^j (\rho^{j-1})' (U^j)^\dagger. \]
 Next, apply the conjugation map \( (E^j)^\dagger (\cdot) E^j \), where \( E^j = U^j U^{j-1} \cdots U^1 \). Using the definition of \( \rho^{j-1} \),
\[
\rho^{j-1} = (E^j)^\dagger \left[ \left( \bigotimes_{a=1}^{j-1} P^a \right) \otimes U^j (\rho^{j-1})' (U^j)^\dagger \right] E^j,
\]we conclude\[\rho^j\preceq\rho^{j-1}.\]
\end{proof}

\begin{lemma}[Lemma B.9 in \cite{bakshi2025learning}]
\label{lemma:tomography-error-agnostic}
Let $\rho, \sigma$ be unnormalized mixed states satisfying $\|\rho-\sigma\|_1\le \eta$. Let $W$ be the span of all eigenvectors of $\sigma$ with eigenvalues exceeding $\eta$. Then, letting $\Pi_W$ denote orthogonal projection onto $W$,
we have that $\|(I-\Pi_W)^\dagger \rho (I-\Pi_W)\|_\infty \le 2\eta$.
\end{lemma}

\begin{lemma}
\label{lemma:bound-fidelity-varphirho-agnostic}
Fix \(j \in \{1,\dots, M\}\). For each \(i = 1, \dots, 2^{M-j}\), let \(\mathbb{H}_{\mathcal{B}^j_i}\) be the Hilbert space associated with \(\mathcal{B}^j_i\). Let \(W^j_i \subseteq \mathbb{H}_{\mathcal{B}^j_i}\) be a subspace, and let \(\Pi_{W^j_i}\) denote the orthogonal projector onto \(W^j_i\). Let \( \mathcal{B}^j \coloneq \bigcup_{k=1}^{2^{M-j}} \mathcal{B}^j_k\) be the collection of all qudits under consideration at step \(j\). Let \((\rho^{j-1})'\) be a density matrix on the Hilbert space associated with \(\mathcal{B}^j\), and let \(\ket{\varphi}\) be a unit vector in the same Hilbert space that has Schmidt rank at most $D^2$ across any bipartition of $ \mathcal{B}^j_i$ and $\mathcal{B}^{j} \setminus \mathcal{B}^j_i$. Define the product projector \(\Pi_{W^j} \coloneq \bigotimes_{i=1}^{2^{M-j}} \Pi_{W^j_i}.\)
Suppose that
\[
    (I - \Pi_{W^j}) \ket{\varphi} = \ket{\varphi},
\]
and that for each \(i = 1, \dots, 2^{M-j}\),
\[
    \left\| (I - \Pi_{W^j_i})\,
    \mathrm{Tr}_{\,(\mathcal{B}^{j} \setminus \mathcal{B}^j_i)}\big[(\rho^{j-1})'\big]\,
    (I - \Pi_{W^j_i}) \right\|_\infty \leq 2\eta.
\]
Then
\[
    \bra{\varphi} (\rho^{j-1})' \ket{\varphi} \leq 2 \eta D^2 \, 2^{M-j}.
\]
\end{lemma}
\begin{proof}
According to Fact~\ref{fact:decompose-projector-agnostic}, any unit vector \(\ket{\varphi}\) satisfying \((I - \Pi_{W^j}) \ket{\varphi} = \ket{\varphi}\) can be decomposed as
\[
    \ket{\varphi} = \sum_{i=1}^{2^{M-j}} \ket{\varphi_i},
\]
where we define
\begin{equation}
\label{eq:decompose-varphi-agnostic}
  \ket{\varphi_i} \coloneq 
    \left( \bigotimes_{k=1}^{i-1} \Pi_{W^j_k} \right) 
    \otimes \left( I_{\mathcal{B}^j_i} - \Pi_{W^j_i} \right) 
    \otimes \left( \bigotimes_{k=i+1}^{2^{M-j}} I_{\mathcal{B}^j_k} \right) 
    \ket{\varphi},  
\end{equation}
and we set \(p_i \coloneq \|\ket{\varphi_i}\|^2\), so that \(\sum_{i=1}^{2^{M-j}} p_i = 1\).

With this decomposition, we have
\begin{align}
\label{eq:phirho'phi-agnostic}
    &\bra{\varphi} (\rho^{j-1})' \ket{\varphi}
    = \sum_{i=1}^{2^{M-j}} \bra{\varphi_i} (\rho^{j-1})' \ket{\varphi_i}
    + \sum_{\substack{i,k=1 \\ i \neq k}}^{2^{M-j}}
        \bra{\varphi_i} (\rho^{j-1})' \ket{\varphi_k}\nonumber\\
    &=\sum_{i=1}^{2^{M-j}} p_i\frac{\bra{\varphi_i}}{\|\bra{\varphi_i}\|} (\rho^{j-1})' \frac{\ket{\varphi_i}}{\|\ket{\varphi_i}\|} 
    + \sum_{\substack{i,k=1 \\ i \neq k}}^{2^{M-j}}
         \sqrt{p_ip_k}\frac{\bra{\varphi_i}}{\|\bra{\varphi_i}\|} \sqrt{(\rho^{j-1})'}\sqrt{(\rho^{j-1})'} \frac{\ket{\varphi_k}}{\|\ket{\varphi_k}\|}\nonumber\\
    &\leq \sum_{i=1}^{2^{M-j}} p_i\frac{\bra{\varphi_i}}{\|\bra{\varphi_i}\|} (\rho^{j-1})' \frac{\ket{\varphi_i}}{\|\ket{\varphi_i}\|} 
    + \sum_{\substack{i,k=1 \\ i \neq k}}^{2^{M-j}}
         \sqrt{p_ip_k}\left\|\frac{\bra{\varphi_i}}{\|\bra{\varphi_i}\|} \sqrt{(\rho^{j-1})'}\right\|\left\|\sqrt{(\rho^{j-1})'} \frac{\ket{\varphi_k}}{\|\ket{\varphi_k}\|}\right\|\nonumber\\
    &=\sum_{i=1}^{2^{M-j}} p_i\frac{\bra{\varphi_i}}{\|\bra{\varphi_i}\|} (\rho^{j-1})' \frac{\ket{\varphi_i}}{\|\ket{\varphi_i}\|} 
    + \sum_{\substack{i,k=1 \\ i \neq k}}^{2^{M-j}}
         \sqrt{p_ip_k}\sqrt{\frac{\bra{\varphi_i}}{\|\bra{\varphi_i}\|} (\rho^{j-1})'\frac{\ket{\varphi_i}}{\|\ket{\varphi_i}\|}}\sqrt{ \frac{\bra{\varphi_k}}{\|\bra{\varphi_k}\|}(\rho^{j-1})'\frac{\ket{\varphi_k}}{\|\ket{\varphi_k}\|}}.
\end{align}
Next, since $\ket{\varphi}$ has Schmidt rank at most $D^2$ across any bipartition of $\mathcal{B}^j_i$ and $\mathcal{B}^j \setminus \mathcal{B}^j_i$. Hence for each $i$ we can write
\[
    \ket{\varphi}=\sum_{l=1}^{D^2}\ket{\alpha^l}_{\mathcal{B}^j_i}\otimes \ket{\beta^l}_{\mathcal{B}^j \setminus \mathcal{B}^j_i},
\]
where $\{\ket{\alpha^l}\}$ and $\{\ket{\beta^l}\}$ are vectors in the respective subsystems.  Since in Equation \ref{eq:decompose-varphi-agnostic} the operator $I_{\mathcal{B}^j_i}-\Pi_{W^j_i}$ acts only on the $\ket{\alpha^l}_{\mathcal{B}^j_i}$ components while the other projectors act only on the $\ket{\beta^l}_{\mathcal{B}^j \setminus \mathcal{B}^j_i}$ components, it follows that each $\ket{\varphi_i}$ inherits the same Schmidt rank bound, i.e., $\mathrm{rank}(\ket{\varphi_i}) \leq D^2$.

Therefore, the normalized state $\frac{\ket{\varphi_i}}{\|\varphi_i\|}$ can be expressed as a superposition of at most $D^2$ product terms across the bipartition $(\mathcal{B}^j_i,\mathcal{B}^j\setminus\mathcal{B}^j_i)$. More precisely, there exist orthonormal vectors $\{\ket{\xi^l_{(W^j_i)^\perp}}_{\mathcal{B}^j_i}\}_{l=1}^{D^2}$ in $(W^j_i)^\perp$ and unit vectors $\{\ket{\mu^l}_{\mathcal{B}^j \setminus \mathcal{B}^j_i}\}_{l=1}^{D^2}$ such that
\[
    \frac{\ket{\varphi_i}}{\|\varphi_i\|} 
    = \sum_{l=1}^{D^2} \sqrt{q_l}\,
      \ket{\xi^l_{(W^j_i)^\perp}}_{\mathcal{B}^j_i} 
      \otimes \ket{\mu^l}_{\mathcal{B}^j \setminus \mathcal{B}^j_i},
\]
where $q_l \geq 0$, $\sum_{l=1}^{D^2} q_l = 1$, and $\|\ket{\xi^l_{(W^j_i)^\perp}}\|=\|\ket{\mu^l}\|=1$ for all $l$. For brevity, we henceforth omit explicit tensor symbols and subsystem labels when the context is clear, writing $\ket{\xi^l_{(W^j_i)^\perp}}\ket{\mu^l}$ instead of $\ket{\xi^l_{(W^j_i)^\perp}}_{\mathcal{B}^j_i}\otimes \ket{\mu^l}_{\mathcal{B}^j \setminus \mathcal{B}^j_i}$.

Substituting this decomposition into the expression above yields
\begin{align}
\label{eq:nphirho'nphi1-agnostic}
    &\frac{\bra{\varphi_i}}{\|\varphi_i\|}
    (\rho^{j-1})'
    \frac{\ket{\varphi_i}}{\|\varphi_i\|}
    \nonumber\\
    &= \sum_{l=1}^{D^2} q_l
    \bra{\xi^l_{(W^j_i)^\perp}} \bra{\mu^l}
    (\rho^{j-1})'\ket{\xi^l_{(W^j_i)^\perp}}\ket{\mu^l} + \sum_{\substack{l,m=1 \\ m \neq l}}^{D^2}
    \sqrt{q_l q_m} \;
    \bra{\xi^l_{(W^j_i)^\perp}} \bra{\mu^l}
    (\rho^{j-1})'\ket{\xi^m_{(W^j_i)^\perp}}\ket{\mu^m}.
\end{align}
For the first term, note that
\begin{align*}
    \sum_{l=1}^{D^2} q_l
    \bra{\xi^l_{(W^j_i)^\perp}} \bra{\mu^l}
    (\rho^{j-1})'\ket{\xi^l_{(W^j_i)^\perp}}\ket{\mu^l}
    &\leq \sum_{l=1}^{D^2} q_l
    \bra{\xi^l_{(W^j_i)^\perp}} \mathrm{Tr}_{\mathcal{B}^j \setminus \mathcal{B}^j_i}\left[
    (\rho^{j-1})'\right]\ket{\xi^l_{(W^j_i)^\perp}}\\
    &\leq \sum_{l=1}^{D^2} q_l 2 \eta \\
    &\leq 2\eta,
\end{align*}
where the first inequality follows from the fact that, for any positive semidefinite operator $\sigma$ on $\mathcal B_i^j\otimes(\mathcal B^j\setminus \mathcal B_i^j)$, $\bra{x}\bra{\mu}\sigma\ket{x}\ket{\mu} \le \bra{x}\operatorname{Tr}_{\mathcal B^j\setminus \mathcal B_i^j}(\sigma)\ket{x}.$ The second inequality follows from the assumption (stated earlier) that
\[
    \left\| (I - \Pi_{W^j_i})\,
    \mathrm{Tr}_{\,(\mathcal{B}^{j} \setminus \mathcal{B}^j_i)}\big[(\rho^{j-1})'\big]\,
    (I - \Pi_{W^j_i}) \right\|_\infty \leq 2\eta \qquad \forall i=1,\dots, 2^{M-j}
\]
For the second term, applying Cauchy--Schwarz inequality, we obtain
\begin{align*}
    &\sum_{\substack{l,m=1 \\ m \neq l}}^{D^2} \sqrt{q_l q_m}
        \bra{\xi^l_{(W^j_i)^\perp}}\bra{\mu^l} (\rho^{j-1})' \ket{\xi^m_{(W^j_i)^\perp}}\ket{\mu^m}\\
    &=\sum_{\substack{l,m=1 \\ m \neq l}}^{D^2}  \sqrt{q_l q_m}
        \bra{\xi^l_{(W^j_i)^\perp}}\bra{\mu^l} \sqrt{(\rho^{j-1})'}\sqrt{(\rho^{j-1})'}  \ket{\xi^m_{(W^j_i)^\perp}}\ket{\mu^m}\\
    &\leq \sum_{\substack{l,m=1 \\ m \neq l}}^{D^2}  \sqrt{q_l q_m}
        \left\|\bra{\xi^l_{(W^j_i)^\perp}}\bra{\mu^l} \sqrt{(\rho^{j-1})'}\right\|  \left\|\sqrt{(\rho^{j-1})'}  \ket{\xi^m_{(W^j_i)^\perp}}\ket{\mu^m}\right\|\\
    &=\sum_{l=1}^{D^2}{\sqrt{q_l} \sqrt{\bra{\xi^l_{(W^j_i)^\perp}}\bra{\mu^l} (\rho^{j-1})' \ket{\xi^l_{(W^j_i)^\perp}}\ket{\mu^l}}}
    \sum_{\substack{m=1 \\ m \neq l}}^{D^2} \sqrt{q_m} {\sqrt{\bra{\xi^m_{(W^j_i)^\perp}}\bra{\mu^m} (\rho^{j-1})' \ket{\xi^m_{(W^j_i)^\perp}}\ket{\mu^m}}}\nonumber\\
    &\leq \sum_{l=1}^{D^2}{\sqrt{q_l } \sqrt{\bra{\xi^l_{(W^j_i)^\perp}} \mathrm{Tr}_{\mathcal{B}^j \setminus \mathcal{B}^j_i}\left[
    (\rho^{j-1})'\right]\ket{\xi^l_{(W^j_i)^\perp}}}}
    \sum_{\substack{m=1 \\ m \neq l}}^{D^2} \sqrt{q_m} {\sqrt{\bra{\xi^m_{(W^j_i)^\perp}} \mathrm{Tr}_{\mathcal{B}^j \setminus \mathcal{B}^j_i}\left[
    (\rho^{j-1})'\right]\ket{\xi^m_{(W^j_i)^\perp}}}}\\
    &\leq \sum_{l=1}^{D^2}{\sqrt{q_l } \sqrt{2\eta}}
    \sum_{\substack{m=1 \\ m \neq l}}^{D^2} \sqrt{q_m} {\sqrt{2\eta}}
\end{align*}
where we do the same trick to the last two inequalities as in the first term.

Substituting these two terms back to Equation \eqref{eq:nphirho'nphi1-agnostic}, we get
\begin{align}
\label{eq:nphirho'nphi2-agnostic}
    &\frac{\bra{\varphi_i}}{\|\varphi_i\|}
    (\rho^{j-1})'
    \frac{\ket{\varphi_i}}{\|\varphi_i\|}
    \leq 2\eta + 2\eta \sum_{l=1}^{D^2}\sqrt{q_l } \sum_{\substack{m=1 \\ m \neq l}}^{D^2} \sqrt{q_m} = 2\eta + 2\eta \sum_{l=1}^{D^2}\sqrt{q_l } \left[\left(\sum_{m=1}^{D^2} \sqrt{q_m}\right)-\sqrt{q_l}\right]\nonumber\\
    &=2\eta + 2\eta \left[\left(\sum_{m=1}^{D^2} \sqrt{q_m}\right)^2
-\sum_{l=1}^{D^2} q_l\right]\leq 2\eta+ 2\eta\left[D^2-1\right]=2\eta D^2,
\end{align}
where in the second to last inequality we use the Cauchy--Schwarz inequality, and in the last inequality follows by the fact that \(W^j_i\) is a subspace of \(\mathbb{H}_{\mathcal{B}^j_i}\).

Now, let's substitute Equation \eqref{eq:nphirho'nphi2-agnostic} back to Equation \eqref{eq:phirho'phi-agnostic}:
\begin{align}
    \bra{\varphi} (\rho^{j-1})' \ket{\varphi} 
&\leq 2\eta D^2  \left(\sum_{i=1}^{2^{M-j}} p_i\right)
    + 2\eta D^2 \left(\sum_{\substack{i,k=1 \\ i \neq k}}^{2^{M-j}}\sqrt{p_ip_k}\right) \nonumber \\
&= 2\eta D^2 + 2\eta D^2 \sum_{i=1}^{2^{M-j}}\sqrt{p_i} \left[\left(\sum_{k=1}^{2^{M-j}}\sqrt{p_k}\right)-\sqrt{p_i}\right]\nonumber\\
&= 2\eta D^2  
    + 2\eta D^2  \left[\left(\sum_{k=1}^{2^{M-j}}\sqrt{p_k}\right)^2-\sum_{i=1}^{2^{M-j}}p_i\right] \nonumber\\
&\leq 2\eta D^2  
    + 2\eta   \left(2^{M-j}-1\right)\nonumber\\
&=2\eta D^2 2^{M-j},
\end{align}
where in the last inequality we use the Cauchy--Schwarz inequality.
\end{proof}

\begin{lemma}
\label{lemma:tomography-error-pass-agnostic}
Fix $j \in \{1,\dots,M\}$. For each \(i = 1, \dots, 2^{M-j}\), let \(\mathbb{H}_{\mathcal{B}^j_i}\) denote the Hilbert space associated with \(\mathcal{B}^j_i\), where \(\mathcal{B}^j_i\) is defined in Algorithm \ref{alg:A4}. Let \(W^j_i \subseteq \mathbb{H}_{\mathcal{B}^j_i}\) be a subspace defined in Definition \ref{definition:Wji-agnostic}, and denote by \(\Pi_{W^j_i}\) the orthogonal projection onto \(W^j_i\). Let \(\mathcal{B}^j \coloneqq \bigcup_{k=1}^{2^{M-j}} \mathcal{B}^j_k\) be the set of all qudits under consideration at step \(j\). Let \((\rho^{j-1})'\) be a density matrix on \(\mathbb{H}_{\mathcal{B}^j}\) defined in Algorithm \ref{alg:A4}, and let \(\rho^j\) be the density matrix defined in Definition~\ref{definition:rhoj-agnostic}.

Suppose that, for each \(i = 1, \dots, 2^{M-j}\), the tomography procedure succeeds in the sense that
\[
    \big\| (I - \Pi_{W^j_i})\,
    \mathrm{Tr}_{\mathcal{B}^{j} \setminus \mathcal{B}^j_i}\big[(\rho^{j-1})'\big]\,
    (I - \Pi_{W^j_i}) \big\|_\infty \;\leq\; 2\eta.
\]
Then, for any $n$-qudit matrix product state $\ket{\phi}$ with bond dimension $D$,
\[
    \big| \bra{\phi} \rho^{j-1} \ket{\phi}
    - \bra{\phi} \rho^{j} \ket{\phi} \big|
    \;\leq\;
    2 \sqrt{\,4 \eta \, D^2 \, 2^{M-j}} \, .
\]
\end{lemma}

\begin{proof}
According to Definition \ref{definition:rhoj-agnostic} and \ref{definition:psij-agnostic}, we can express the fidelity $\bra{\phi} \rho^{j} \ket{\phi} $ as
\begin{align}
    \bra{\phi} \rho^{j} \ket{\phi} 
    &= \bra{\phi}  (E^j)^\dagger \left[\left(\bigotimes_{a=1}^{j}P^a\right)\otimes(\rho^{j})' \right]E^j\ket{\phi} \nonumber\\
    &= \bra{\phi^{j-1}}  (U^j)^\dagger \left[\left(\bigotimes_{a=1}^{j}P^a\right)\otimes(\rho^{j})' \right]U^j\ket{\phi^{j-1}} \nonumber\\
    &= \bra{\phi^{j-1}}  \left(\bigotimes_{a=1}^{j-1}P^a\right) (U^j)^\dagger\left[ P^j\otimes(\rho^{j})' \right]U^j \left(\bigotimes_{a=1}^{j-1}P^a\right)\ket{\phi^{j-1}} \nonumber\\
   &=    \bra{\psi^{j-1}} (U^j)^\dagger\left[ P^j\otimes(\rho^{j})'\right] U^j \ket{\psi^{j-1}}
\end{align}
Use Equation \eqref{eq:rhoj'-agnostic} and Fact \ref{fact:U-PiW-relation-agnostic}, we can rewrite the fidelity $\bra{\phi} \rho^{j} \ket{\phi} $ as 
\begin{equation}
    \bra{\phi} \rho^{j} \ket{\phi} =\bra{\psi^{j-1}}\Pi_{W^j_1}\otimes\dots\otimes\Pi_{W^j_{2^{M-j}}}(\rho^{j-1})' \;\Pi_{W^j_1}\otimes\dots\otimes\Pi_{W^j_{2^{M-j}}}\ket{\psi^{j-1}}
\end{equation}
For brevity, we denote \(\Pi_{W^j} \coloneq \bigotimes_{i=1}^{2^{M-j}} \Pi_{W^j_i}.\) Then together with Fact \ref{fact:fidelity-rep-agnostic}, this implies
\begin{align}
      \left| \bra{\phi} \rho^{j-1} \ket{\phi}
    - \bra{\phi} \rho^{j} \ket{\phi} \right|
    &= \left|\bra{\psi^{j-1}} (\rho^{j-1})' \ket{\psi^{j-1}}-\bra{\psi^{j-1}}  \Pi_{W^j}(\rho^{j-1})' \,\Pi_{W^j}\ket{\psi^{j-1}}\right| \nonumber\\
&\leq \left|\bra{\psi^{j-1}}  (\rho^{j-1})' \,(I-\Pi_{W^j})\ket{\psi^{j-1}}\right|+\left|\bra{\psi^{j-1}}  (I-\Pi_{W^j})(\rho^{j-1})' \,\Pi_{W^j}\ket{\psi^{j-1}}\right| \nonumber\\
&\leq \left\|\ket{\psi^{j-1}}\right\|  \left\|(\rho^{j-1})' \,(I-\Pi_{W^j})\ket{\psi^{j-1}} \right\|+
\left\|\Pi_{W^j}\ket{\psi^{j-1}}\right\| \left\|(\rho^{j-1})' \,(I-\Pi_{W^j})\ket{\psi^{j-1}} \right\| \nonumber \\
&\leq 2 \left\|(\rho^{j-1})' \,(I-\Pi_{W^j})\ket{\psi^{j-1}} \right\|
\end{align}
Let $\ket{\widetilde{\varphi}} = (I - \Pi_{W^j}) \ket{\psi^{j-1}}$, and define the normalized vector $\ket{\varphi} \coloneq \frac{\ket{\widetilde{\varphi}}}{\|\widetilde{\varphi}\|}$. Then
\begin{align}
2 \left\| (\rho^{j-1})' \,(I-\Pi_{W^j})\ket{\psi^{j-1}} \right\|
&= 2 \left\| (\rho^{j-1})' \,\ket{\varphi} \right\| \,\big\|\widetilde{\varphi}\big\| \nonumber\\
&\le 2 \big\|\widetilde{\varphi}\big\| \sqrt{\bra{\varphi} [(\rho^{j-1})']^2 \ket{\varphi}} \nonumber\\
&\le 2 \big\|\widetilde{\varphi}\big\| \sqrt{\bra{\varphi} (\rho^{j-1})' \ket{\varphi}} \nonumber\\
&\le 2 \sqrt{\bra{\varphi} (\rho^{j-1})' \ket{\varphi}} .
\end{align}
For each $j \in \{1,\dots,M\}$ and $i \in \{1,\dots, 2^{M-j}\}$, define the root set for the block $\mathcal{B}^j_i$ at the first layer by
\[
S(\mathcal{B}^j_i) \coloneq \bigcup_{a=0}^{2^{j-1}-1} \mathcal{B}^1_{2(j-1)i - a}.
\]
Since $\ket{\phi}$ is a matrix product state of bond dimension $D$, for the bipartition $(S(\mathcal{B}^j_i), [n]\setminus S(\mathcal{B}^j_i))$, we can write
\[
\ket{\phi} = \sum_{l=1}^{D^2} \ket{\alpha_l}_{S(\mathcal{B}^j_i)} \otimes \ket{\beta_l}_{[n] \setminus S(\mathcal{B}^j_i)}.
\]
Due to the binary tree structure of the cumulative unitary $E^j \coloneq U^{j} U^{j-1} \cdots U^{1}$, where $U^a = \bigotimes_{b=1}^{2^{M-a}} U^a_b$, no single unitary $U^a_b$ acts across the bipartition. Each acts either entirely on $S(\mathcal{B}^j_i)$ or entirely on $[n]\setminus S(\mathcal{B}^j_i)$. Therefore, the Schmidt rank across this bipartition is preserved:
\[
E^j \ket{\phi} = \sum_{l=1}^{D^2} \ket{\alpha'_l}_{S(\mathcal{B}^j_i)} \otimes \ket{\beta'_l}_{[n] \setminus S(\mathcal{B}^j_i)}.
\]
As a consequence, the projected (residual) state at level $j$,
\[
\ket{\psi^{j-1}} \coloneq 
\left(\bigotimes_{a=1}^j \bigotimes_{b=1}^{2^{M-a}} \bra{0^{f(a,b)}}_{\mathcal{B}^a_b\setminus \widetilde{\mathcal{B}}^a_b}\right) \Pi^{(j)} E^j \ket{\phi},
\]
also has Schmidt rank at most $D^2$ across $(\mathcal{B}^j_i, \mathcal{B}^j \setminus \mathcal{B}^j_i)$.

For a fixed bipartition \((\mathcal B_i^j,\mathcal B^j\setminus \mathcal B_i^j)\), write \(\Pi_{W^j}=\Pi_{W_i^j}\otimes \Pi_{\mathrm{rest}},\)
where \(\Pi_{\mathrm{rest}}=\bigotimes_{k\neq i}\Pi_{W_k^j}.\) Then
\[
I-\Pi_{W^j}
=
(I-\Pi_{W_i^j})\otimes I
+
\Pi_{W_i^j}\otimes (I-\Pi_{\mathrm{rest}}).
\]
Each term is local with respect to this bipartition and therefore maps a state of Schmidt rank at most \(D^2\) to a state of Schmidt rank at most \(D^2\). Hence their sum has Schmidt rank at most \(2D^2\). Thus \(\ket{\widetilde\varphi}=(I-\Pi_{W^j})\ket{\psi^{j-1}}
\) and its normalized version \(\ket{\varphi}\), whenever well-defined, have Schmidt rank at most \(2D^2\) across this bipartition. Moreover, \((I-\Pi_{W^j})\ket{\varphi}=\ket{\varphi}.\) Therefore, by Lemma \ref{lemma:bound-fidelity-varphirho-agnostic}, we conclude
\begin{align}
\left| \bra{\phi} \rho^{j-1} \ket{\phi}
    - \bra{\phi} \rho^{j} \ket{\phi}\right|
&\le 2 \sqrt{\bra{\varphi} (\rho^{j-1})' \ket{\varphi}} 
\le 2\sqrt{4\eta D^2 2^{M-j}}.
\end{align}
% % The projector $(I-\Pi_{W^j})$ only removes components in $W^j$ and does not introduce new correlations across this bipartition. Hence, both $\ket{\widetilde{\varphi}} = (I - \Pi_{W^j}) \ket{\psi^{j-1}}$ and its normalized version $\ket{\varphi}$ maintain Schmidt rank at most $D^2$. Moreover, note that $(I-\Pi_{W^j})\ket{\varphi} = \ket{\varphi}$. Therefore, by Lemma \ref{lemma:bound-fidelity-varphirho-agnostic}, we conclude
% \begin{align}
% \left| \bra{\phi} \rho^{j-1} \ket{\phi}
%     - \bra{\phi} \rho^{j} \ket{\phi}\right|
% &\le 2 \sqrt{\bra{\varphi} (\rho^{j-1})' \ket{\varphi}} 
% \le 2\sqrt{2\eta D^2 2^{M-j}}.
% \end{align}
\end{proof}

\begin{theorem}[Restatement of Theorem \ref{theorem:2}]
Given access to copies of an $n$-qudit state $\rho$, as well as parameters $\epsilon , \delta\in (0,1) $, and a bond dimension parameter $D$. Algorithm~\ref{alg:A4} outputs a description of a state $\ket{\hat{\phi}}$, such that, with probability $1-\delta$,
\[
\bra{\hat{\phi}} \rho \ket{\hat{\phi}} \;\geq\; \max_{\ket{\phi}\in \rm \text{MPS}(D)}\bra{\phi} \rho \ket{\phi} - \epsilon
\]
The algorithm requires \(N=O\!\Biggl(
\frac{\displaystyle D^{12} n^{7}\, d^{4} (\ln d)^{7}\, \log\!\tfrac{n}{\delta}}
{\displaystyle
\epsilon^{12}\!\Bigl[
  \ln(\ln d)
  + \ln\!\tfrac{128\,n\,D^2}{(\sqrt2-1)^2\epsilon^2}
\Bigr]^{7}}
\Biggr)\) copies of $\rho$ and runs in time $\text{poly}\left(D,n,\frac{1}{\epsilon},\log(\frac{1}{\delta})\right)$.
\end{theorem}
\begin{proof}
Suppose that the state tomography procedure succeeds for all $\sigma^j_i$ throughout the algorithm. Then, for each $\sigma^j_i$, the tomography produces an estimate $\hat{\sigma}^j_i$ satisfying
\[
\|\sigma^j_i - \hat{\sigma}^j_i\|_1 \;\leq\; \frac{(\sqrt{2}-1)^2 \epsilon^2 p}{128 D^2 n}.
\]
In the next step, Algorithm \ref{alg:A4} invokes Algorithm \ref{alg:A3} to select eigenvectors corresponding to eigenvalues exceeding $\eta$ from the estimated density matrix $\hat{\sigma}^j_i$ to construct a disentangling unitary accordingly. We denote \(\widetilde{W}^j_i\) to be the subspace spanned by the eigenvectors of \(\hat{\sigma}^j_i\) with eigenvalues exceeding $\eta$. From Lemma \ref{lemma:tomography-error-agnostic}, we know that 

\[\|(I-\Pi_{\widetilde{W}^j_i})^\dagger \sigma^j_i (I-\Pi_{\widetilde{W}^j_i})\|_\infty \leq \frac{(\sqrt{2}-1)^2 \epsilon^2 p}{64 D^2 n}.
\]

Since we have $\widetilde{W}^j_i \subseteq W^j_i$ \footnote{In Algorithm \ref{alg:A3}, we denote $m$ the number of eigenvector exceeding $\eta$. Since the trace of $\hat{\sigma}^j_i$ is 1, $m$ must be smaller than $1/\eta$. Then, by the definition of \(t \coloneq \left\lceil \log_d m \right\rceil\) and \(p \coloneq \left\lceil \log_d (1/\eta) \right\rceil\), we get dim$(\widetilde{W}^j_i)\le$dim$(W^j_i)$}, where $W^j_i$ is defined in Definition \ref{definition:Wji-agnostic}. Equivalently, their orthogonal complements satisfy
$$(W^j_i)^\perp \subseteq (\widetilde{W}^j_i)^\perp$$
Therefore, it follows that
\[\|(I-\Pi_{W^j_i})^\dagger \sigma^j_i (I-\Pi_{W^j_i})\|_\infty \leq \frac{(\sqrt{2}-1)^2 \epsilon^2 p}{64D^2 n}.
\]
Let $\ket{\phi} \in (\mathbb{C}^d)^{\otimes n}$ be a matrix product state with bond dimension $D$ on an $n$-qudit system. Applying Lemma \ref{lemma:tomography-error-pass-agnostic}, we get
\[
    \big| \bra{\phi} \rho^{j-1} \ket{\phi}
    - \bra{\phi} \rho^{j} \ket{\phi} \big|
    \;\leq\;
    \frac{(\sqrt{2}-1)\epsilon}{2} \sqrt{\,\frac{ p}{ 2n} \, 2^{M-j}} \, .
\]
Using the triangular inequality, we derive
\[\big| \bra{\phi} \rho^{0} \ket{\phi}
    - \bra{\phi} \rho^{M} \ket{\phi} \big|
    \;\leq\;
    \sum_{j=1}^{M}\frac{(\sqrt{2}-1)\epsilon}{2} \sqrt{\,\frac{ p}{ 2n} \, 2^{M-j}}\le\frac{\epsilon}{2} \sqrt{\,\frac{ p}{2 n} 2^M\,}\le\frac{\epsilon}{2} ,\]
where in the last inequality we use the fact that $n>p 2^{M-1}$ (see line 3 in Algorithm \ref{alg:A4}). This implies 
\begin{equation}
\label{proof:t6-agnostic}
\bra{\phi} \rho^{M} \ket{\phi} \geq \bra{\phi} \rho^{0} \ket{\phi} -\frac{\epsilon}{2}. 
\end{equation}
For any matrix product state $\ket{\phi} \in (\mathbb{C}^d)^{\otimes n}$ with bond dimension $D$, we define its level-$j$ projected states $\ket{\psi^{j}} \in \mathbb{H}_{\mathcal{\widetilde{B}}^j_1}\otimes\dots\otimes\mathbb{H}_{\widetilde{\mathcal{B}}^j_{2^{M-j}}}$ through Definition \ref{definition:psij-agnostic}. By Fact~\ref{fact:fidelity-rep-agnostic}, we have
\begin{equation}
\label{proof-t1-agnostic}
\bra{\phi}\rho^{M}\ket{\phi} = \bra{\psi^{M}}(\rho^{M})'\ket{\psi^{M}}.
\end{equation}
Similarly, for the reconstructed state \(\ket{\hat{\phi}} = (E^j)^\dagger \left( \bigotimes_{j=1}^{M}\bigotimes_{i=1}^{2^{M-j}} \ket{0^{f(j,i)}}_{\mathcal{B}^j_i\setminus \widetilde{\mathcal{B}}^j_i}
\otimes \ket{\hat{\psi}} \right),\) we obtain
\begin{equation}
\label{proof-t2-agnostic}
\bra{\hat{\phi}}\rho^{M}\ket{\hat{\phi}} = \bra{\hat{\psi}}(\rho^{M})'\ket{\hat{\psi}}.
\end{equation}
If the final tomography on $(\rho^M)'$ succeeds, it yields an estimate $(\hat{\rho}^M)'$ such that 
$\|(\rho^M)' - (\hat{\rho}^M)'\|_1 \leq \epsilon/4$. By H\"older's inequality for Schatten $p$-norms, for any state $\ket{\varphi} \in \mathbb{H}_{\mathcal{\widetilde{B}}^M_1}$,
\begin{equation}
\label{proof-t3-agnostic}
\left| \bra{\varphi}(\rho^M)'\ket{\varphi} - \bra{\varphi}(\hat{\rho}^M)'\ket{\varphi} \right| \leq \epsilon/4.    
\end{equation}
Applying this bound gives
\begin{equation}
\label{proof-t4}
\bra{\hat{\psi}}(\rho^{M})'\ket{\hat{\psi}}
\geq \bra{\hat{\psi}}(\hat{\rho}^{M})'\ket{\hat{\psi}} - \epsilon/4
\geq \bra{\psi^M}(\hat{\rho}^{M})'\ket{\psi^M} - \epsilon/4
\geq \bra{\psi^M}(\rho^{M})'\ket{\psi^M} - \epsilon/2.
\end{equation}
Here, we use Equation \eqref{proof-t3-agnostic} in the first and last inequalities, while the second step follows from the fact that $\ket{\hat{\psi}}$ is the largest eigenvector of $(\hat{\rho}^M)'$. Combining with Equations \eqref{proof-t1-agnostic} and \eqref{proof-t2-agnostic}, we obtain
\begin{equation}
\label{proof:t5}
\bra{\hat{\phi}}\rho^{M}\ket{\hat{\phi}} \geq \bra{\phi}\rho^{M}\ket{\phi}-\frac{\epsilon}{2}.
\end{equation}
Together with Equation \eqref{proof:t6-agnostic}, this implies
\begin{equation}
\label{proof-t7-agnostic}
\bra{\hat{\phi}}\rho^{M}\ket{\hat{\phi}} \geq \bra{\phi} \rho^{0} \ket{\phi} -\epsilon
\end{equation}
Finally, applying Lemma~\ref{lemma:Monotonicity-lost-agnostic} to Equation \eqref{proof-t7-agnostic}, we obtain
\begin{equation}
\bra{\hat{\phi}}\rho^{0}\ket{\hat{\phi}} \geq \bra{\hat{\phi}}\rho^{M}\ket{\hat{\phi}} \geq \bra{\phi} \rho^{0} \ket{\phi} -\epsilon.
\end{equation}
Since $\rho^0\coloneq \rho$ by Definition \ref{definition:rhoj-agnostic}, the desired inequality  is established:
\begin{equation}
\bra{\hat{\phi}}\rho\ket{\hat{\phi}} \geq \bra{\phi} \rho \ket{\phi} -\epsilon.
\end{equation}
Next, we compute the sample and time complexity of Algorithm \ref{alg:A4}. From Lemma \ref{lemma:subnorm-tomo-rank-agnostic}, taking the dimension $d^{r-i}$ to be $d^{2p}$, the error parameter to be $\eta$, and the failure probability to be $\delta/n$, we can perform tomography on each sub-normalized state $\sigma^j_i$ with $O\left(\frac{d^{4p}}{\eta^{2}}
\log\!\frac{n}{\delta}
\right)$ copies of sample, and runtime that is polynomial in the same parameters. In total, the algorithm performs sub-normalized tomography $\sum_{j=1}^M 2^{M-j}=2^M-1=O(n/p)$ times. Therefore, the sample complexity needed to perform tomography on all $\sigma^j_i$ is $$N=O\left(\frac{nd^{4p}}{p\eta^{2}}
\log\!\frac{n}{\delta}
\right).$$
Notice that $d^{p}<d/\eta$ and $\eta =\frac{(\sqrt{2}-1)^2 \epsilon^2 p}{128 D^2 n}$. Using the asymptotic notation of $p = \Theta\left(\ln (nD^2\ln d /\epsilon^2)\right)$, we can write 
\[N=O\left(\frac{nd^{4}}{p\eta^{6}}
\log\!\frac{n}{\delta}
\right)= O\!\Biggl(
\frac{\displaystyle D^{12} n^{7}\, d^{4} (\ln d)^{7}\, \log\!\tfrac{n}{\delta}}
{\displaystyle
\epsilon^{12}\!\Bigl[
  \ln(\ln d)
  + \ln\!\tfrac{128\,n\,D^2}{(\sqrt2-1)^2\epsilon^2}
\Bigr]^{7}}
\Biggr).\]
See Appendix \ref{appendix-2} for a more detailed derivation. The last call of tomography on $(\rho^M)'$ takes the dimension $d^{r-i}$ to be $d^{p}$, the error parameter to be $\tau=\epsilon/4$, and the failure probability to be $\delta/n$. The copies of sample needed for the last call of tomography is $K= O\left(\frac{d^{2p}}{\epsilon^{2}}
\log\!\frac{n}{\delta}
\right)$. We show in Appendix \ref{appendix-3} that the complexity bound $K$ is asymptotically dominated by $N$. Therefore, the overall sample complexity is  
$$N= O\!\Biggl(
\frac{\displaystyle D^{12} n^{7}\, d^{4} (\ln d)^{7}\, \log\!\tfrac{n}{\delta}}
{\displaystyle
\epsilon^{12}\!\Bigl[
  \ln(\ln d)
  + \ln\!\tfrac{128\,n\,D^2}{(\sqrt2-1)^2\epsilon^2}
\Bigr]^{7}}
\Biggr),$$
and the whole algorithm runs in polynomial time with respect to the same parameters.
\end{proof}

\appendixsection{Implementation on Nearest-Neighbor Lattices}
\label{appendix-higher-dim}

\begin{theorem*}[Restatement of Theorem \ref{thm:2d}]
If quantum gates are restricted to nearest-neighbor interactions on a square lattice, the depth of the quantum circuit generated by Algorithm~\ref{alg:A2} is $O(\sqrt{n/p})$.
\end{theorem*}
\begin{proof}
Without loss of generality, we assume that $\sqrt{n/p}=2^{M/2}$ \footnote{If this is not the case, we can replace $n$ by the smallest $n' \ge n$ such that $\sqrt{n'/p}=2^{M/2}$. The following argument then applies to $n'$ without any change.}. Note that we may arrange the qubits on a square lattice in any convenient way. To prove the lemma, we view a bunch of $p$ qubits as a single lattice point and adopt the following construction to form a $2^{M/2}\times 2^{M/2}$ square lattice. For simplicity, we assume that the circuit depth $M$ is even. 

First, we place $\widetilde{\mathcal{B}}^{M}_1$ at the bottom-right lattice site of the entire square lattice, which serves as the anchor for the $0$-th level. 

Next, we partition the lattice into $4$ blocks, each of size $2^{M/2-1}\times2^{M/2-1}$. In each block, we place $\widetilde{\mathcal{B}}^{M-2}_i$ at its bottom-right lattice site, which we define as anchors for the $1$-st level. The blocks are indexed in row-major order, i.e., from left to right and then from top to bottom, with $i=1,\dots,4$. 

We then iterate this procedure. At level $j \ge 1$, for each block indexed by $i$ (whose bottom-right anchor is $\widetilde{\mathcal{B}}^{M-2(j-1)}_i$), we partition the block into $2\times2$ subblocks, each of size $2^{M/2-j}\times2^{M/2-j}$. In each subblock we place $\widetilde{\mathcal{B}}^{M-2j}_{\,k+4(i-1)}$ at the bottom-right lattice site as the next-level anchor, with $k=1,\dots,4$. The four subblocks are again indexed in row-major order. At the $M/2$-th level, we obtain a $2^{M/2}\times 2^{M/2}$ square lattice (see Figure \ref{fig:recursive-proof-2d}). Each lattice site is associated with a patch of $p$ qubits, which we label by $\widetilde{\mathcal{B}}^{0}_{\ell}$, where $\ell=1,\dots,2^M$.

Now, we have placed all the qubits on a $2^{M/2}\times 2^{M/2}$ square lattice. Using the level structure described above, we then implement the algorithm iteratively over $M/2$ levels, from $j=M/2$ down to $j=1$. 

Each level $j$ consists of two steps. First, we apply the disentangling unitaries horizontally on pairs of $p$-qudit blocks $(\widetilde{\mathcal{B}}^{M-2j}_{k-1},\widetilde{\mathcal{B}}^{M-2j}_{k})$, where $k$ ranges over even integers in $\{1,\dots,2^{2j}\}$. These horizontal unitaries disentangle the qubits at the left lattice sites (i.e., $\widetilde{\mathcal{B}}^{M-2j}_{k-1}$ for even $k$) to the all-zero state. In the second step, we apply the disentangling unitaries vertically on pairs of $p$-qudit blocks $(\widetilde{\mathcal{B}}^{M-2j}_{k-2},\widetilde{\mathcal{B}}^{M-2j}_{k})$ where $k$ ranges over multiples of $4$ in $\{1,\dots,2^{2j}\}$. These vertical unitaries disentangle the qubits at the top lattice site to the all-zero state. As a result, only the sites $\widetilde{\mathcal{B}}^{M-2j}_{k}$ where $k$ are multiples of $4$ in $\{1,\dots,2^{2j}\}$ remain. Notice that the site $\widetilde{\mathcal{B}}^{M-2j}_{k}$ is exactly the same site with the label $\widetilde{\mathcal{B}}^{M-2(j-1)}_{k/4}$ in the next level. Each level corresponds to a pair of successive iterations in Algorithm~\ref{alg:A2}.

The collection of these surviving lattice sites then forms the input to the next level. After executing $M/2$ levels, only a single lattice site containing $p$ qubits remains. We then perform full state tomography on this site to complete the two-dimensional version of Algorithm~\ref{alg:A2}.

Now, we analyze the circuit depth under the restriction to nearest-neighbor quantum gates. At level $j$, before applying the horizontal disentangling unitaries, we need to move the qubits from the left lattice sites (i.e., $\widetilde{\mathcal{B}}^{M-2j}_{k-1}$ for even $k$) to the lattice sites immediately to the left of their right partners. Similarly, before applying the vertical disentangling unitaries, we need to move the qubits from the top lattice sites (i.e., $\widetilde{\mathcal{B}}^{M-2j}_{k-2}$ for $k$ being multiples of $4$) to the lattice sites immediately above their bottom partners. Therefore, at level $j$, this requires $O(2^{M/2-(j-1)})$ SWAP gates within each block $\widetilde{\mathcal{B}}^{M-2(j-1)}_{k}$, where $k=1,\dots,2^{2(j-1)}$. Since these blocks are disjoint, the SWAP gates acting on different blocks at the same level can be executed in parallel. Therefore, the total circuit depth is $\sum_{j=1}^{M/2}2^{M/2-(j-1)}=O(2^{M/2})=O(\sqrt{n/p}).$

\end{proof}
\begin{figure}[H]
\includegraphics[width=\textwidth]{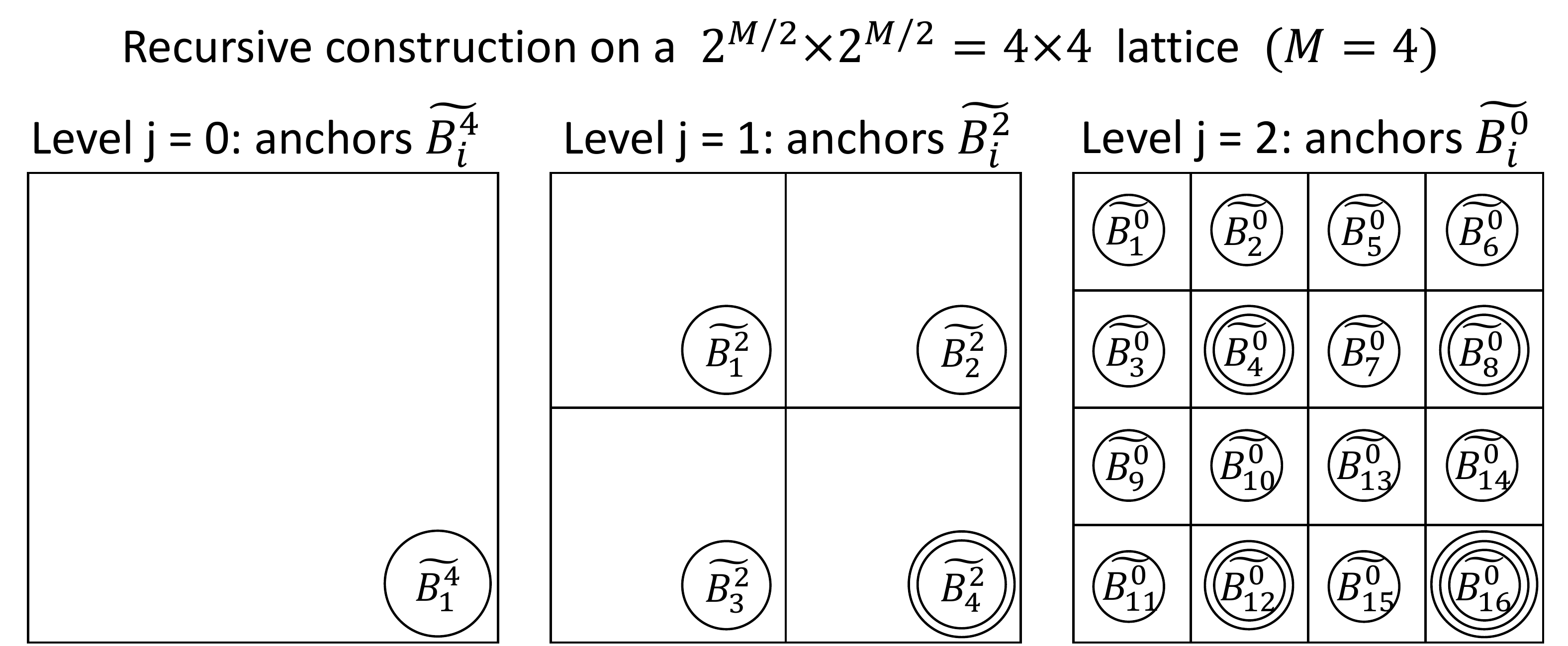}
\caption{Illustration of the recursive block structure on a $2^{M/2} \times 2^{M/2}$ square lattice with $M=4$. At level $j$, each block of size $2^{M/2-(j-1)} \times 2^{M/2-(j-1)}$ is partitioned into four subblocks, each of size $2^{M/2-j} \times 2^{M/2-j}$. In each subblock, the bottom-right site (corresponding to the maximal-coordinate corner in each block) is designated as the anchor $\widetilde{\mathcal{B}}^{M-2j}_i$ (circled). These anchors form a coarser lattice for the next level.}
\end{figure}

\begin{corollary*}[Restatement of Corollary~\ref{coro:higher-dim}]
Suppose quantum gates are restricted to nearest-neighbor interactions on a $q$-dimensional hypercubic lattice. Then the quantum circuit generated by Algorithm~\ref{alg:A2} can be implemented with depth $O\!\left(q\,(n/p)^{1/q}\right).$
\end{corollary*}

\begin{proof}
The proof is a direct generalization of Theorem~\ref{thm:2d}. 
Without loss of generality, assume that $(n/p)^{1/q}=2^{M/q}$.\footnote{If this is not the case, we can replace $n$ by the smallest $n' \ge n$ such that $(n'/p)^{1/q}=2^{M/q}$ for some integer $M$. The following argument then applies to $n'$ without any essential change.}
We view each patch of $p$ qubits as a single lattice point and arrange the $n/p$ patches on a $q$-dimensional hypercubic lattice of side length $2^{M/q}$ in each coordinate direction.

We recursively define a level structure on this lattice. At level $0$, the entire hypercube has a distinguished anchor site, chosen to be the corner $(2^{M/q},\dots,2^{M/q})$. At level $j \ge 1$, each block from level $j-1$ is partitioned into $2^q$ equal subblocks of side length $2^{M/q-j}$. In each subblock, we designate as its anchor the corner with maximal coordinates in all directions. After $M/q$ levels, every lattice site is identified with one patch $\widetilde{\mathcal{B}}^0_\ell$, where $\ell=1,\dots,2^M$.

We now implement the algorithm level by level, from level $M/q$ down to level $1$. 
Each level consists of $q$ successive steps, one for each coordinate direction. 
In the $r$-th step, where $r\in\{1,\dots,q\}$, we apply disentangling unitaries to pairs of neighboring subblocks that differ only in the $r$-th coordinate, so that after this step only one half of the sites in that coordinate direction remain active. After performing all $q$ steps, only one corner site remains in each block, which becomes the anchor of the corresponding block at the next coarser level. Thus, each level corresponds to $q$ successive iterations of Algorithm~\ref{alg:A2}.

It remains to bound the depth under the nearest-neighbor constraint. Consider level $j$. Inside each block labeled by the level $j-1$ anchor, the active sites form a hypercube of side length $L_j = 2^{M/q-(j-1)}.$ Before applying the disentangling unitaries in a given coordinate direction, we move the qubits from one half of the active sites to the neighboring sites adjacent to their partner blocks along that direction. This routing can be implemented by nearest-neighbor SWAP gates with depth $O(L_j)$ inside each block. Since there are $q$ such directions in each level, the total routing depth per level is $O(qL_j)$. 

Because distinct blocks at the same level are disjoint, the routing operations in different blocks can be performed in parallel. The disentangling unitaries themselves contribute only constant depth per step, so the total depth of level $j$ is $O(qL_j)$. Summing over all levels, the total depth is $\sum_{j=1}^{M/q} O\!\left(q \, 2^{M/q-(j-1)}\right)
= O(q \, 2^{M/q})
= O\!\left(q \, (n/p)^{1/q}\right).$
This proves the claim.
\end{proof}
\appendixsection{Sample complexity lower bound on learning product states}
\label{sec:lowerbound}

To establish a lower bound, we prove that for product state learning, the sampling complexity lower bound for producing an approximate description of the given quantum state to fidelity $\epsilon$-close to optimum is $\Omega(nd/\epsilon)$, where $n$ is the number of $d$-dimensional systems. \ 
% In particular, we show that learning product states to fidelity at least $\mathrm{OPT}-\epsilon$ requires at least $\Omega(nd/\epsilon)$ copies, where $n$ is the number of registers. \ 
% We assume that the given state is a product state, i.e., $\ket{\psi}=\ket{\psi_1}\otimes\ldots\otimes\ket{\psi_n}$. \ 

\begin{theorem*}[Restatement of Theorem~\ref{theorem:lowerbound}]
For integer $n\geq 1$, $d\geq 2$ and $\epsilon\in[0,2/3]$, consider learning a product state $\rho \;=\; \bigotimes_{i=1}^n \rho_i$,
where each $\rho_i$ is a state on a $d$-dimensional Hilbert space. Any quantum algorithm that outputs a product state $\hat\rho$ satisfying $F(\rho,\hat\rho)\geq 1-\epsilon$ must use at least $\Omega(nd/\epsilon)$ copies of $\rho$. 
\end{theorem*}
\begin{proof}
It suffices to consider the special case in which $\rho$ is a pure product state.

The sample complexity for approximating a single-qudit Haar random state to fidelity at least $1-\epsilon$ requires $\Omega(d/\epsilon)$ samples. This follows from a direct worst-case to average-case reduction. Suppose there exists a learning algorithm $\mathcal{B}$ which, given $o(d/\epsilon)$ samples of a single-qudit Haar random state, outputs a estimate with fidelity at least $1-\epsilon$. Then we can construct an algorithm $\mathcal{B}^*$ that learns \emph{arbitrary} single-qudit states using $o(d/\epsilon)$ samples, contradicting the $\Omega(d/\epsilon)$ lower bound for learning $d$-dimensional states to fidelity $1-\epsilon$ due to Yuen~\cite{Yuen_2023}.

$\mathcal{B}^*$ can be described as follows: given an arbitrary single-qudit state $\ket{\phi}$, one first samples a single-qudit Haar-random unitary $U_1$ and apply $U_1$ on $\ket{\phi}$ to prepare a single-qudit Haar-random state $U_1\ket{\phi}$, runs $\mathcal{B}$ on $U\ket{\phi}$ using $o(d/\epsilon)$ samples to obtain a state $\hat\rho$ such that $F(\hat\rho,U\ket{\phi})\geq 1-\epsilon$, and then outputs $U^\dag \ket{\varphi} U$. By unitary invariance of fidelity, this estimate achieves fidelity at least $1-\epsilon$ with $\ket{\phi}$, completing the reduction and yielding the desired contradiction to the sample complexity lower bound.

Now, consider a product state of $n$ single-qudit Haar random state sampled independently, $\ket{\psi}=\otimes_{i=1}^n \ket{\psi_{i}}_{Haar}$. Suppose there exists a quantum algorithm $\mathcal{A}$ that takes $o(\frac{nd}{\epsilon})$ copies of state $\ket{\psi}$ and outputs another product state $\hat\rho$ such that $F(\hat\rho,\ket{\psi})\geq 1-\epsilon$. Then, we can obtain an algorithm $\mathcal{A}^*$ to learn a single-qudit Haar random state with fidelity at least $1-\epsilon/n$ using at most $o(nd/\epsilon)$ samples.

Note that $F(\hat\rho,\ket{\psi})\geq 1-\epsilon$ implies that $F_1\cdot F_2\cdots F_n \geq 1-\epsilon$, 
where $F_i=\bra{\psi_i}_{Haar} \Tr_{\bar{i}}(\hat\rho)\ket{\psi_i}_{Haar}$ for $i\in [n]$ and $\bar{i}=\{1,2,\dots,i-1,i+1,\dots,n\}$. Let $i^*$ be the position that has the largest $F_{i^*}$, by an averaging argument, $F_{i^*}^n \geq F_1\cdot F_2\cdots F_n \geq 1-\epsilon$. Then, for $\epsilon \leq 2/3$, we can obtain that 
\begin{align}
\label{eq:fid_lowerbound1}
    F_{i^*} \geq (1-\epsilon)^{1/n} \geq 1-2\epsilon/n.
\end{align}

%$\mathcal{A}^*$ is described as follows: The input is a single-qudit Haar random state $\ket{\phi}$. The algorithm first samples $n$ single-qudit Haar random states independently, denoted as $\ket{\psi}=\otimes_{i=1}^n \ket{\psi_{i}}_{Haar}$, runs $\mathcal A$, checks the fidelity of each qudit, identifies $i^*$ which has the largest fidelity $F_{i^*}\geq 1-\epsilon/n$. Then, $\mathcal{A}^*$ replaces $\ket{\psi_{i^*}}_{Haar}$ with $\ket{\phi}$ and applies $\mathcal A$ to learn $\hat\rho$ that approximates to $\ket{\psi^*}=\otimes_{i\neq i^*}^n \ket{\psi_{i}}_{Haar}\ket{\phi}$ with fidelity at least $1-\epsilon$. Finally, the algorithm outputs $\rho^*=\Tr_{\bar{i^*}}(\hat\rho)$. 

$\mathcal{A}^*$ is described as follows:
Given a single-qudit Haar-random input state $\ket{\phi}$, the algorithm first samples $n$ independent single-qudit Haar-random states, preparing $\ket{\psi} \;=\; \bigotimes_{i=1}^n \ket{\psi_i}_{\mathrm{Haar}}$.
It then runs $\mathcal{A}$ on copies of $\ket{\psi}$, evaluates the per-qudit fidelities, and identifies an index $i^*$ achieving the largest fidelity so that $F_{i^*} \geq 1-2\epsilon/n$ by Equation~\ref{eq:fid_lowerbound1}. (This step will take many copies of $\ket{\psi}$ to obtain desired precision of the per-qudit fidelities; however, it does not require any copy of the input $\ket{\phi}$.) Then $\mathcal{A}^*$ replaces the state $\ket{\psi_{i^*}}_{\mathrm{Haar}}$ with $\ket{\phi}$ and applies $\mathcal{A}$ to learn a state $\hat{\rho}$ that approximates
\[
\ket{\psi^{*}} \;=\; (\bigotimes_{i< i^*} \ket{\psi_i}_{\mathrm{Haar}}) \otimes \ket{\phi} \otimes (\bigotimes_{i> i^*}\ket{\psi_i}_{\mathrm{Haar}})
\]
to fidelity at least $1-\epsilon$. Finally, it outputs the reduced state $\rho^{*} = \Tr_{\overline{i^*}}(\hat{\rho})$. Since $F_{i^*}\ge 1-2\epsilon/n$ and $\mathcal{A}^*$ uses only $o(nd/\epsilon)$ copies, this contradicts the $\Omega(nd/\epsilon)$ sample-complexity lower bound for learning a single-qudit Haar-random state to fidelity at least $1-2\epsilon/n$ (setting the $\Omega(d/\epsilon')$ bound with $\epsilon'=\epsilon/n$). Therefore, any algorithm that learns $\ket{\psi}=\bigotimes_{i=1}^n \ket{\psi_i}_{\mathrm{Haar}}$ to fidelity $\ge 1-\epsilon$ must use $\Omega(nd/\epsilon)$ copies. Consequently, learning an arbitrary $n$-qudit product state to fidelity at least $1-\epsilon$ also requires $\Omega(nd/\epsilon)$ copies, since the average-case lower bound extends to the worst case.

\end{proof}
\appendixsection{Single-qubit measurement is hard for MPS learning}
\label{sec:single-qubit}

\paragraph{Notation.}
Let $\mathbb{Z}_2 = \{0,1\}$ denote the field with two elements. For $\ell \in \mathbb{N}$, let $\mathbb{Z}_2^\ell$ denote the set of binary vectors of length $\ell$.  For vectors $a,s \in \mathbb{Z}_2^\ell$, we denote their inner product over $\mathbb{Z}_2$ by $$\langle a, s \rangle = \sum_{i=1}^{\ell} a_i s_i \bmod 2.$$
We write $x \xleftarrow{\$} \mathcal{X}$ to denote that $x$ is sampled uniformly at random from the set $\mathcal{X}$. For $\tau \in (0,1/2)$, we denote by $\mathrm{Ber}_\tau$ the Bernoulli distribution over $\mathbb{Z}_2$ satisfying
$$\Pr_{e \sim \mathrm{Ber}_\tau}[e=1] = \tau.$$

\begin{definition}[Search LPN$_{\tau,\ell}$ \cite{pietrzak2012cryptography}]
\label{def:LPN}
Let $\ell \in \mathbb{N}$ and $\tau \in (0,1/2)$. A secret vector $s \xleftarrow{\$} \mathbb{Z}_2^\ell$ is chosen uniformly at random. An adversary is given $q$ independent samples of the form
$$(a_i, b_i) \in \mathbb{Z}_2^\ell \times \mathbb{Z}_2,$$
from the distribution $(a_i,b_i) \sim D_{\text{LPN}}^{r,\ell}$ where
$$a_i \xleftarrow{\$} \mathbb{Z}_2^\ell,
\quad
b_i = \langle a_i, s \rangle \oplus e_i,
\quad
e_i \sim \mathrm{Ber}_\tau
\quad
for \quad i=1,\dots, q.$$
The goal is to recover the secret $s$.
\end{definition}
\begin{assumption}[Search LPN$_{\tau,\ell}$ Assumption~\cite{pietrzak2012cryptography}]
\label{assumption:lpn}
Let $\ell \in \mathbb{N}$ and $\tau \in (0,1/2)$. 
For every quantum polynomial-time adversary $\mathcal{A}$, given $q = \mathrm{poly}(\ell)$ samples 
$(a_i,b_i)$ generated as in Definition~\ref{def:LPN}, the probability that $\mathcal{A}$ outputs the secret $s$ is negligible in $\ell$, i.e.,
$$\Pr\big[\mathcal{A}((a_1,b_1),\dots,(a_q,b_q)) = s \big] \le \mathrm{negl}(\ell),$$
where the probability is taken over the distribution $(a_i,b_i) \sim D_{\text{LPN}}^{r,\ell}$ and the quantum algorithm $\mathcal{A}$.
\end{assumption}

% \begin{definition}[Parity-encoded MPS family]
% For each secret string $s \in \mathbb{Z}_2^\ell$, define a $2\ell$-qubit state $\ket{\psi_s}$ as follows. First prepare
% $\frac{1}{\sqrt{2^\ell}}\sum_{x\in\mathbb Z_2^\ell}\ket{x},$
% where the first $\ell$ qubits encode the input string $x=(x_1,\dots,x_\ell)$. Next, define the prefix parity variables $z_i(x)=\bigoplus_{j=1}^i x_j s_j,$ where $z_0=0.$ Using CNOT gates, coherently compute the register $\ket{z(x)}=\ket{z_1(x),\dots,z_\ell(x)}.$ Then apply Hadamard gates to all parity registers. The final state is $\ket{\psi_s}=\frac{1}{\sqrt{2^\ell}}
% \sum_{x\in\mathbb Z_2^\ell}
% \ket{x}\otimes H^{\otimes \ell}\ket{z(x)}.$ We define the family $\mathcal F_\ell:=\{\ket{\psi_s}:s\in\mathbb Z_2^\ell\}.$
% \end{definition}

\begin{lemma}
\label{lemma:mps(2)}
For every $s\in\mathbb Z_2^\ell$, the state $\ket{\psi_s}$ generated by the circuit in Figure~\ref{Fig-single-qubit} is a matrix product state with bond dimension at most $2$.
\end{lemma}

\begin{proof}
The prefix variable satisfies $z_i=z_{i-1}\oplus x_i s_i.$ Thus the evolution of the parity register depends only on the previous parity bit $z_{i-1}$ and the current input bit $x_i$. This induces a sequential tensor-network structure where the virtual bond stores the current parity value, which takes only two possible values:
$0$ or $1$. Therefore the bond dimension is at most $2$.
\end{proof}

\begin{theorem*} [Restatement of Theorem~\ref{theorem:single}] 
Assume the $\mathrm{LPN}_{\tau,\ell}$ assumption holds for some constant 
$\tau \in (0,1/2-\gamma)$, where $\gamma>0$ is a constant. Let 
$\{|\psi_s\rangle : s\in\{0,1\}^{\ell}\}$ be the family of MPS$(2)$ states 
defined in Figure~\ref{Fig-single-qubit}. Then there is no polynomial-time algorithm 
that, using only computational-basis single-qubit measurements on 
$\operatorname{poly}(\ell)$ copies of $|\psi_s\rangle$, outputs a classical 
description of a hypothesis state $\widehat{\rho}$ from which $\widehat{\rho}$ 
can be efficiently prepared and $
F(\widehat{\rho},|\psi_s\rangle\langle\psi_s|)\ge 1-\epsilon$.  

%Let $\ket{\psi_s}$ be the quantum state generated by the circuit in Figure~\ref{Fig-single-qubit}, where $s\in\{0,1\}^{\ell}$ is the hidden secret. Suppose a learner is restricted to computational-basis single-qubit measurements and is allowed to measure $q=\mathrm{poly}(\ell)$ independent copies of $\ket{\psi_s}$.  Under the LPN$_{\tau,\ell}$ assumption, for some $\tau\in(0,1/2-\gamma)$, there does not exist polynomial-time algorithms that can output a classical description from which one can efficiently prepare the hypothesis state. 
\end{theorem*}

\begin{proof}
%Let $\ket{\psi_s}$ be the quantum state generated by the circuit in Figure~\ref{Fig-single-qubit}, where $s\in\{0,1\}^{\ell}$ is the hidden secret. We show that there no polynomial-time, non-adaptive computational-basis single-qubit measurements learners that can output a classical description from which one can efficiently prepare the hypothesis state.

First, the string $x\in \mathbb{Z}_2^n$ is uniformly random; indeed, each of the
corresponding qubits is first prepared in the state $H\ket{0}=(\ket{0}+\ket{1})/\sqrt{2}$.
Subsequent CNOT gates, if any, with these qubits as controls and $\ket{0}$ ancillas
as targets do not change the marginal distribution of the control qubits. For each input $x\in\mathbb Z_2^\ell$, the circuit computes $z_\ell(x)=\langle x,s\rangle \bmod 2.$ After applying $R_y(2\theta)$ to the final parity register and measuring in the computational basis, the observed bit $w_\ell$ satisfies
$$
w_\ell=z_\ell(x)\oplus e,
\qquad e\sim\mathrm{Bernoulli}(\sin^2\theta).
$$
Choosing $\theta$ such that $\sin^2\theta=\tau$, we obtain
$$
w_\ell=\langle x,s\rangle\oplus e,
\qquad e\sim\mathrm{Bernoulli}(\tau).
$$
Therefore, each observed pair $(x,w_\ell)$ is distributed exactly as an LPN$_{\tau,\ell}$ sample.

For each $i\in\{1,2,\dots,\ell-1\}$, immediately before the final Hadamard
gate on the $w_i$-register, the value of $w_i$ is determined by the previously
generated variables. More precisely, there exists a Boolean function
$f_i$ such that $w_i=f_i(x_1,\dots,x_i,w_1,\dots,w_{i-1})$
at this point in the circuit. Thus, without loss of generality, the state can
be written as
\[
\sum_{x,w} a_{x,w}
\ket{x_1,\dots,x_i,w_1,\dots,w_{i-1}}
\ket{\sum_{j=1}^{i} x_is_i}
\ket{x_{i+1},\dots,x_\ell,w_{i+1},\dots,w_\ell},
\]
where the middle register is the $w_i$-register.

Applying the Hadamard gate to this middle register gives
\[
\sum_{x,w} \frac{a_{x,w}}{\sqrt{2}}
\ket{x_1,\dots,x_i,w_1,\dots,w_{i-1}}
\Big(
\ket{0}+(-1)^{\sum_{j=1}^{i} x_is_i}\ket{1}
\Big)\otimes
\ket{x_{i+1},\dots,x_\ell,w_{i+1},\dots,w_\ell}.
\]
Therefore, conditioned on any fixed values of the other registers, measuring
$w_i$ in the computational basis gives $0$ and $1$ with equal probability.
Hence each $w_i$, for $i\in\{1,\dots,\ell-1\}$, is uniformly random after the final Hadamard gate and thus is independent of the other computational-basis measurement outcomes.

Suppose that we obtain a classical description of a state $\widehat{\rho}$ 
from which $\widehat{\rho}$ can be efficiently prepared, and suppose that
\[
    F(\widehat{\rho},|\psi_s\rangle\langle\psi_s|)\ge 1-\epsilon .
\]
We show that this allows us to recover the secret $s$ in polynomial time with
non-negligible probability.

Consider $\ket{\psi_s}$. By applying Hadamard gates $H$ to the qubits $w_1, w_2, \dots, w_{\ell-1}$, we obtain the values $(x_1s_1, x_1s_1+x_2s_2, \dots, \sum_{i=1}^{\ell-1} x_is_i)$. By measuring these qubits and $x$, we can directly recover $s_1, s_2, \dots, s_{\ell-1}$. For $s_\ell$, we can also uncompute $R_y(2\theta)$ to obtain $\sum_{i=1}^{\ell} x_is_i$ and then identify $s_{\ell}$ after computing $x$ and $s_1, s_2, \dots, s_{\ell-1}$. Therefore, if we can prepare $\ket{\psi_s}$, we can recover $s$ with probability $1$.

% %Ignoring the qubits $w_1,w_2,\ldots,w_{\ell-1}$, we can write the rest
% %part of $\ket{\psi_s}$ as
% %\[
% %    \sum_{x\in \{0,1\}^{\ell}}\frac{1}{\sqrt{2^{\ell}}} \ket{x}
% %    \left(
% %    \sqrt{1-\tau}\ket{\langle x,s\rangle}
% %    +\sqrt{\tau}\ket{1\oplus\langle x,s\rangle}
% %    \right)_{w_\ell}.
% %\]
% %By applying a Hadamard gate $H$ to the qubit $w_\ell$, we obtain
% \[
%     \sum_{x\in \{0,1\}^{\ell}}\frac{1}{\sqrt{2^{\ell}}} \ket{x}
%     \left(
%     \frac{\sqrt{1-\tau}+\sqrt{\tau}}{\sqrt{2}}\ket{0}
%     +
%     \frac{\sqrt{1-\tau}-\sqrt{\tau}}{\sqrt{2}}
%     (-1)^{\langle x,s\rangle}\ket{1}
%     \right)_{w_\ell}.
% \]
% Therefore, if measuring $w_\ell$ gives outcome $1$, then the post-measurement
% state on the $x$-register is
% \[
%     \frac{1}{\sqrt{2^{\ell}}}
%     \sum_{x\in \{0,1\}^{\ell}} (-1)^{\langle x,s\rangle}\ket{x}.
% \]
% Applying $H^{\otimes \ell}$ to this state and measuring in the computational
% basis outputs $s$ with probability $1$. The probability that measuring $w_\ell$
% gives outcome $1$ on the ideal state $\ket{\psi_s}$ is
% \[
%     c_\tau := \frac{\left(\sqrt{1-\tau}-\sqrt{\tau}\right)^2}{2}.
% \]

Now consider the learned state $\widehat{\rho}$. Since
\[
    F(\widehat{\rho},|\psi_s\rangle\langle\psi_s|)\ge 1-\epsilon,
\]
we can obtain that $D\bigl(\widehat{\rho},|\psi_s\rangle\langle\psi_s|\bigr)
    \le \sqrt{\epsilon}$, 
where $D(\rho,\sigma):=\frac12\|\rho-\sigma\|_1$ denotes trace distance.
Thus, for any measurement, the total variation distance between the outcome
distributions obtained from $\widehat{\rho}$ and from
$|\psi_s\rangle\langle\psi_s|$ is at most $\sqrt{\epsilon}$.

Applying the algorithm above to $\widehat{\rho}$ therefore recovers
$s$ with probability at least $1-\sqrt{\epsilon}$, which is non-negligible. Hence a high-fidelity efficiently preparable
$\widehat{\rho}$ allows us to recover the secret $s$ in polynomial time
with non-negligible probability. This completes the proof.

%If a polynomial-time learner could recover $s$ from $q=\mathrm{poly}(\ell)$ such measurement samples, then it would solve Search LPN$_{\tau,\ell}$ with non-negligible probability, contradicting the Search LPN$_{\tau,\ell}$ assumption. Hence no such learner exists.
\end{proof}

\begin{corollary}
 Assume the $\mathrm{LPN}_{\tau,\ell}$ assumption holds for 
$\tau \in (0,1/2-\gamma)$, where $\gamma>0$ is non-negligible. Then there is no polynomial-time non-adaptive algorithm 
that, using only single-qubit measurements on 
$\operatorname{poly}(\ell)$ copies of $|\psi_s\rangle$, outputs a classical 
description of a hypothesis state $\widehat{\rho}$ from which $\widehat{\rho}$ 
can be efficiently prepared and such that $
F(\widehat{\rho},|\psi_s\rangle\langle\psi_s|)\ge 1-\epsilon$
with non-negligible probability.
\end{corollary}

\begin{proof}
Consider any non-adaptive single-qubit measurement strategy specified by local unitaries $U_1,\dots,U_{2\ell},$ where the learner measures the $i$-th qubit in the basis defined by $U_i$. Equivalently, this measurement procedure can be viewed as first applying $U_1^\dagger\otimes\cdots\otimes U_{2\ell}^\dagger$ to the state and then performing computational-basis measurements. Now define a modified state family
$$
\ket{\phi_s} = \left(
U_1^\dagger\otimes\cdots\otimes U_{2\ell}^\dagger
\right) \ket{\psi_s}.
$$

Since local unitaries do not increase MPS bond dimension, Lemma~\ref{lemma:mps(2)} implies that $\ket{\phi_s}\in \mathrm{MPS}(2).$ Measuring $\ket{\phi_s}$ in the computational basis is statistically equivalent to measuring $\ket{\psi_s}$ using the original non-adaptive single-qubit measurement strategy. Therefore, if a polynomial-time learner could efficiently learn $\mathrm{MPS}(2)$ under arbitrary non-adaptive single-qubit measurements, then it could also recover the hidden secret $s$ for the family $\{\ket{\phi_s}\}$. This contradicts Theorem~\ref{theorem:single}. Hence learning $\mathrm{MPS}(2)$ remains hard in the worst case under non-adaptive single-qubit measurements.
\end{proof}

\appendixsection{Solving the self-consistent equation for $p$}
\label{appendix-1}
We would like to solve the self-consistent equation
\[
p = \Bigg\lceil \log_d \!\left( \frac{64 n}{p}\, \frac{D^2}{(\sqrt{2}-1)^2 \epsilon^2} \right) \Bigg\rceil.
\]
Ignoring the ceiling for the moment, we can rewrite this as
\[
p = \log_d \!\left( \frac{64\,n\,D^2}{p (\sqrt{2}-1)^2 \epsilon^2} \right).
\]
Define
\[
B := \frac{64\,n\,D^2}{(\sqrt{2}-1)^2 \epsilon^2},
\]
so that the equation becomes
\[
p = \log_d B - \log_d p.
\]
Multiplying through by $\ln d$ gives
\[
\ln p + p \ln d = \ln B.
\]
Setting $x := p \ln d$, we obtain the equivalent form
\[
x e^x = B \ln d,
\]
whose solution is expressed in terms of the Lambert $W$ function:
\[
p = \frac{1}{\ln d}\; W\!\left( \ln d \cdot B \right).
\]
Thus, the integer solution of the original equation is
\[
p = \left\lceil \frac{1}{\ln d}\; W\!\left( \ln d \cdot \frac{64\,n\,D^2}{(\sqrt{2}-1)^2 \epsilon^2} \right) \right\rceil.
\]

\paragraph{Bounds on $p$.}
For $z>e$, one has the standard bounds
\begin{equation}
\label{eq:W(z)-bounds}
\ln z - \ln\ln z < W(z) < \ln z,
\end{equation}
which follow from the monotonicity of $f(w)=we^w$.  

Let $z := \ln d \cdot B$. Dividing the inequality \eqref{eq:W(z)-bounds} by $\ln d > 0$ gives
\[
\frac{\ln z - \ln\ln z}{\ln d}
\;<\;
\frac{W(z)}{\ln d}
\;<\;
\frac{\ln z}{\ln d}.
\]
Since $p = \lceil W(z)/\ln d \rceil$, applying the ceiling function to the above inequality yields
\[
\left\lceil \frac{\ln z - \ln\ln z}{\ln d}\right\rceil
\;\le\;
p
\;\le\;
\left\lceil \frac{\ln z}{\ln d}\right\rceil.
\]
Hence, in the usual asymptotic notation,
\[
p = \Theta\!\left(\frac{\ln z}{\ln d}\right)
= \Theta\!\left(\frac{\ln(\ln d\cdot B)}{\ln d}\right).
\]
Moreover, one may record the next-order correction:
\[
p = \frac{\ln z}{\ln d} + O\!\Big(\frac{\ln\ln z}{\ln d}\Big),
\]
so that writing \(p \approx \dfrac{\ln(\ln d\cdot B)}{\ln d}\) is justified as a leading-order approximation (the omitted term is of smaller order).

\appendixsection{Existence condition for integer solution $p$}
\label{appendix-Existence}
Let $d>1$ and $B>0$ be given, and consider the self-consistent equation
\begin{equation}\label{eq:main-ceil}
p = \left\lceil \log_d\!\left(\frac{B}{p}\right) \right\rceil,\qquad p\in\mathbb{Z}_{>0}.
\end{equation}
Using the equivalence $\lceil x\rceil = p \iff p-1 < x \le p$, equation \eqref{eq:main-ceil} is equivalent to
\begin{equation}\label{eq:pd-ineq}
p-1<\log_d\!\left(\frac{B}{p}\right) \le p.
\end{equation}
Applying the increasing map $t\mapsto d^t$ (since $d>1$) to the three parts of \eqref{eq:pd-ineq} yields the following strict double inequality
\begin{equation}\label{eq:star}
\boxed{\; p\,d^{\,p-1}<B\le p\,d^{\,p} \;}\tag{$\ast$}
\end{equation}
Hence, the integer $p\ge1$ solves \eqref{eq:main-ceil} if and only if it satisfies the inequalities \eqref{eq:star}.

\paragraph{Rewriting the existence condition using the Lambert $W$ function:} Define
\[ a:=\frac{1}{\ln d}\,W(B\ln d),\qquad b:=\frac{1}{\ln d}\,W(B d\ln d), \]
where $W$ denotes the principal branch $W_0$ of the Lambert $W$ function (this is appropriate because its argument is positive).
A standard computation (putting $x=p\ln d$ and using $x e^x\mapsto W$ as the inverse) shows that the two inequalities in \eqref{eq:star} are equivalent to
\begin{equation}\label{eq:ab}
\boxed{\; a\le p< b \;}
\end{equation}
Therefore the integer solution exists if and only if the half-open interval $[a,b)$ contains an integer.

\begin{lemma}[Disjointness of the integer intervals]
\label{lemma:disjointness}
For each integer $p\ge1$ define the interval
\[
I_p := \big(p d^{p-1},\; p d^{p}\big] .
\]
Then $I_p\cap I_q=\varnothing$ for every pair of distinct integers $p\neq q$.
\end{lemma}

\begin{proof}
It suffices to show $I_p\cap I_{p+1}=\varnothing$ for every $p\ge1$, since the general statement for arbitrary distinct integers follows by iterating this adjacent gap. Compute the right endpoint of $I_p$ and the left endpoint of $I_{p+1}$:
\[
\text{right}(I_p)=p d^{p},\qquad \text{left}(I_{p+1})=(p+1)d^{p}.
\]
Because $p+1>p$ and $d^{p}>0$, we have
\[
\text{left}(I_{p+1})=(p+1)d^{p} > p d^{p} = \text{right}(I_p).
\]
Thus the entire interval between the right endpoint of $I_p$ and the left endpoint of $I_{p+1}$ is empty of any $I$-intervals, so $I_p$ and $I_{p+1}$ do not overlap.

For arbitrary distinct integers $p+1<q$ the same reasoning applied repeatedly gives
\[
\text{right}(I_p)=p d^{p} < (p+1)d^{p} = \text{left}(I_{p+1}) < \cdots < (q-1)d^{q-2} = \text{left}(I_{q-1}) < q d^{q-1} = \text{left}(I_q),
\]
so $I_p$ and $I_q$ are separated by nonempty gaps and hence disjoint. This completes the proof.
\end{proof}

\begin{lemma}[Length of the interval $[a,b)$]
\label{lemma:length-of-the-interval}
For every $B>0$ and $d>1$ we have
\[ 0< b-a < 1. \]
\end{lemma}
\begin{proof}
Put $z:=B\ln d>0$ and consider the function $f(t)=W(t)$. Its derivative for $t>0$ is
\[ f'(t)=\frac{W(t)}{t(1+W(t))}, \]
which satisfies $0<f'(t)<1/t$ because $W(t)>0$ and $1+W(t)>1$. Therefore
\begin{align*}
W(d z)-W(z) &= \int_{z}^{dz} f'(t)\,dt < \int_{z}^{dz} \frac{1}{t}\,dt = \ln d.
\end{align*}
Dividing both sides by $\ln d$ yields $b-a<1$. Strict positivity of $b-a$ follows from monotonicity of $W$ on $(0,\infty)$ and the strict inequality $dz>z$.
\end{proof}

\paragraph{Uniqueness and a practical test:} The intervals corresponding to different integers are disjoint (see Lemma \ref{lemma:disjointness}), hence if a solution exists it is unique. Moreover, since $b-a<1$ (see Lemma \ref{lemma:length-of-the-interval}), at most one integer can lie in $[a,b)$; consequently the following test is necessary and sufficient:
\begin{enumerate}
\item Compute $a=\dfrac{1}{\ln d}W(B\ln d)$ and set the candidate integer $p_{\mathrm{cand}}=\lceil a\rceil$.
\item Test the inequality $p_{\mathrm{cand}}\,d^{p_{\mathrm{cand}}-1}<B\le p_{\mathrm{cand}}\,d^{p_{\mathrm{cand}}}$. If it holds, $p_{\mathrm{cand}}$ is the unique solution; otherwise \eqref{eq:main-ceil} has no integer solution. Since $a=\tfrac{1}{\ln d}W(B\ln d)$ satisfies $a d^a = B$, the upper bound $B \le p_{\mathrm{cand}}\,d^{p_{\mathrm{cand}}}$ always holds by monotonicity. Therefore it suffices to test the strict inequality
\[
p_{\mathrm{cand}}\,d^{p_{\mathrm{cand}}-1}<B,
\]
which is equivalent to the condition $\lceil a\rceil < b$. 
If it holds, $p_{\mathrm{cand}}$ is the unique solution; otherwise \eqref{eq:main-ceil} has no integer solution.
\end{enumerate}

\paragraph{Integral values of $a$:} In this work, we have \(B \;=\; \frac{64\,n\,D^2}{(\sqrt{2}-1)^2\,\epsilon^2}.\) Then for any positive integer $m$ one can choose
\[
\epsilon^2 \;=\; \frac{64\,n\,D^2}{(\sqrt{2}-1)^2\, m\, d^m},
\qquad
\epsilon = \sqrt{\frac{64\,n\,D^2}{(\sqrt{2}-1)^2\, m\, d^m}},
\]
so that
\[
B = m d^m.
\]
In this case we obtain
\[
a = \frac{1}{\ln d}\, W(B\ln d) = m,
\]
an integer, and hence
\[
p_{\mathrm{cand}} = \lceil a\rceil = m.
\]
Consequently the interval test reduces to
\[
m d^{m-1} < B = m d^m \le m d^m,
\]
which clearly holds. Thus $p=m$ is the unique solution, and no ``gap'' phenomenon can occur.

However, in our case, the accuracy parameter is required to satisfy \(0<\epsilon \le 1\). This imposes an additional constraint on the admissible integers \(m\). 
\[
0<\frac{64\,n\,D^2}{(\sqrt{2}-1)^2 m d^m}\le 1 \quad\Longrightarrow\quad m d^m \ge \frac{64\,n\,D^2}{(\sqrt{2}-1)^2}.
\]
Since the function $m\mapsto m d^{m}$ is strictly increasing for $m\ge1$, there always exists an integer $m$ (indeed many) satisfying \(m d^{m}\ge \frac{64\,n\,D^2}{(\sqrt{2}-1)^2}\). Thus mathematically it is always possible to pick an integer \(m\) so that the corresponding \(\epsilon\) lies in \((0,1]\).

\appendixsection{Substituting the expression for \(p\) into the sample-complexity}
\label{appendix-2}
We start from
\[
N = O\!\left(\frac{nd^{4p}}{p\eta^{2}}\,
\log\!\frac{n}{\delta}\right),
\qquad
\eta = \frac{(\sqrt2-1)^2\,\epsilon^2\,p}{64\,D^2 n}.
\]
\paragraph{Step 1. Rewriting in terms of $\eta$.}
Using the inequality $d^p < d/\eta$ and simplifying, one finds
\[
N = O\!\left(\frac{nd^4}{p\eta^{6}}\,
\log\!\frac{n}{\delta}\right).
\]
\paragraph{Step 2. Expanding $\eta$.}
Since
\[
\eta = c_1\frac{\epsilon^2 p}{D^2 n},\qquad 
c_1:=\frac{(\sqrt2-1)^2}{64},
\]
we obtain
\[
\frac{1}{p\eta^{6}}
= \frac{1}{c_1^{6}}\;\frac{D^{12} n^{6}}{\epsilon^{12}}\;\frac{1}{p^{7}}.
\]
\paragraph{Step 3. Eliminating $p$.}
Recall that in Appendix \ref{appendix-1} the self-consistent equation gives
\[
p = \Theta\!\left(\frac{\ln z}{\ln d}\right),\qquad
z := \ln d \cdot B,
\quad
B:=\frac{64\,n\,D^2}{(\sqrt2-1)^2\epsilon^2}.
\]
Define
\[
L := \ln z = \ln\!\Big(\ln d \cdot \tfrac{64\,n\,D^2}{(\sqrt2-1)^2\epsilon^2}\Big).
\]
Therefore
\[
\frac{1}{p^{7}} = \Theta\!\left(\frac{(\ln d)^{7}}{L^{7}}\right).
\]
Substituting this gives
\[
\frac{1}{p\eta^{6}}
= \Theta\!\left(
\frac{D^{12} n^{6}}{\epsilon^{12}}\;\frac{(\ln d)^{7}}{L^{7}}
\right).
\]

\paragraph{Step 4. Final expression.}
Collecting all terms, we obtain
\[
N = O\!\left(
\frac{D^{12} n^{7}}{\epsilon^{12}}\;\frac{d^4(\ln d)^{7}}{L^{7}}\,
\log\frac{n}{\delta}
\right),
\]
where
\[
L = \ln\!\Big(\ln d \cdot \tfrac{64\,n\,D^2}{(\sqrt2-1)^2\epsilon^2}\Big).
\]
\paragraph{Step 5. Simplified $\widetilde O$ form.}
Up to polylogarithmic factors, this can be expressed as
\[
N = \widetilde O\!\left(
\frac{D^{12} n^{7}}{\epsilon^{12}} d^4(\ln d)^{7}\log\!\frac{n}{\delta}\right).
\]
\appendixsection{Complexity Comparison}
\label{appendix-3}
Consider the two complexity expressions:  
\[
K = O\!\left(\frac{d^{2p}}{\epsilon^{2}} \, 
\log\frac{n}{\delta}\right),
\qquad
N = O\!\left(\frac{n d^{4p}}{p \eta^{2}} \,
\log\frac{n}{\delta}\right),
\qquad
\eta = \frac{(\sqrt{2}-1)^2 \,\epsilon^{2} p}{64 D^2 n}.
\]
We analyze the ratio
\[
\frac{N}{K} 
= O\!\left(\frac{\tfrac{n d^{4p}}{p \eta^{2}} }
{\tfrac{d^{2p}}{\epsilon^{2}}}\right)
= O\!\left(
\frac{n \epsilon^{2} d^{2p}}{p \eta^{2}}\right).
\]

Next, substitute the expression of $\eta$:
\[
\frac{N}{K} 
= O\!\left(\frac{n \epsilon^{2} d^{2p}}{p \cdot \frac{p^{2} \epsilon^{4}}{n^{2} D^{4}}}\right)
= \Theta\!\left( \frac{n^{3} D^{4} d^{2p}}{p^{3} \epsilon^{2}} \right).
\]
For typical parameter ranges, $d^{2p}$ grows exponentially in $p$ and dominates the polynomial $1/p^{3}$. Therefore, $N/K \to \infty$ as $n$ increases and $\epsilon \to 0$. Therefore,
\[
\frac{N}{K} \gg 1 \quad \implies \quad 
K = O\!\left(\frac{d^{2p}}{\epsilon^{2}}  \log\frac{n}{\delta}\right) 
< 
O\!\left(\frac{n d^{4p}}{p \eta^{2}}\log\frac{n}{\delta}\right) = N.
\]
In other words, the complexity bound $K$ is asymptotically dominated by $N$.

% \newpage
% \input{checklist.tex}
\end{document}